%% file: sizeShapeAparicioGargalloRoca.tex
\pdfoutput=1
\documentclass[preprint,12pt]{elsarticle}

\journal{Computer-Aided Design}
\usepackage{graphicx}
\usepackage{subfigure}
\usepackage{multicol}
\usepackage{algorithm}
\usepackage{amsmath}
\usepackage{amsfonts}
\usepackage{algpseudocode}
\usepackage{subfigure,enumerate}
\usepackage{mathrsfs}
\usepackage{dsfont}
\usepackage{hyperref}
\usepackage{tikz}
\usetikzlibrary {arrows.meta}
\DeclareMathOperator*{\argmin}{argmin}

\input{definitions.tex}
\graphicspath{{./figures/}}

\begin{document}
	
	\begin{frontmatter}
		
		\title{Defining metric-aware size-shape measures to validate and optimize curved high-order meshes}

		\author{Guillermo Aparicio-Estrems}
		\ead{guillermo.aparicio@bsc.es}
		\author{Abel Gargallo-Peir\'{o}}
		\ead{abel.gargallo@bsc.es}
		\author{Xevi Roca\corref{mycorrespondingauthor}}
		\cortext[mycorrespondingauthor]{Corresponding author}
		\ead{xevi.roca@bsc.es}
		\address{Barcelona Supercomputing Center, Carrer de Jordi Girona, 29, 31, Barcelona 08034, Spain}

		\begin{abstract}
			%%%
			We define a regularized size-shape distortion (quality) measure for curved high-order elements on a Riemannian space. To this end, we measure the deviation of a given element, straight-sided or curved, from the stretching, alignment, and sizing determined by a target metric. The defined distortion (quality) is suitable to check the validity and the quality of straight-sided and curved elements on Riemannian spaces determined by constant and point-wise varying metrics. The examples illustrate that the distortion can be minimized to curve (deform) the elements of a given high-order (linear) mesh and try to match with curved (linear) elements the point-wise stretching, alignment, and sizing of a discrete target metric tensor. In addition, the resulting meshes simultaneously match the curved features of the target metric and boundary. Finally, to verify if the minimization of the metric-aware size-shape distortion leads to meshes approximating the target metric, we compute the Riemannian measures for the element edges, faces, and cells. The results show that, when compared to anisotropic straight-sided meshes, the Riemannian measures of the curved high-order mesh entities are closer to unit. Furthermore, the optimized meshes illustrate the potential of curved $r$-adaptation to improve the accuracy of a function representation.
			%%%%
		\end{abstract}
		
		
	\end{frontmatter}
	
	%\linenumbers
	
	%% main text
	\section{Introduction}
	\label{sec:intro}
	Recently, there has been an increased interest to modify the coordinates and topology of a high-order mesh to match curved anisotropic solution features with high-order meshes. This interest has been awakened because these modified curved high-order meshes promise to reduce the error of the approximation to solution for the same number of degrees of freedom, especially when the solution has curved anisotropic features. To this end, existing interior mesh curving approaches exploit the non-constant Jacobian of high-order meshes to match the target curved anisotropic features of the solution using coordinate modifications \cite{dobrev2019target,sanjaya2016improving,coupez:BasisFrameworkHighOrderAnisotropicMeshAdaptation,marcona2017variational,zahr2020implicit} and local cavity modifications \cite{dobrev2021hr,rochery2021p2,zhangthesis,ekelschot2019parallel,feuillet2019embedded}.
	
	To exploit existent high-order goal-oriented \cite{yano2012,fidkowski2011review} and interpolation-oriented \cite{loseille2011continuous,coulaud:VeryHighOrderAnisotropic} error estimators, curved high-order mesh optimization approaches \cite{knupp2012introducing,rochery2021p2,zhang2018curvilinear,ekelschot2019parallel,sanjaya2016improving,aparicio2018defining,aparicio2022metricinterpolation,aparicio2023combining} consider an objective function that accounts for the discrete metric, an inner product represented by a positive-definite symmetric matrix, obtained from the error estimator. These approaches enforce either curved edges of length one (unitary) \cite{rochery2021p2,zhangthesis} or curved elements featuring the stretching and alignment (direction of the metric eigenvectors) of the target metric, the prescribed one, at a reference \cite{sanjaya2016improving,ekelschot2019parallel} or a physical
	\cite{aparicio2018defining,aparicio2022metricinterpolation,aparicio2023combining} mesh. Alternatively, instead of a metric, it is possible to match a pointwise target deformation matrix \cite{dobrev2019target,knupp2012introducing}.
	Although this alternative approach has proven to successfully adapt to solution curved meshes, it cannot be directly used to exploit discrete metrics obtained from error estimators. Hence, none of the previous approaches enforce unitary Riemannian measures for the mesh edges as well as for the face areas and the cell volumes.
%	Unfortunately, no approaches enforce unitary Riemannian measures for the mesh edges as well as for the face areas and the cell volumes.
	
	Enforcing unitary Riemannian metrics for all mesh entities is critical when the metric varies pointwise. Without this feature, the resulting mesh might not reduce the error as expected. This issue is so because the differential measure at each point of the curved high-order mesh might not match the stretching, alignment, and sizing of the prescribed pointwise metric.
	
	Accordingly, we aim to enforce unitary Riemannian measures for all the mesh entities. To this end, the main contribution of this work is to define a differentiable point-wise size-shape distortion measure that accounts for the stretching, alignment, and sizing of the target metric. Moreover, to check if the Riemannian measures are unitary, we detail how to compute metric-aware measures of the mesh entities \textit{i.e.}, Riemannian lengths, areas, and volumes accounting for the target metric. Finally, we verify whether minimizing the metric-aware size-shape distortion leads to meshes with Riemannian measures closer to unity for the element edges, faces, and cells.
	
	To define the differentiable metric-aware size-shape distortion, the main novelty is to propose a differentiable multiplicative combination of an existent metric-aware shape distortion \cite{aparicio2018defining} and a new differentiable metric-aware size distortion. Regarding size-shape distortion measures, there are related works for linear and curved-high-order meshes yet targeting a deformation matrix. For linear meshes, to obtain a distortion measure that accounts for shape and size, it is proposed to multiply a shape and a non-differentiable size distortion \cite{knupp:algebraicQuality}. The size distortion considers dilation volumes. For curved high-order meshes targeting a deformation matrix, existing differentiable distortion measures account for stretching, alignment, and sizing \cite{knupp2012introducing,dobrev2019target}. It is also possible to use a weighted sum of a shape and a reciprocal of a size quality surrogate that depends on a parameter. The quality surrogate considers a normalized difference of volume dilation and its reciprocal.
	The main difference between these approaches and our approach is that we use a target metric to exploit existent error estimators. Another difference is that our differentiable size distortion considers squares of the $d$ roots of a normalized summation of the volume dilation and its reciprocal.

	For metric-based applications, when compared to previous works, our approach has three important advantages. First, considering a metric instead of another object allows us to exploit existent high-order goal-oriented and interpolation-oriented error estimators. This extends the range of applicability of the metric-aware distortion measure. Second, our mesh distortion enforces unitary metric-aware measures of the mesh entities i.e., Riemannian lengths, areas, and volumes. This is important because to ensure that a mesh approximates a target metric, at least all Riemannian measures must be considered. Third, we enforce unitary measures not only in the elementwise sense but also in the pointwise sense. This has greater impact for pointwise varying metrics and meshes of high polynomial degree, where the flexibility of the non-constant Jacobian allows elements to match sharp and curved features imposed by the metric.		
	
	The rest of the paper is organized as follows. First, in Section \ref{sec:prelimaries}, we introduce the shape measures for high-order Euclidean elements. Next, in Section \ref{sec:linear}, we present the new size-shape measures for linear elements equipped with constant metrics.
	Then, in Section \ref{sec:pointwise}, we extend the size-shape measures to curved high-order elements equipped with point-wise varying metrics.
	Following, in Section \ref{sec:results}, we present several examples to illustrate the capabilities of the proposed measure. To finalize, in Section \ref{sec:conclusions}, we present the main conclusions and sum up the future work to develop.
	
	\section{Preliminaries: shape measures for high-order Euclidean elements}
	\label{sec:prelimaries}
	In this section, we present the Jacobian-based shape quality measures for linear and high-order elements defined in the Euclidean space \cite{knupp:algebraicQuality,XRNAGPJSR11:qualityPlanarHO,AGPXRNJPGJSR:IMR13ewc}. In addition, we introduce the required notation for Riemannian elements that is, elements equipped with a metric.
	
\begin{figure}[t]
		\centering
		\includegraphics[width=0.7\textwidth]{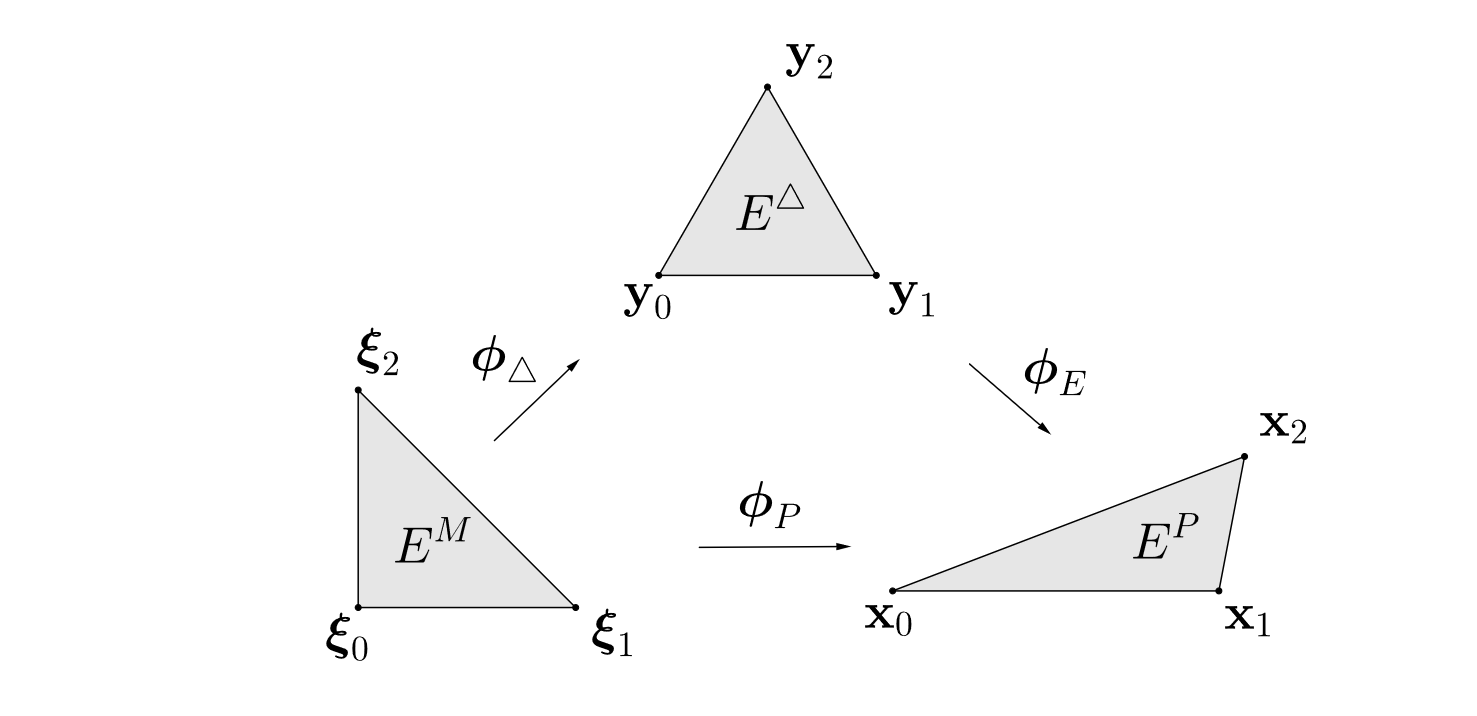}
		\caption{Mappings between the master, the ideal, and the physical elements in the linear case.}
		\label{fig:mappingRefIdealPhysical}
	\end{figure}

	To define and compute a Jacobian-based measure for linear Euclidean elements in $\mathds{R}^d$, three elements are required \cite{knupp:algebraicQuality}: the master, the ideal, and the physical, see Figure \ref{fig:mappingRefIdealPhysical} for 2D simplices. The master $(\zmaster)$ is the element from which the iso-parametric mapping is defined. The ideal element $\left(\zideal\right)$ represents the target configuration which, in the Euclidean case, is an equilateral element $\left(\zequilater\right)$. Since in the Euclidean case $\zideal = \zequilater$, we do not require $\zideal$ as an additional element in Figure \ref{fig:mappingRefIdealPhysical}. The physical $(\zphysical)$ is the element to be measured.
	
	First, we obtain the mappings between the ideal and the physical elements through the master element.
	By means of these mappings, we determine a mapping between the ideal and physical elements by the composition
	\begin{equation*}
		\zequilaterphysicalmap :\zequilater\xrightarrow{\zequilatermap^{-1}} \zmaster \xrightarrow{\zphysicalmap} \zphysical.
	\end{equation*}
	The Jacobian of the affine mapping $\zequilaterphysicalmap$, denoted by $\zJacobianidealphysical$, encodes the deviation of the physical element with respect to the equilateral one.
	
	We define the shape distortion measure $\zshape$ of the physical element as \cite{knupp:algebraicQuality}
	\begin{equation}\label{eq:shape}
		\zshape(\zJacobianidealphysical) = \frac{1}{d}\frac{\textbf{S}^2}{\sigma^{2/d}},
	\end{equation}
	where $\textbf{S}$ and $\sigma$ are the Frobenius norm and the determinant of $\zJacobianidealphysical$, respectively.
	This distortion measure quantifies the shape deviation between the physical and ideal elements.
	
	The matrix $\zJacobianidealphysical$ is computed for linear triangles as
	\begin{equation*}
		\zJacobianidealphysical = \zJacobianphysical\ \zJacobianequilater^{-1} = \left(\begin{array}{cc}
			x_1 - x_0 & \frac{2x_2 - x_1 - x_0}{\sqrt{3}}\\
			y_1 - y_0 & \frac{2y_2 - y_1 - y_0}{\sqrt{3}}
		\end{array}\right),
	\end{equation*}
	where
	\begin{equation}\label{eq:matrixcomputation}
		\zJacobianphysical = \left(\begin{array}{cc}
			x_1 - x_0 & x_2 - x_0\\
			y_1 - y_0 & y_2 - y_0
		\end{array}\right),\quad \text{and}\quad
		\zJacobianequilater = \left(\begin{array}{cc}
			1 & \frac{1}{2}\\
			0 & \frac{\sqrt{3}}{2}
		\end{array}\right),
	\end{equation}
	being $\zx_i = (x_i,y_i)$ the coordinates of the physical element $\zphysical$.
	These matrices are written for the master element $\zmaster$ with node coordinates
	$\left\lbrace \zxi_0 = \left(0,0\right),\right.$
	$\left.\zxi_1 = \left(1,0\right),\ \ \zxi_2 = \left(0,1\right)\right\rbrace$
	, and the ideal element $\zideal$ determined by the nodes
	$\left\lbrace\zy_0 = \left(0,0\right),\ \ \zy_1 = \left(1,0\right),\ \ \zy_2 = \left(1/2,\sqrt{3}/2\right)\right\rbrace$.
	
	The distortion measure, Equation \eqref{eq:shape}, quantifies the shape deviation between the physical and ideal shapes.
	The measure gets value 1 when the physical element is a scaled equilateral element. It is important to note that it is invariant under translations, rotations, and symmetries. Moreover, it can be regularized by enforcing infinite values for non-positive Jacobians \cite{AGPXRNJPGJSR:IMR13ewc,gargallo:generation3Doptimization,gargallo:highorderSurfaces}, so it detects inverted elements.
	From the distortion measure, we define the shape quality measure of an element as
	\begin{equation}\label{eq:qualityshape}
		q_{\mathrm{shape}} = \frac{1}{\zshape},
	\end{equation}
	which takes values in the interval $[0,1]$, being 0 for degenerated elements and 1 for the ideal element and its symmetric analogs.
	
	For high-order \cite{gargallo:generation3Doptimization,AGPXRNJPGJSR:IMR13ewc,gargallo:highorderSurfaces} and multi-linear \cite{gargallo:2015hexesResearchNote} elements $\zphysical$ with non-constant Jacobian, we reinterpret a distortion measure $\eta$ as a point-wise measure $\zdistortionoperator$ as advocated in  \cite{knupp2012introducing,roca2012defining}. In particular, we define
	\begin{equation*}
		\zdistortionoperator(\zy) := \eta(\zJacobianidealphysical(\zy)),\quad \forall \zy\in\zequilater.
	\end{equation*}
	Furthermore, we define the elemental distortion \cite{XRNAGPJSR11:qualityPlanarHO,AGPXRNJPGJSR:IMR13ewc} as
	\begin{equation}\label{eq:elemental}
		\eta_{\zphysical}:= \frac{\int_{\zequilater} \zdistortionoperator (\zy)\ d\zy}{\int_{\zequilater} 1\ d\zy},
	\end{equation}
	and its quality $q_{\zphysical}$ follows from Equation \eqref{eq:qualityshape}.
	\section{Size and size-shape measures for linear elements and constant metric}
	\label{sec:linear}
	
	Herein, we present the measures for linear elements equipped with constant metrics.
	First, in Section \ref{sec:size}, we define a quality measure that quantifies the size deviation of Euclidean elements.
	Second, in Section \ref{sec:sizeshape}, we extend the quality measure to linear simplices equipped with a constant metric.
	Finally, in Section \ref{sec:behavior}, we illustrate the behavior of the proposed measure.
	
	\subsection{Differentiable size and size-shape distortion for linear Euclidean elements}
	\label{sec:size}
	
	The shape distortion measure of Section \ref{sec:prelimaries} quantifies the shape deviation between the physical and ideal elements.
	However, it does not take into account the size deviation between the physical and ideal elements.
	For this reason, we define an additional distortion measure that takes into account sizing.
	In particular, we define the size distortion measure $\zsize$ of the physical element as
	\begin{equation}\label{eq:size}
		\zsize(\zJacobianidealphysical) = \left(\frac{1}{2}\left(\sigma + \frac{1}{\sigma} \right)\right)^{2/d},
	\end{equation}
	where $\sigma := \det\left( \zJacobianidealphysical \right)$.
	\begin{figure}[t!]
		\centering
		\setlength{\tabcolsep}{-10pt}
		\begin{tabular}{cc}
			\subfigure[]{\label{fig:etasize}
				\includegraphics[width=0.55\textwidth]
				{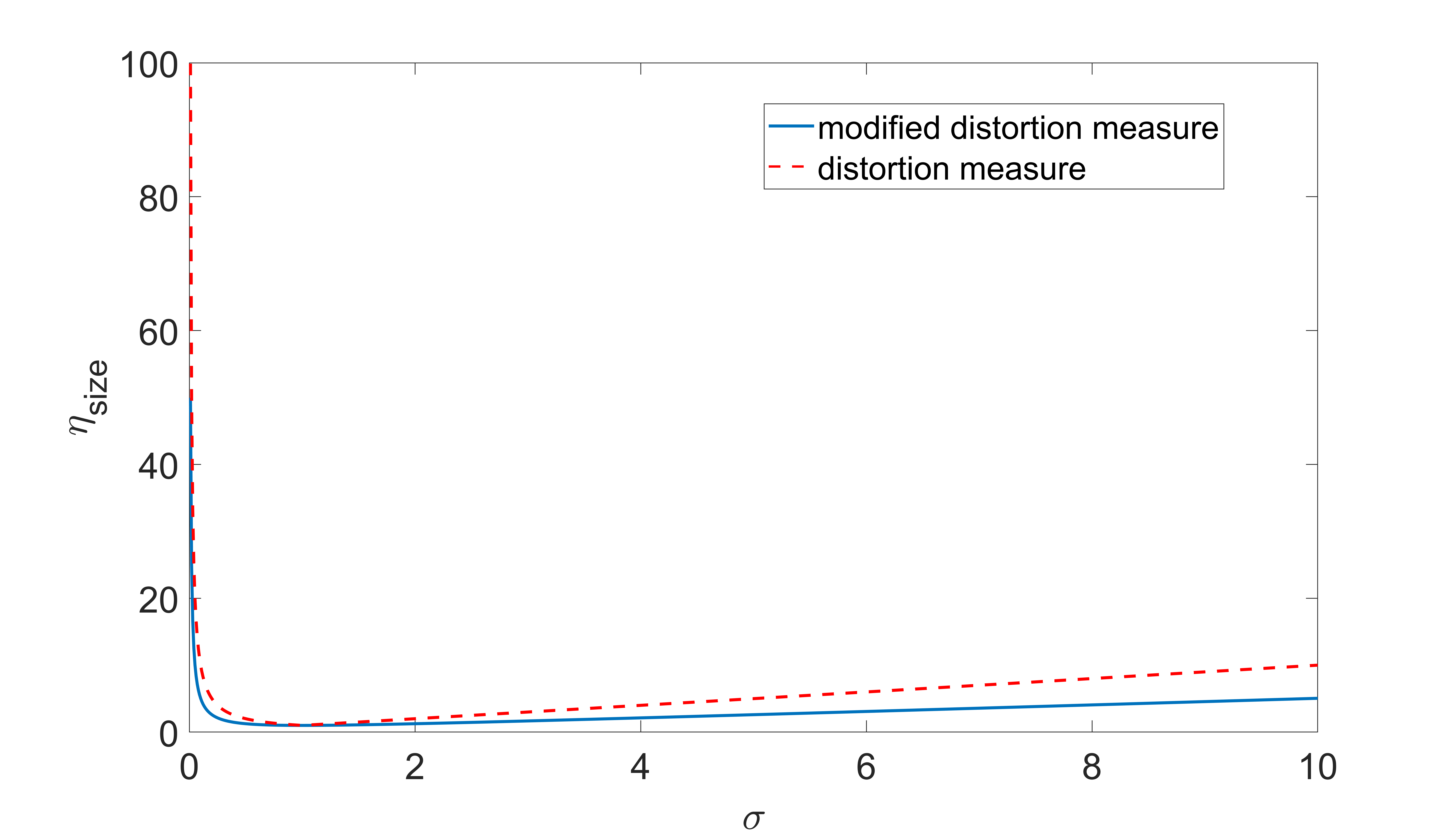}
			}
			&
			\subfigure[]{\label{fig:qsize}
				\includegraphics[width=0.55\textwidth]
				{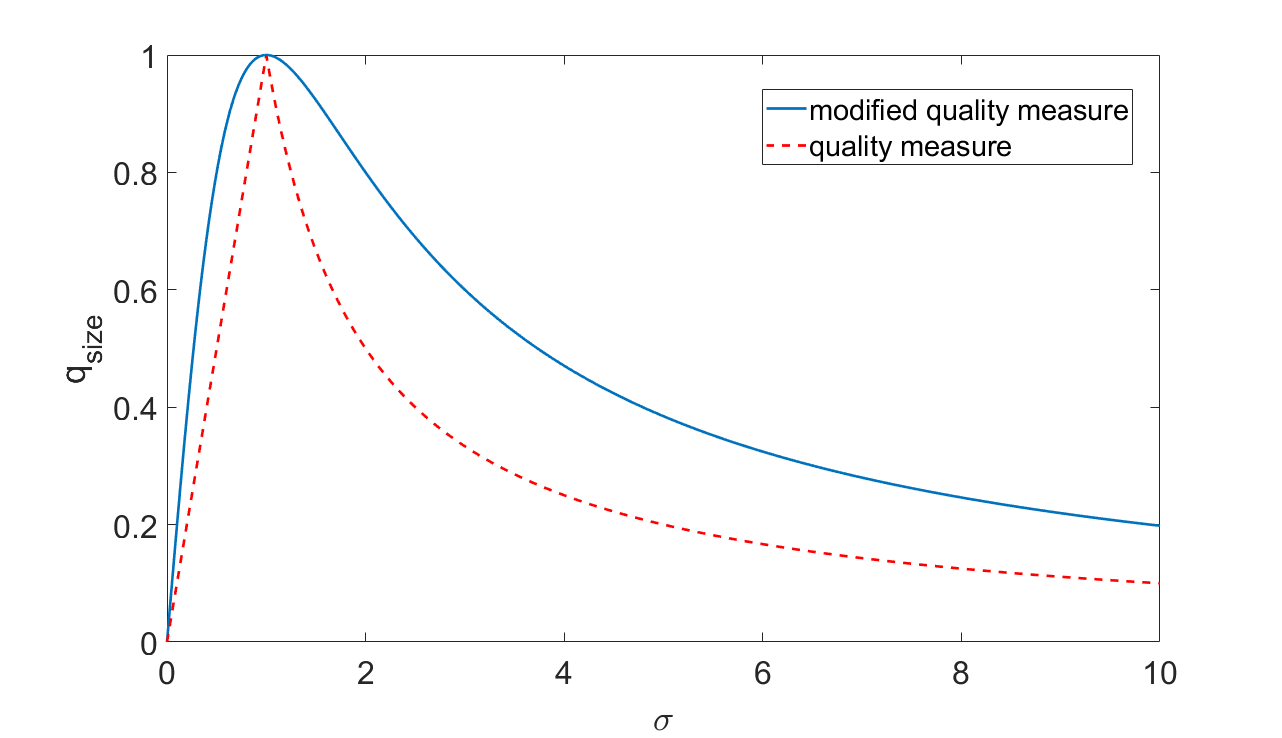}  	 }
			\\
		\end{tabular}
		\caption{Plots of the original and modified: (a) size distortion measure and (b) size quality measure.}
		\label{fig:sizeplot}
	\end{figure}
	This distortion measure quantifies the size deviation between the physical and ideal elements.
	We expect the size distortion measure to behave as
	\begin{equation*}
		\mu(\sigma) = \max(\sigma,\sigma^{-1})^{2/d}.
		%= e^{\frac{2}{d}|\log{\sigma}|}.
	\end{equation*}
	Note that, the base $\max(\sigma,\sigma^{-1})$ is the standard size measure of \cite{knupp:algebraicQuality}.
	
	However, we cannot use the function $\mu$ in a continuous optimization procedure since it is not differentiable. To overcome this drawback, we propose to replace $\mu(\sigma)$ by the size distortion measure, see Equation \eqref{eq:size}, a continuous and differentiable function that holds the same minimum and the same asymptotic behavior.
	
	Figure \ref{fig:sizeplot} shows the size distortion and the size quality measures using the original, $\mu(\sigma)$, and the modified function, $\zsize$, in terms of $\sigma$.
	It is worth to notice that using the modification presented in Equation \eqref{eq:size}, the size distortion measure $\zsize$ is still a distortion measure that is, orientation-invariant, positive, and transpose-invariant \cite{knupp:algebraicQuality}.
	
	Finally, we define the distortion measure $\eta$ of the physical element by
	\begin{equation}\label{eq:distortion}
		\eta(\zJacobianidealphysical) = \zshape(\zJacobianidealphysical)\ \zsize(\zJacobianidealphysical).
	\end{equation}
	The distortion measure combines $\zshape$ and $\zsize$, see \cite{knupp:algebraicQuality} for more details. Thus, it quantifies both the size and the shape of the element.
	
%	Note that the size-shape distortion measure is applicable to all types of high-order elements. This is because they are based on high-order iso-parametric mappings characterizing the shape of an element and of an equilateral configuration.
%	Although in this paper we only implement the size-shape distortion measure for simplicial elements, note that it is applicable to all types of high-order element shapes. This is so because the distortion only depends on a high-order iso-parametric mapping characterizing the shape of an element from a regular configuration.
	
	\subsection{Size-shape distortion for linear elements and constant metric}
	\label{sec:sizeshape}
	
	%To define a measure that quantifies the quality of a given element, we need to define an ideal element that represents the desired configuration, as detailed in Section \ref{sec:prelimaries}.
	%In the unitary-Euclidean case, where the metric $\zmetric$ is represented by the identity matrix $\textbf{Id}$, the ideal element is the equilateral element with edges of unit length $\zequilater$. For non-unitary metrics, we describe how to obtain the ideal configuration. Then, we measure the distortion of the physical element by comparing it with the ideal element.
	
	To define a measure that quantifies the quality of a given element, we need to define an ideal element that represents the desired configuration, as detailed in Section \ref{sec:prelimaries}.
	In the unitary-Euclidean case, where the metric $\zmetric$ is represented by the identity matrix $\textbf{Id}$, the ideal element $\zideal$ corresponds to the equilateral element $\zequilater$, the one with unit length edges. For non-unitary metrics, we describe how to obtain the ideal configuration. Then, we measure the distortion of the physical element by comparing it with the ideal element.
	
\begin{figure}
		\centering
		\includegraphics[width=1.0\textwidth]{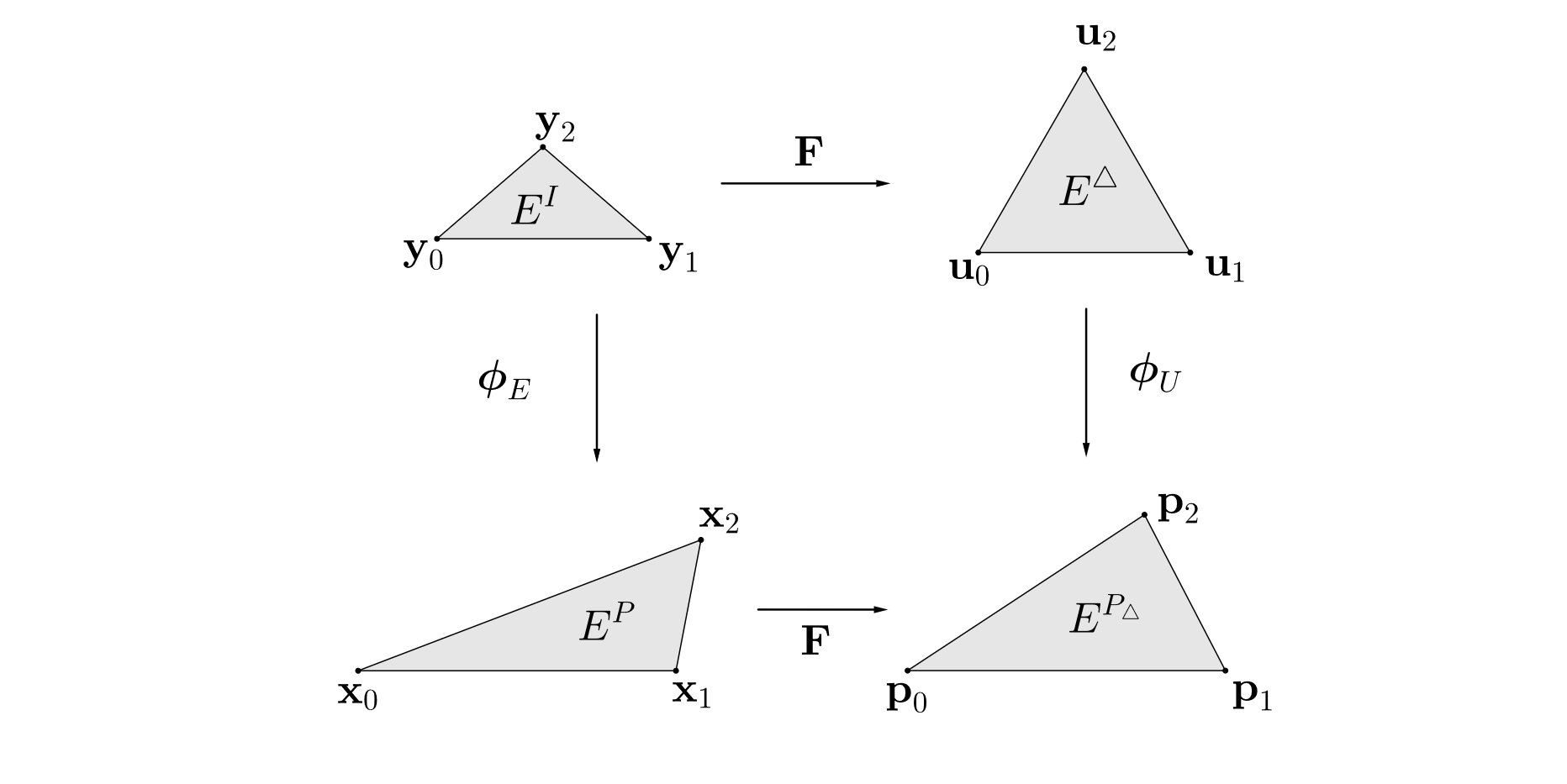}
		\caption{Mappings between the equilateral, the ideal, the physical, and the unitary physical triangles.}
		\label{fig:xevisDiagram1}
	\end{figure}

	We define the ideal element as the element with edges of unit length under the desired metric. To compute this configuration, we first decompose $\zmetric$ as follows
	\begin{equation}\label{eq:metricdecompositionvar}
		\zmetric = \zfield^\mathrm{T}\ \zfield.
	\end{equation}
	Matrix $\zfield$ can be interpreted as a linear mapping between the space with metric $\zmetric$ and the space with unitary metric $\textbf{Id}$. 
	Thus, we define the anisotropic ideal $\zideal$ as the preimage by $\zfield$ of the equilateral element, see Figure \ref{fig:xevisDiagram1}.
	In particular, let $\zu_i,\ i = 0,1,2$ be the nodes of the equilateral element $\zequilater$.
	Then, we define the nodes $\zy_i,\ i = 0,1,2$ of the ideal element $\zideal$ as
	\begin{equation*}\label{eq:idealnodes}
		\zy_i = \zfield^{-1}\ \zu_i,\ \ i = 0,1,2.
	\end{equation*}
	A direct consequence of the above definition is that the ideal triangle has unit edge lengths in the metric sense.
	Once the ideal triangle is defined, we measure the deviation between the ideal and physical elements.
	Similarly to the approaches for a unitary metric, see Section \ref{sec:prelimaries}, in this section we define the distortion between the ideal $\zideal$ and physical $\zphysical$ elements in terms of the mapping between those elements, $\zidealphysicalmap$.
	
	A priori, we do not know how to compare elements considering the target metric.
	Nevertheless, we know how to compare elements in the unitary sense, see Section \ref{sec:prelimaries}, and thus, we map both elements $\zideal$ and $\zphysical$ to the same Euclidean space using $\zfield$, see Figure \ref{fig:xevisDiagram1}. 
	Then, we compare the image elements $\zequilater$ and $\zisotropicphysical$ using the distortion measure presented in Equation \eqref{eq:distortion}.
	
	Let $\zisotropicphysical$ be the image of the physical triangle $\zphysical$ by $\zfield$.
	By construction, the image by $\zfield$ of the ideal triangle is the equilateral triangle. 
	We measure the distortion between the ideal $\zideal$ and physical $\zphysical$ elements in terms of the distortion of the mapping between the $\zequilater$ and $\zisotropicphysical$.
	
	Finally, we define the distortion between the physical triangle $\zphysical$ and the ideal triangle $\zideal$ with respect to the desired metric as the distortion of the matrix $\zJacobianequilaterisotropicphysical$:
	\begin{equation}\label{eq:distortionAnisoPhiU}
		\eta_{\zmetric}(\zJacobianidealphysical) := \eta(\zJacobianequilaterisotropicphysical).
		%= \eta_{\zmetric}(\zfield^{-1} \ \zJacobianequilaterisotropicphysical \ \zfield)
	\end{equation}
	
	%\begin{remark}
	\label{rem:rotationInvariant}
	The distortion presented in Equation \eqref{eq:distortionAnisoPhiU} is well defined.
	This is because the measure does not depend on the symmetries of $\zisotropicphysical$.
	We show first the case for rotations. The rotation of angle $\theta$ of $\zisotropicphysical$ is the triangle $\tilde{E}^{P_\triangle}$ composed by the nodes $\tilde{\zy}_i = \zrotation\ \zy_i,\ i = 0,1,2$. Then	
	\begin{equation*}
		\textbf{D}\tilde{\zphi}_{U} = \zrotation \ \zJacobianequilaterisotropicphysical,
	\end{equation*}	
	where $\tilde{\zphi}_{U}$ is the mapping between the equilateral triangle $\zequilater$ and $\tilde{E}^{P_\triangle}$.
	Consequently, we have	
	\begin{align}\label{eq:rotationInvariance}
		\textbf{D}\tilde{\zphi}_{U}^\mathrm{T}\ \textbf{D}\tilde{\zphi}_{U} = \zJacobianequilaterisotropicphysical^\mathrm{T} \ \zrotation^{\mathrm{T}}\ \zrotation  \ \zJacobianequilaterisotropicphysical = \zJacobianequilaterisotropicphysical^\mathrm{T} \  \zJacobianequilaterisotropicphysical.
	\end{align}
	From Equations \eqref{eq:rotationInvariance}, \eqref{eq:distortionAnisoPhiU}, and \eqref{eq:distortion} we conclude that the corresponding distortions are equal.
	The case for reflections follows analogously since any symmetry $\boldsymbol{\Sigma}$ satisfies that $\boldsymbol{\Sigma}^\mathrm{T}\  \boldsymbol{\Sigma} = \zid$.
	%\end{remark}
	\begin{figure}
		\centering
		\includegraphics[width=0.8\textwidth]{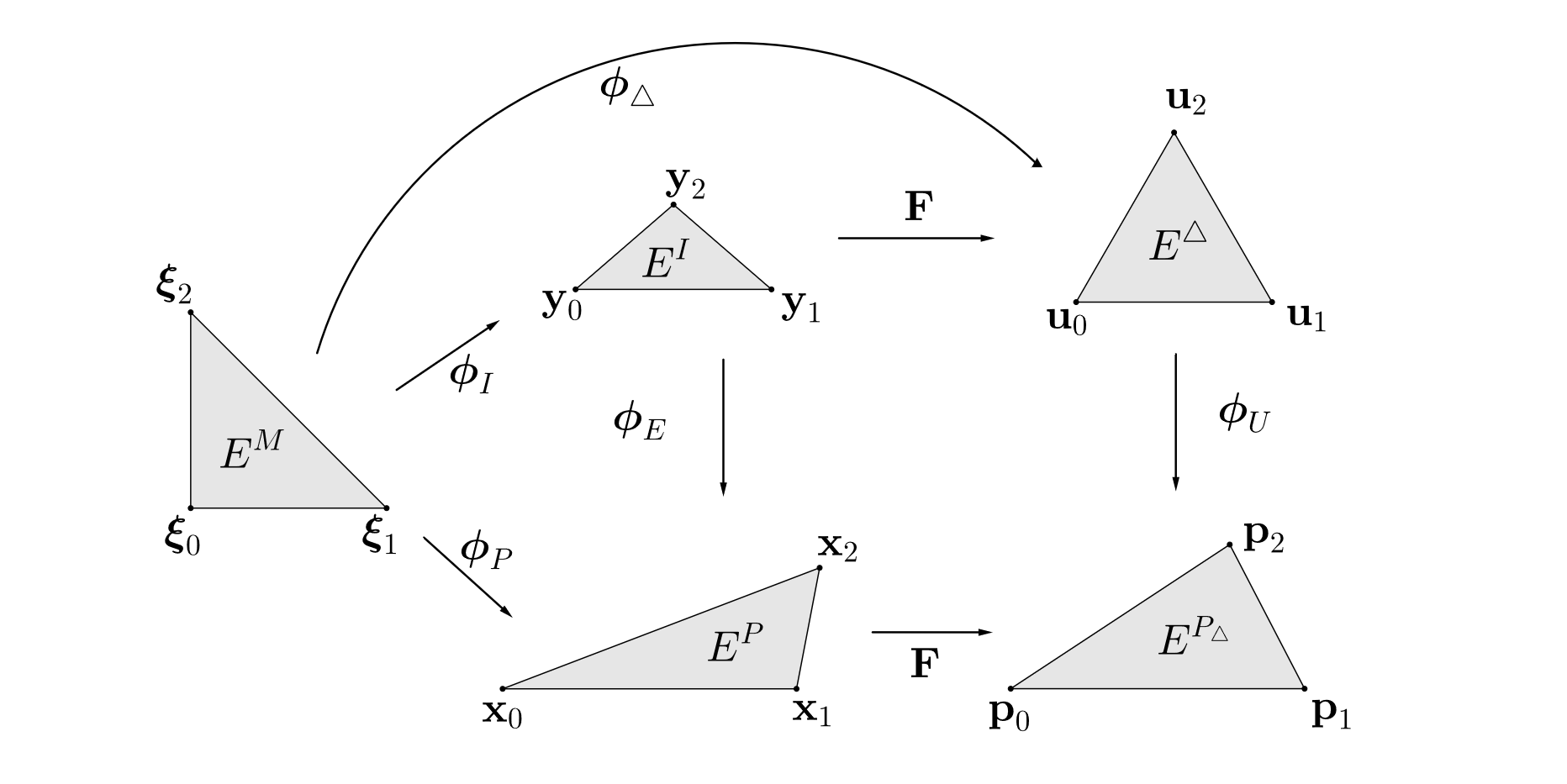}
		\caption{Mappings between the master, the equilateral, the ideal, the physical, and the unitary physical triangles.}
		\label{fig:xevisDiagram2}
	\end{figure}
	%\label{rem:noDecomposition}
	%To compute the quality measure in Equation \eqref{eq:distortionAnisoPhiU} the explicit decomposition of the metric shown in Equation \eqref{eq:metricdecompositionvar} is not required.
	
	Next, we show how to compute the distortion presented in Equation \eqref{eq:distortionAnisoPhiU} without decomposing it using matrix $\zfield$.
	First, in Figure \ref{fig:xevisDiagram2}, we include the master element in the diagram of mappings of Figure \ref{fig:xevisDiagram1}.
	Let $\zequilatermap$ be the mapping between the master and the equilateral triangle.
	This mapping is equivalent to the composition of the mappings $\zidealmap$ and $\zfield$, but it can be directly computed from the coordinates of the master and equilateral triangles, as previously done for the isotropic case in Section \ref{sec:prelimaries}.
	Taking into account the computation of $\zJacobianequilater$ in terms of the node coordinates in Equation \eqref{eq:matrixcomputation}, the distortion measure $\eta_{\zmetric}(\zJacobianidealphysical) $ can be rewritten without decomposing $\zmetric$.
	Note that, \textit{a priori}, the right-hand side in Equation \eqref{eq:distortionAnisoPhiU} depends on $\zfield$ since
	\begin{equation}\label{eq:mappingF}
		\zJacobianequilaterisotropicphysical = \zJacobianisotropicphysical \ \zJacobianequilater^{-1} = \zfield \ \zJacobianphysical \ \zJacobianequilater^{-1}.
	\end{equation}
	Manipulating Equation \eqref{eq:distortionAnisoPhiU}, one realizes that there is no explicit dependence on $\zfield$:
	\begin{align*}
%		\label{eq:simplificationF}
		\zJacobianequilaterisotropicphysical^\mathrm{T}\ \zJacobianequilaterisotropicphysical =&
		\left(\zJacobianequilater\right)^{-\mathrm{T}}\ \zJacobianphysical^\mathrm{T}\
		\zfield^\mathrm{T}\ \zfield\ \zJacobianphysical\ \left(\zJacobianequilater\right)^{-1}\\
		=& \left(\zJacobianequilater\right)^{-\mathrm{T}}\ \zJacobianphysical^{\mathrm{T}}\
		\zmetric\ \zJacobianphysical\ \left(\zJacobianequilater\right)^{-1}.
	\end{align*}
	Thus, we obtain an expression for the distortion that does not require to decompose the metric $\zmetric$.
	In particular, we define the a Riemannian analog for the Frobenius norm $\zSmetric$ and determinant $\zsigmametric$ as follows
	%\begin{equation*}\label{eq:frob&size}
	%\zSmetric:=\sqrt{\ztr\left(\zconjugate\right)},\quad \textrm{and}\quad \zsigmametric:=\sqrt{\det\left(\zconjugate\right)}.
	%\end{equation*}
	\begin{align}\label{eq:frob&size}
		\zSmetric&:=\sqrt{\ztr\left(\zconjugate\right)},\quad \textrm{and}\\
		\zsigmametric&:=\sqrt{\det\left(\zconjugate\right)}.
	\end{align}
	Finally, analogously to the Euclidean size-shape distortion measure of Equation \eqref{eq:distortion}, we define the Riemannian size-shape distortion as
	\begin{equation}\label{eq:distortionmetric}
		\eta_{\zmetric}(\zJacobianidealphysical) = \eta_{\zmetric,\mathrm{shape}}(\zJacobianidealphysical) \ \eta_{\zmetric,\mathrm{size}}(\zJacobianidealphysical),
	\end{equation}
	where the corresponding shape, $\eta_{\zmetric,\mathrm{shape}}$, and size, $\eta_{\zmetric,\mathrm{size}}$, distortion measures are given by
	\begin{equation*}
		\eta_{\zmetric,\mathrm{shape}}(\zJacobianidealphysical) = \frac{1}{\zdim}\frac{\zSmetric^2}{\zsigmametric^{2/\zdim}}, \quad \textrm{and}\quad \eta_{\zmetric,\mathrm{size}}(\zJacobianidealphysical) = \left(\frac{1}{2}\left(\zsigmametric + 
		\frac{1}{\zsigmametric}\right)\right)^{2/\zdim}.
	\end{equation*}
	
	\subsection{Behavior of the quality measures: shape, size, and size-shape}
	\label{sec:behavior}
	
	In this section, we illustrate the behavior of the shape quality measure corresponding to the distortion measure, presented in Equation \eqref{eq:distortionAnisoPhiU}, for linear anisotropic triangles equipped with a constant metric. 
	We first show the level curves of the quality measure of a triangle when we fix two nodes and we let the third node to move in $\zR^2$, in Section \ref{sec:levelsets}.
	Second, in Section \ref{sec:alignment}, we analyze the behavior of the measure with respect to the alignment of the element with the metric.
	
	\subsubsection{Level sets for one moving vertex}
	\label{sec:levelsets}
	
	To show the behavior of the level curves of the shape, size, and size-shape quality measures we consider two cases, the Euclidean or isotropic case when $\zmetric = \textbf{Id}$ and the anisotropic case when $\zmetric$ has two different eigenvalues.
	
	For each quality measure and each metric, we illustrate the behavior of the quality measure by plotting the level sets in terms of a free node of the triangle. 
	We consider the anisotropic metric given by
	\begin{equation}\label{eq:metricbehavior}
		\zmetric = \left(\begin{array}{cc}
			1&0\\0&\frac{1}{h^2}
		\end{array}\right),\quad h = 1/3.
	\end{equation}
	This metric is aligned with the canonical axes and features a stretching ratio of 1 against 3. Specifically, it is devised to ensure that vectors $(1,0)$ and $(0,h)$ have unit length under the metric. The ideal element $\zideal$ is expected to be an element of height $h$ and base 1.
	In each test, we consider a free node, keeping the rest of nodes fixed at their original location, and we compute the quality of the element in terms of the location of this node. The free node considered is the vertex node $\zx_2$.
	
	\begin{figure*}[t!]
		\centering
		\begin{tabular}{ccc}
			\subfigure[]{\label{fig:shapecontour}
				\includegraphics[width=0.25\textwidth]
				{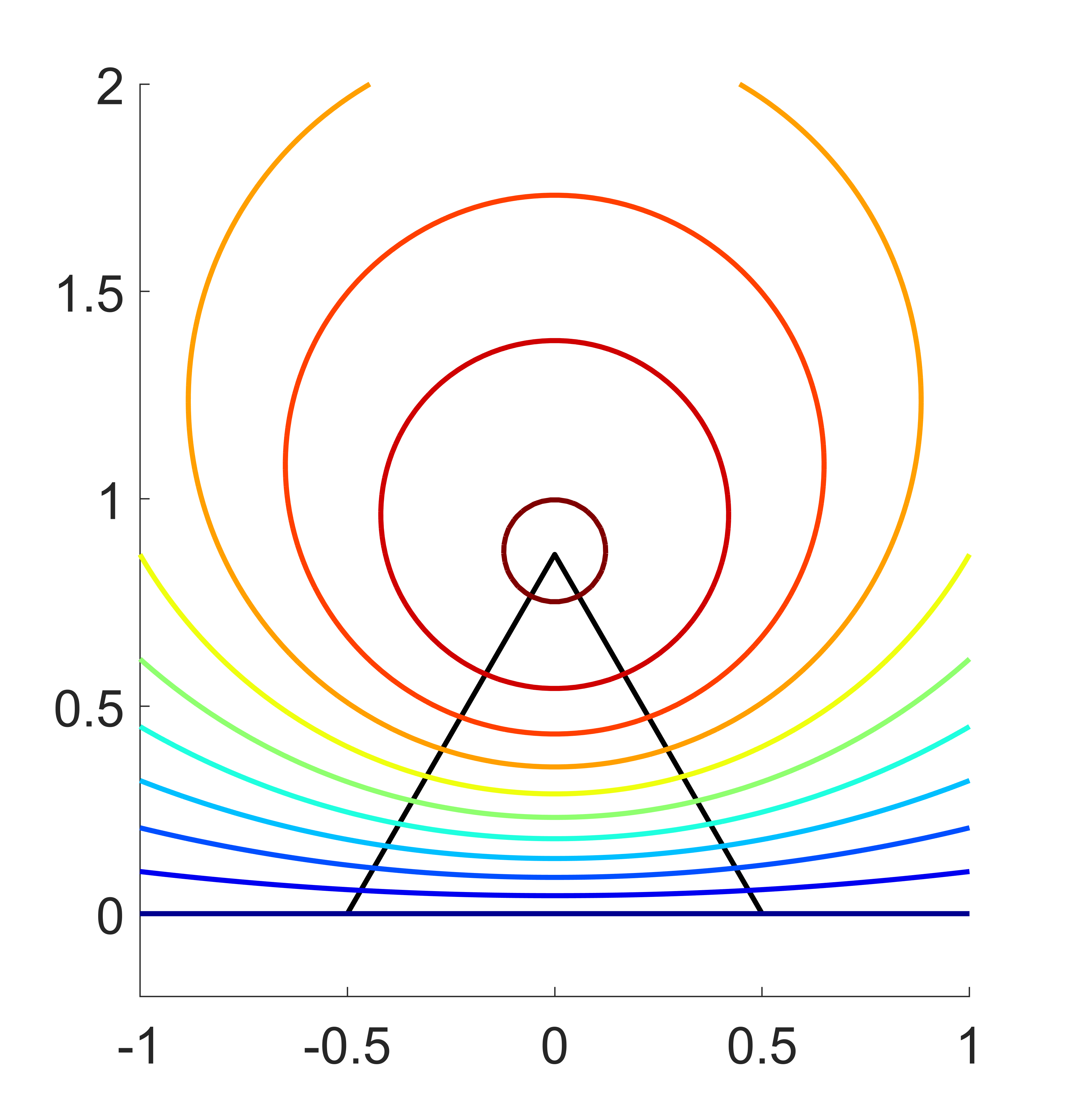}}
			&
			\subfigure[]{\label{fig:sizecontour}
				\includegraphics[width=0.25\textwidth]
				{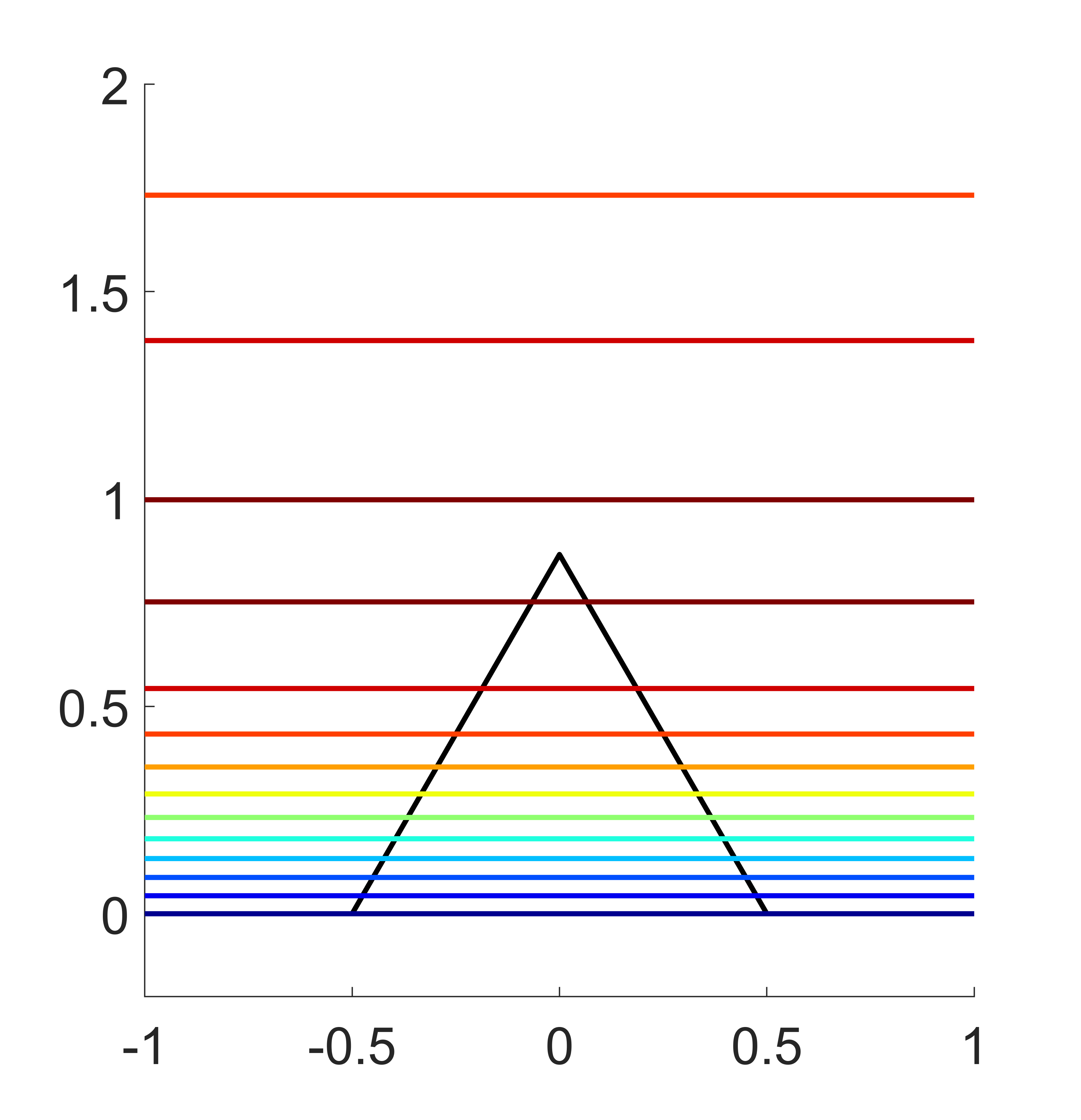}}
			&
			\subfigure[]{\label{fig:sizeshapecontour}
				\includegraphics[width=0.25\textwidth]
				{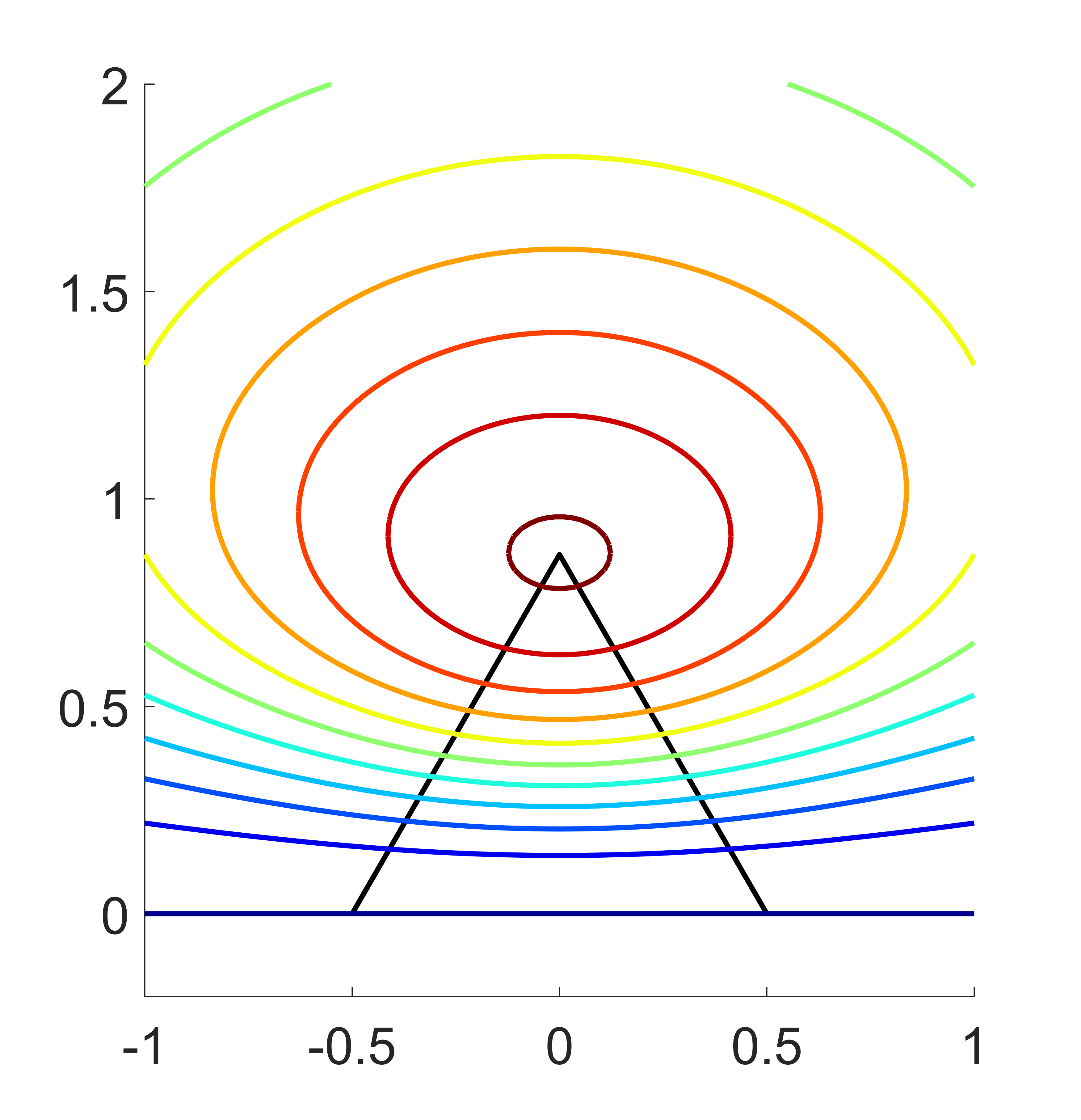}}
			\\
			\subfigure[]{\label{fig:shapemetriccontour}
				\includegraphics[width=0.25\textwidth]
				{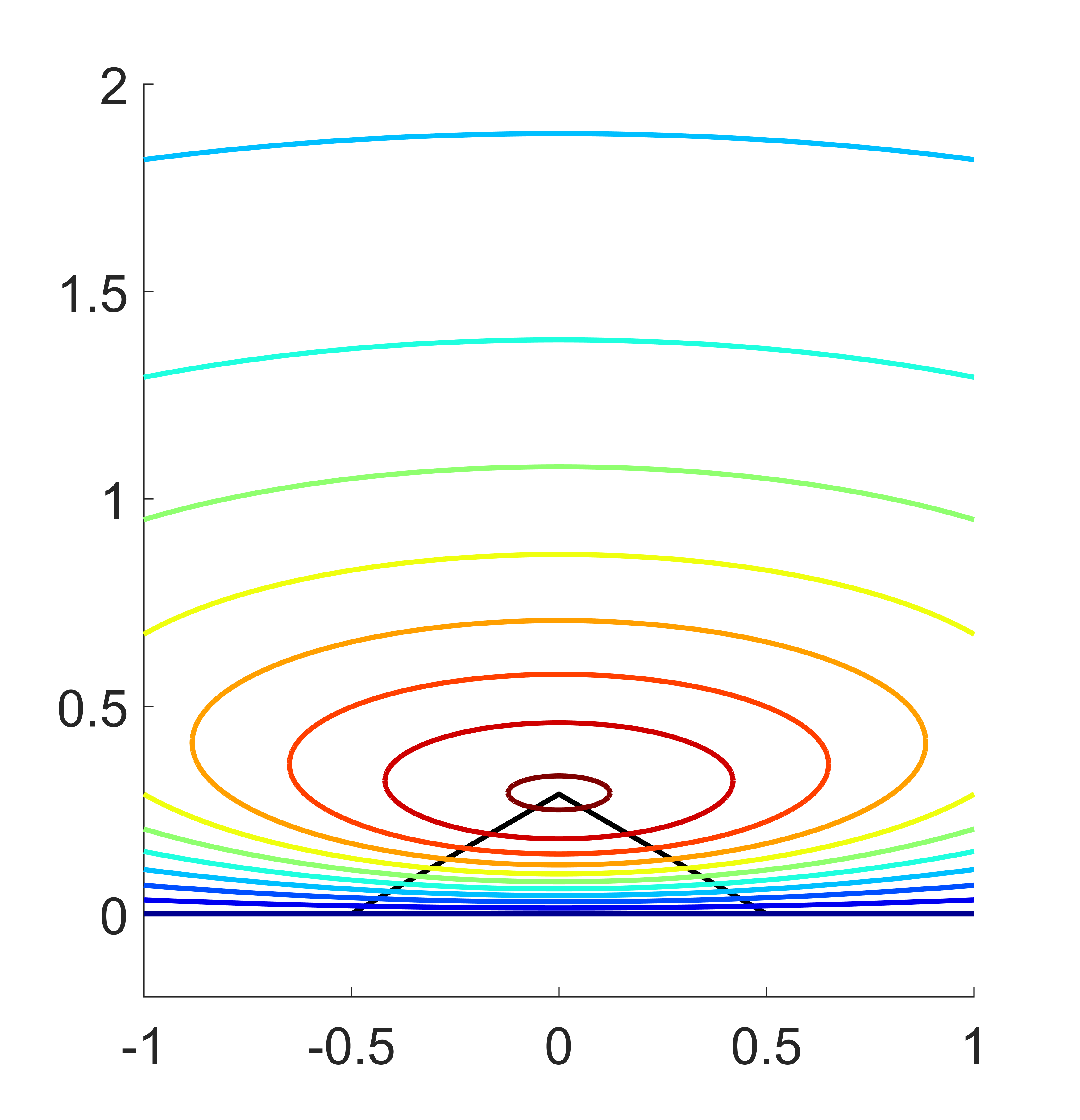}}
			&
			\subfigure[]{\label{fig:sizemetriccontour}
				\includegraphics[width=0.25\textwidth]
				{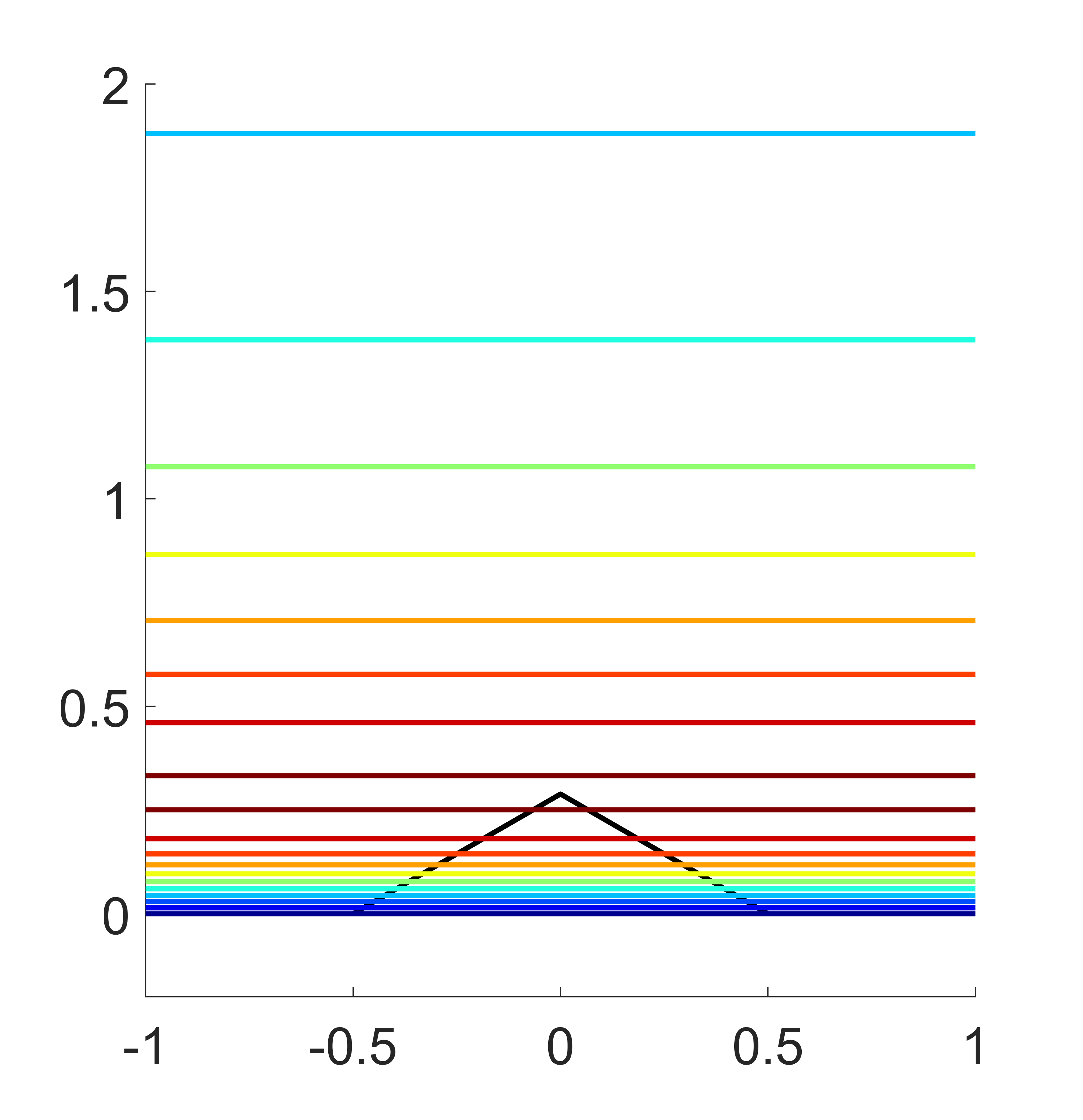}}
			&
			\subfigure[]{\label{fig:sizeshapemetriccontour}
				\includegraphics[width=0.25\textwidth]
				{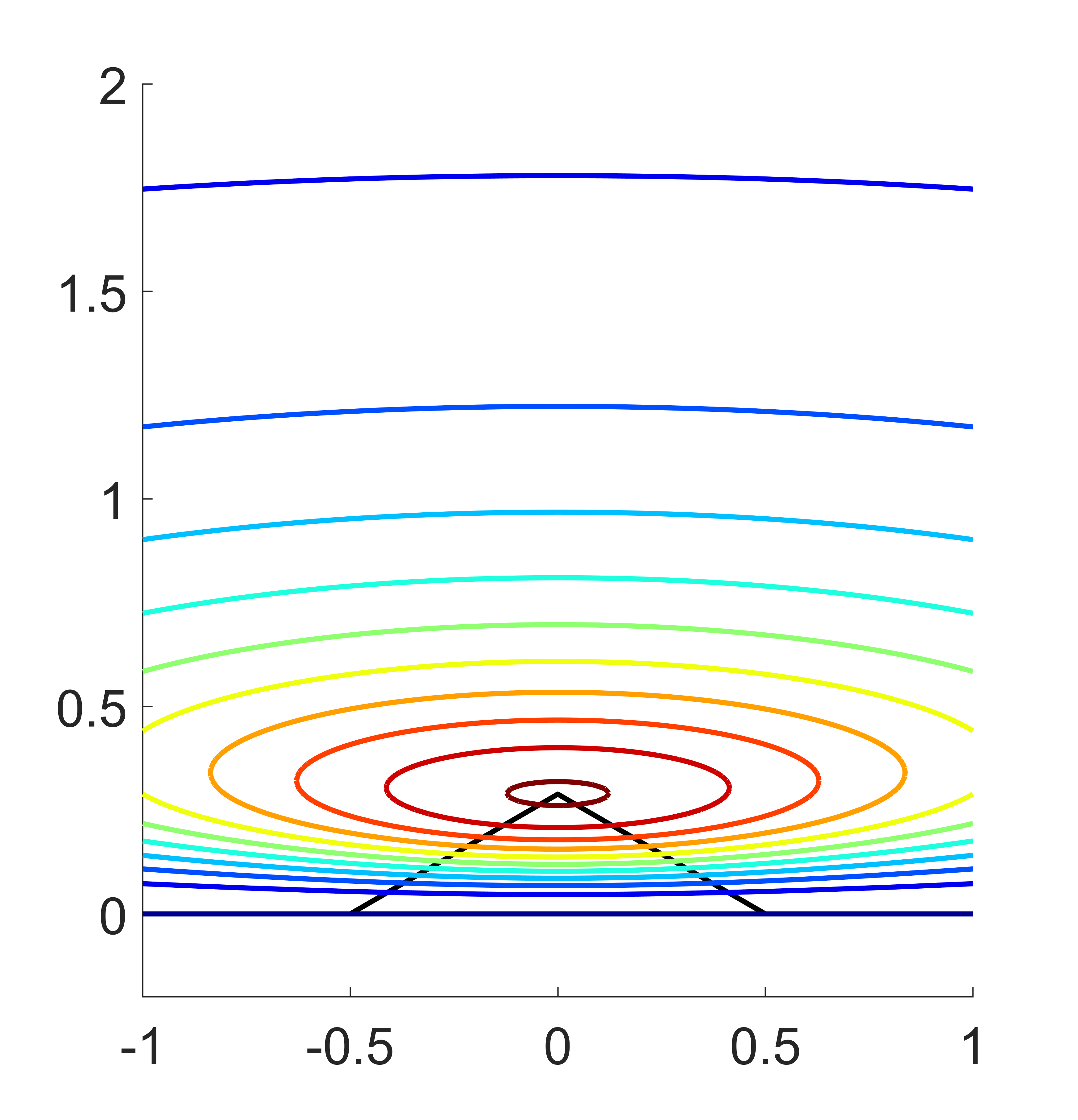}}
			\\
		\end{tabular}
		\\
		\includegraphics[width=0.25\textwidth]
		{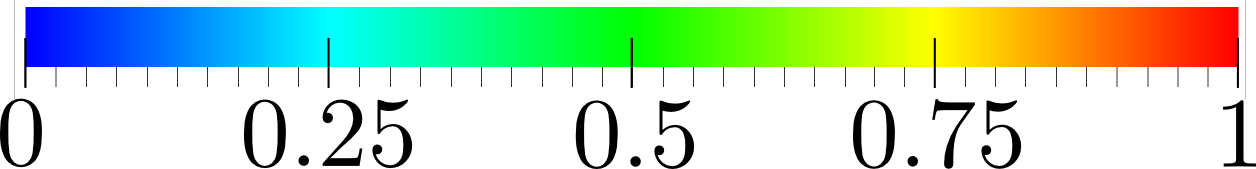} 
		\caption{Level sets for the quality measures with different metrics: \textbf{(a,d)} shape, \textbf{(b,e)} size, and \textbf{(c,f)} size-shape; \textbf{(a,b,c)} isotropic and \textbf{(d,e,f)} anisotropic metrics.}
		\label{fig:contour}
	\end{figure*}
	In Figure $\ref{fig:contour}$, we show the contour plots of the quality for each test when the free node is allowed to move in a region of $\zR^2$. The locus of the points where the element has positive Jacobian, the feasible region, is independent of the metric and corresponds to the half-plane $y>0$. 
	
	As expected, for each metric the optimal node location is different. Furthermore, we can observe that the level sets and the height of the ideal triangle corresponding to the metric of Equation \eqref{eq:metricbehavior} are more stretched than in the isotropic case. Similarly, the level sets of the quality measure become more stretched as the anisotropy of the metric increases.
	
	Similarly, for each quality measure, the optimal node location is also different. First, for the shape quality, the level curves are circular in the Euclidean case and elliptic in the metric case. Second, for the size quality, we observe that the level sets are straight horizontal lines. This is because the size quality depends only on the height of the triangle since the base is fixed. In the metric case, the spacing between the straight lines are more stretched than in the Euclidean case. Third, for the size-shape quality, the level curves are more stretched in the metric case than in the Euclidean case. Moreover, we observe that the level curves of the size-shape quality are more stretched than the ones of the shape quality. This indicates that the size-shape quality is more restrictive, in terms of variation, than the shape one.
	
	\subsubsection{Influence of element alignment}
	\label{sec:alignment}
	
	Next, we illustrate how the quality measure depends on the alignment between the anisotropy axes and the element. 
	We compute the quality measure of a sequence of physical elements generated rotating the ideal element. We consider the metric presented in Equation \eqref{eq:metricbehavior}.
	
	Let $\zrotation$ be the rotation at the origin of angle $\theta\in[0,2\pi)$ which is given by
	\begin{equation*}
		\zrotation = \left(\begin{array}{cc}
			\cos{\theta} &\  -\sin{\theta}\\
			\sin{\theta} &\  \cos{\theta}
		\end{array}\right).
	\end{equation*}
	We define the physical element as the ideal element rotated $\theta$ radians, with nodes  $\zx_i = \zrotation\  \zy_i,\ i = 0,1,2$. For each $\theta$ we compute the quality of the corresponding physical element.
	
	\begin{figure}[t!]
		\centering
		\begin{tabular}{ccccc}
			\subfigure[]{\label{fig:rotation0}
				\includegraphics[width=0.1\textwidth]
				{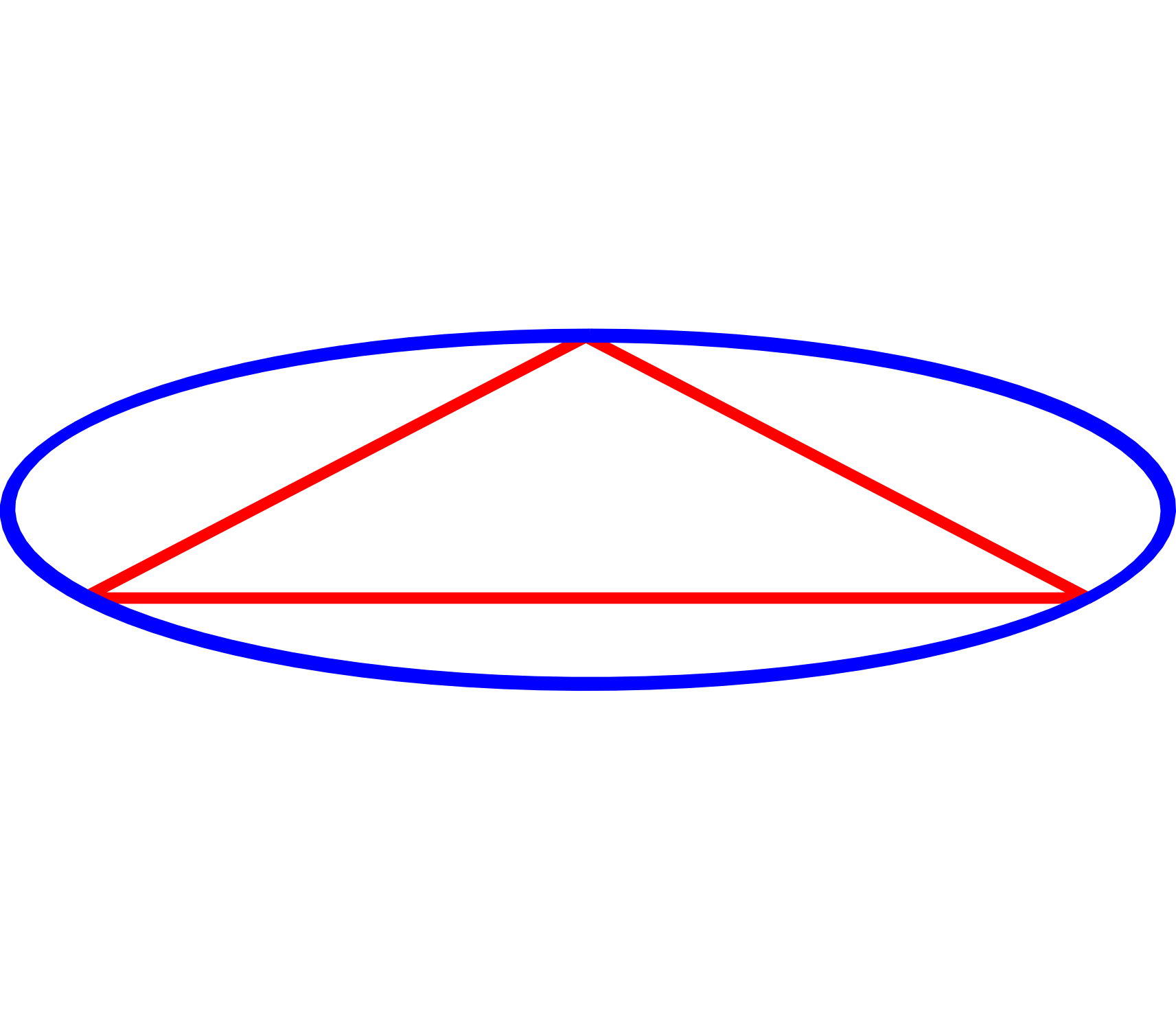}}
			&\subfigure[]{\label{fig:rotation1}
				\includegraphics[width=0.1\textwidth]
				{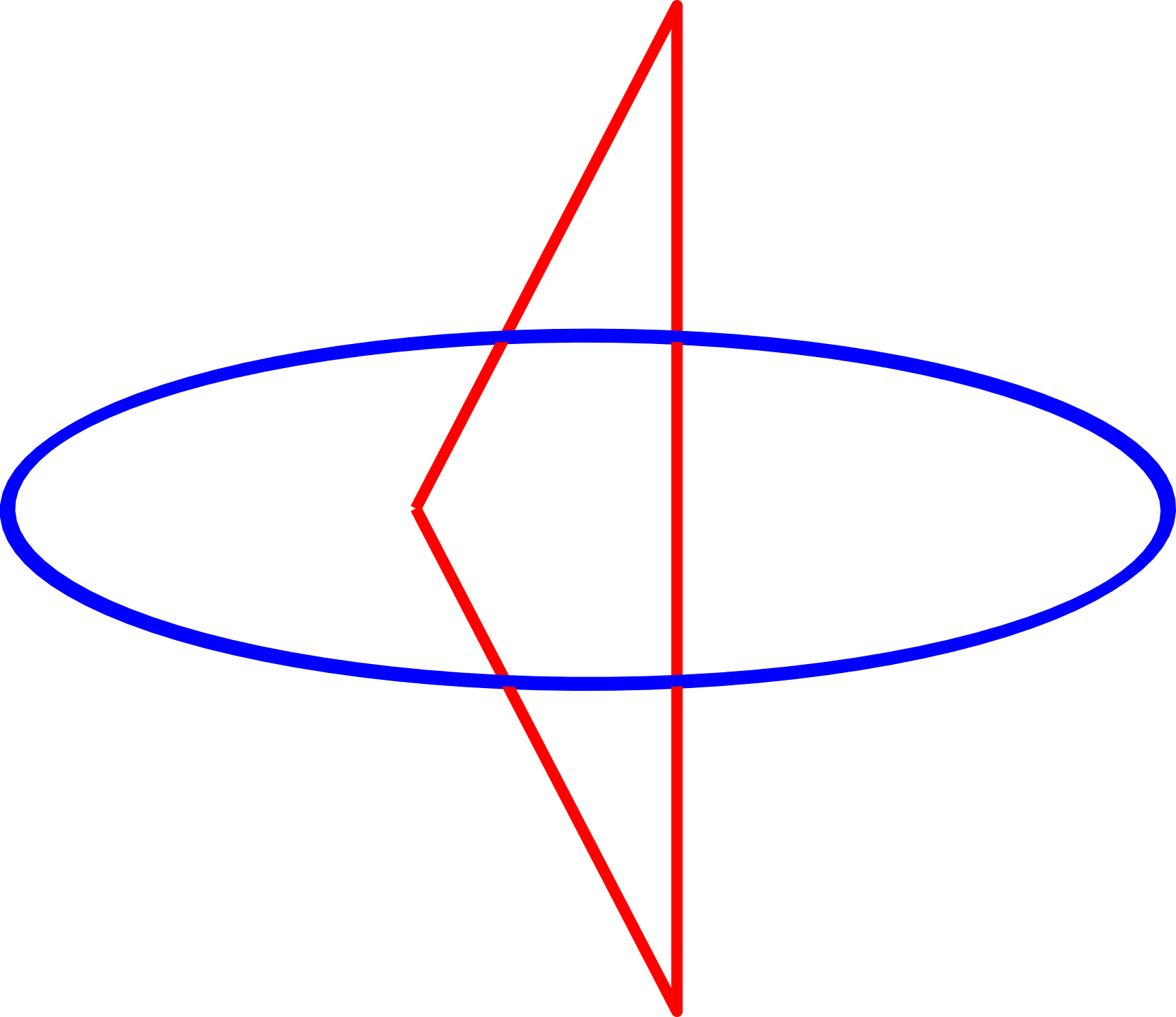}  	 }
			&\subfigure[]{\label{fig:rotation2}
				\includegraphics[width=0.1\textwidth]
				{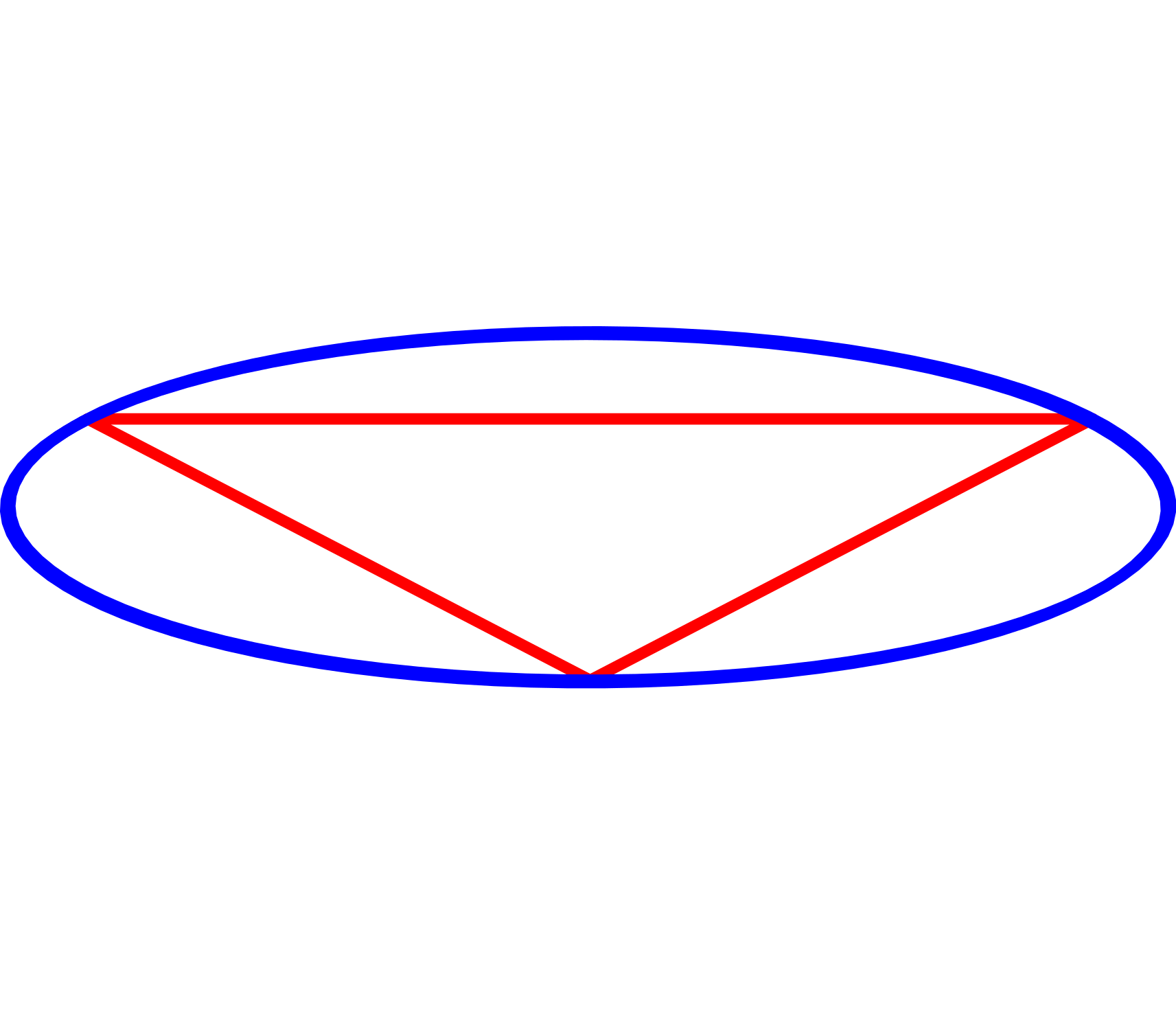}  	 }
			&\subfigure[]{\label{fig:rotation3}
				\includegraphics[width=0.1\textwidth]
				{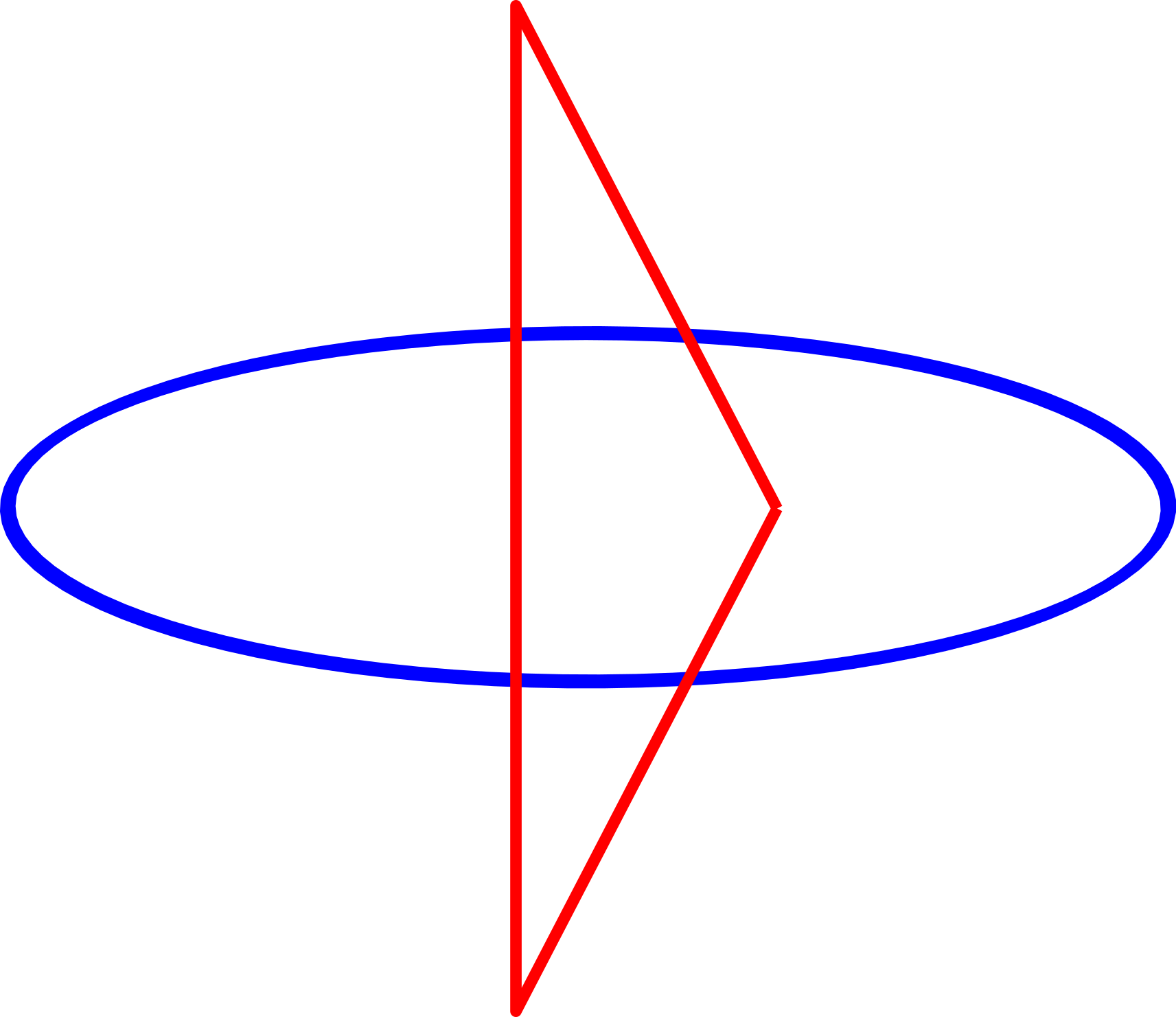}  	 }
			&\subfigure[]{\label{fig:rotation4}
				\includegraphics[width=0.1\textwidth]
				{./fig1}  	 }
			\\
			\multicolumn{5}{c}{
				\includegraphics[width=0.7\textwidth]{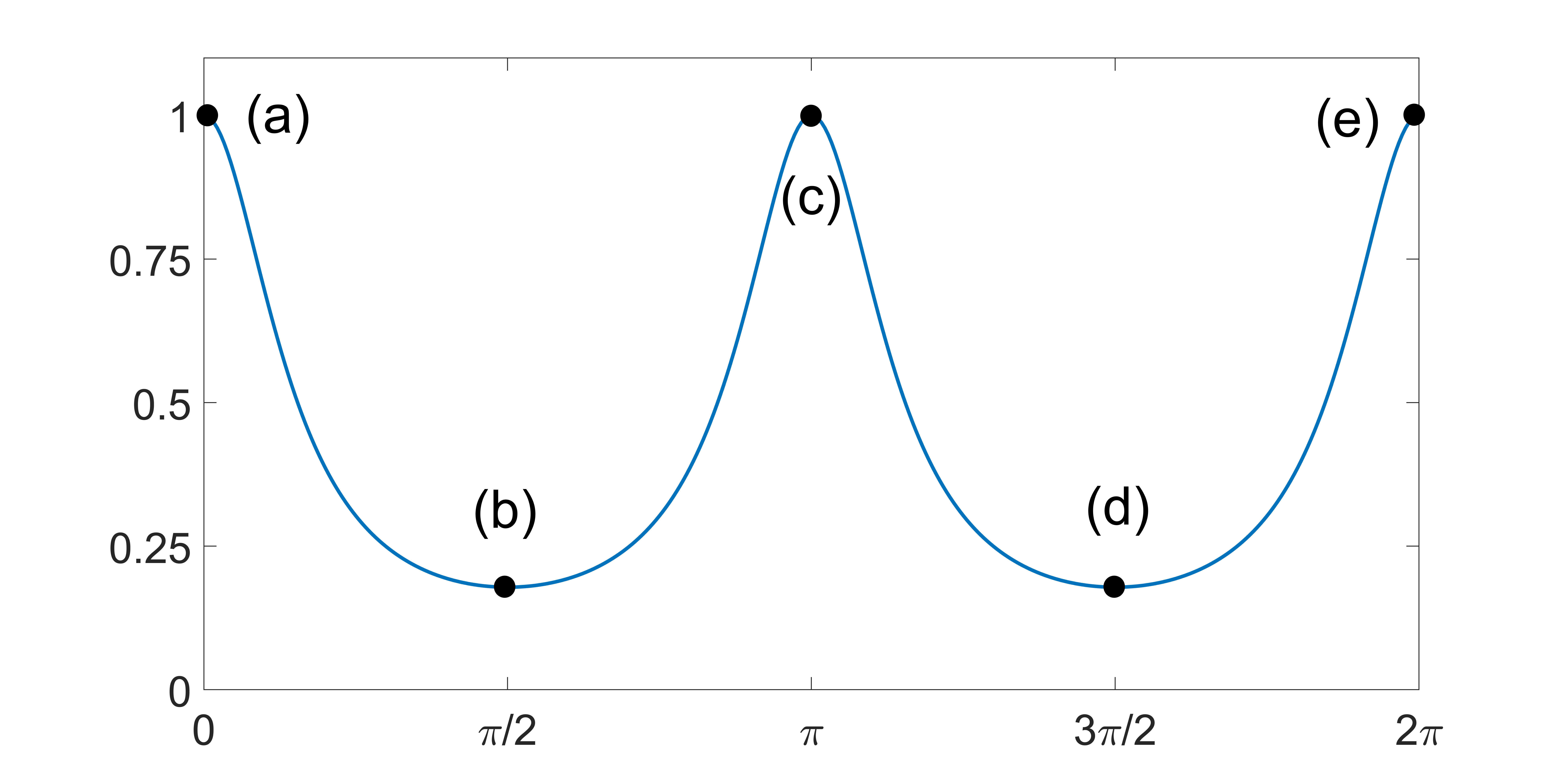}}
		\end{tabular}
		%	\caption{Shape quality measure of physical elements which are rotations of the ideal element.}
		\caption{Influence of alignment in the shape quality measure. First row, physical elements which are rotations of the ideal element in radians: \textbf{(a)} 0; \textbf{(b)} $\pi/2$; \textbf{(c)} $\pi$; \textbf{(d)} $3\pi/2$; and \textbf{(e)} $2\pi$. Second row, shape quality measure in terms of the rotation angle and corresponding mark for rotated elements \textbf{(a,b,c,d,e)}.}
		\label{fig:distortion_rotation}
	\end{figure}
	In Figure \ref{fig:distortion_rotation}, we plot the quality of each physical element with respect to the angle of the rotation applied to the ideal element to generate it. We represent the angle of rotation $\theta$ in the $x$-axis and the quality measure in the $y$-axis. We mark the cases $\theta = 0,\ \pi/2,\ \pi,\ 3\pi/2,\ \mathrm{and}\  2\pi$ with a black dot and we show the corresponding rotations of the ideal element in Figures \ref{fig:rotation0}, \ref{fig:rotation1}, \ref{fig:rotation2}, \ref{fig:rotation3}, and \ref{fig:rotation4}, respectively. We map a rotation of the unit circle in the Euclidean space to the same ellipse in the metric space, see Figures \ref{fig:rotation0}-\ref{fig:rotation4}.
	We highlight that independently of the applied rotation, the ellipse remains constant. An element with quality one must have the nodes on the ideal ellipse.
	
	In the isotropic case, rotations of the equilateral triangle have quality 1. 
	In the anisotropic case, when two axes correspond to different eigenvalues of the metric, we observe that the quality oscillates having two maxima and two minima in $[0,2\pi)$. 
	The maxima are obtained in $\theta = 0$ and $\theta=\pi$ and the minima at $\theta = \frac{\pi}{2}$ and $\theta=\frac{3\pi}{2}$. 
	When $\theta = 0$ the rotation $\zrotation$ is the identity and $\zphysical = \zideal$. 
	When $\theta = \frac{\pi}{2}$ then the axes are interchanged (up to sign) and the quality at $\theta = \frac{\pi}{2}$ attains a minimum. The minima are attained when both axes are interchanged (up to sign) and the maxima are attained when the axes coincide with the eigenvectors of the metric (up to sign).
	
	\section{Measures for curved high-order meshes with varying metric}
	\label{sec:pointwise}
	
	Herein, we define the point-wise measures for curved high-order meshes equipped with point-wise varying metrics.
	First, in Section \ref{sec:extension}, we present the point-wise size-shape distortion measure for high-order elements equipped with point-wise varying metrics.
	Then, in Section \ref{sec:intrinsicmetric}, we present the Riemmanian measure of mesh entities.
	
	\subsection{Size-shape distortion for curved high-order elements on varying metric}
	\label{sec:extension}
	In Section  \ref{sec:linear}, we presented the distortion measure for linear elements equipped with a constant metric. 
	For high-order elements, the Jacobian of the mapping is not constant. 
	In this section, we describe the analogous formulation for high-order elements and for linear elements equipped with a non-constant metric field.
	
	The point-wise distortion measure for an element $\zphysical$ equipped with a metric $\zmetric$, at a point $\zu\in\zequilater$ is defined as
	\begin{equation*}\label{eq:euclideandistortionoperator}
		\zeuclideandistortionoperator(\zu) := \eta(\zJacobianequilaterisotropicphysical(\zu)).
	\end{equation*}
	Following Equation \eqref{eq:elemental}, the distortion measure for an element $\zphysical$ equipped with a metric $\zmetric$ is defined as
	\begin{equation}\label{eq:HOdistortion}
		\eta_{\left(\zphysical,\zmetric\right)} = \frac{\int_{\zequilater} \zeuclideandistortionoperator(\zu)\ \text{d}\zu}{\int_{\zequilater} 1\ \text{d}\zu}.
	\end{equation}
	Equation \eqref{eq:HOdistortion} can be written in terms of $\zxi$ on the master element. That is, the Jacobian of the map $\zequilaterisotropicphysicalmap$ can be written in terms of $\zxi$ as:
	\begin{equation*}
		\zJacobianequilaterisotropicphysical(\zequilatermap(\zxi)) = \zfield(\zphysicalmap(\zxi))\ \zJacobianphysical(\zxi)\  \left(\zJacobianequilater(\zxi)\right)^{-1},
	\end{equation*}
	where
	\begin{equation*}\label{eq:metricdecompositionpoint-wise}
		\zmetric(\zphysicalmap(\zxi)) = \zfield(\zphysicalmap(\zxi))^\mathrm{T}\ \zfield(\zphysicalmap(\zxi)).
	\end{equation*}
	Then, Equation \eqref{eq:HOdistortion} reads
	\begin{equation*}\label{eq:HOdistortionxi}
		\eta_{\left(\zphysical, \zmetric \right)} = \frac{\int_{\zmaster} \zeuclideandistortionoperator(\zequilatermap(\zxi))\ |\det\zJacobianequilater(\zxi)|\ \text{d}\zxi}{\int_{\zmaster} |\det\zJacobianequilater(\zxi)|\ \text{d}\zxi}.
	\end{equation*}
	Similarly to Equation \eqref{eq:distortionmetric}, the decomposition of the metric is not required:
	\begin{equation*}
		\zJacobianequilaterisotropicphysical(\zequilatermap(\zxi))^\mathrm{T}\ \zJacobianequilaterisotropicphysical(\zequilatermap(\zxi)) = \textbf{A}(\zxi)^\mathrm{T}\  \zmetric(\zphysicalmap(\zxi))\  \textbf{A}(\zxi),
	\end{equation*}
	where
	\begin{equation*}
		\textbf{A}(\zxi) := \zJacobianphysical(\zxi)\  \left(\zJacobianequilater(\zxi)\right)^{-1}.
	\end{equation*}
	Using the above equation we obtain the final expression on each point $\zxi$ of the master element:
	\begin{equation}\label{eq:computationvariable}
		\zeuclideandistortionoperator(\zequilatermap(\zxi)) = \eta_\zmetric\left(\textbf{A}(\zxi)\right).
	\end{equation}
	Here, $\eta_\zmetric$ is defined in Equation \eqref{eq:distortionmetric} where
	%\begin{equation*}\label{eq:vol}
	%\textbf{S}_\zmetric\left(\textbf{A}(\zxi)\right):=\sqrt{\ztr\left(\zconjugatexi\right)},\quad \text{and}\quad \sigma_\zmetric\left(\textbf{A}(\zxi)\right):=\det{\zJacobianphysical(\zxi)}\det{\zJacobianequilater(\zxi)^{-1}}\sqrt{\det{\zmetric(\zphysicalmap(\zxi))}}.
	%\end{equation*}
	\begin{align}\label{eq:vol}
		\textbf{S}_\zmetric\left(\textbf{A}(\zxi)\right)&:=\sqrt{\ztr\left(\zconjugatexi\right)},\quad \text{and}\\ \sigma_\zmetric\left(\textbf{A}(\zxi)\right)&:=\det{\zJacobianphysical(\zxi)}\det{\zJacobianequilater(\zxi)^{-1}}\sqrt{\det{\zmetric(\zphysicalmap(\zxi))}}.
	\end{align}
	%In order to detect inverted elements \cite{garanzha:regularization, storti:globalSmoothUntangl,escobar:untanglingSmoothing,gargallo:generation3Doptimization} we regularize the determinant in the denominator of Equation \eqref{eq:computationvariable} to
	
	Although in this work we only implement the size-shape distortion measure for simplicial elements, note that the proposed size-shape distortion measure applies to all types of high-order elements. For a new type, according to our diagram, we only need to determine the master and the regular element in the Euclidean space.
	
	In order to detect inverted elements \cite{garanzha:regularization, storti:globalSmoothUntangl,escobar:untanglingSmoothing,gargallo:generation3Doptimization} we regularize the determinant $\sigma_{\zmetric}$ to
	\begin{equation*}
		\sigma_{0,\zmetric} = \frac{1}{2}(\sigma_{\zmetric} + |\sigma_{\zmetric}|).
	\end{equation*}
	Then, we define the point-wise regularized size-shape distortion measure of a physical element $\zphysical$ as
	\begin{equation*}
		\zeuclideandistortionoperatorreg(\zu) := \eta_0(\zJacobianequilaterisotropicphysical(\zu)) := \frac{1}{d}\frac{\zSmetric^2}{\sigma_{0,\zmetric}^{2/d}}\  \left(\frac{1}{2}\left(\sigma_{0,\zmetric} + \frac{1}{\sigma_{0,\zmetric}}\right)\right)^{2/d},
	\end{equation*}
	and its corresponding quality as
	\begin{equation}\label{eq:pointquality}
		\zeuclideanqualityoperatorreg(\zu) = \frac{1}{\zeuclideandistortionoperatorreg(\zu)}.
	\end{equation}
	Finally, we regularize the elemental distortion of Equation \eqref{eq:HOdistortion} as
	\begin{equation*}\label{eq:distortionpoint-wise}
		\eta_{0,\left(\zphysical, \zmetric \right)} := \frac{\int_{\zequilater} \zeuclideandistortionoperatorreg(\zu)\ \text{d}\zu}{\int_{\zequilater} 1\ \text{d}\zu},
	\end{equation*}
	and its corresponding quality as
	\begin{equation}\label{eq:qualityreg}
		q_{0,\left(\zphysical, \zmetric \right)} = \frac{1}{\eta_{0,\left(\zphysical, \zmetric \right)}}.
	\end{equation}
	
	We can improve the mesh configuration by means of relocating the nodes of the mesh according to a given distortion measure \cite{aparicio2018defining,aparicio2019imr,aparicio2020severoochoa,gargallo:PhDdissertation}.
	For example, in \cite{aparicio2018defining} it is proposed the optimization of the distortion (quality) of a mesh $\zmesh$ equipped with a target metric $\zmetric$ that describes the desired alignment and stretching of the mesh elements. 
	To optimize the given mesh $\zmesh$, we define the mesh distortion by
	\begin{equation}\label{eq:minfun}
		\mathcal{F}\left(\zmesh \right) := \sum_{\zphysical\in\zmesh} \int_{\zequilater} \left(\mathcal{N}_0\zequilaterisotropicphysicalmap(\zy)\right)^2\ \text{d}\zy,
	\end{equation}
	which allows to pose the following global minimization problem
	\begin{equation}\label{eq:argmin}
		\zmesh^* := \argmin_{\zmesh} \mathcal{F}\left(\zmesh\right),
	\end{equation}
	to improve the mesh configuration according to $\mathcal{F}$.
	In particular, herein, the degrees of freedom of the minimization problem in Equation \eqref{eq:argmin} correspond to the spatial coordinates of the mesh nodes.
	\subsection{Riemmanian measure of mesh entities}\label{sec:intrinsicmetric}
	Next, we propose a method to compare how a mesh matches a target metric. To this end, we present the point-wise and element-wise size measure of the mesh entities according to a Riemannian metric.
	
	We consider the Riemannian measure of the mesh entities relative to the target metric $\zmetric$.
	On the one hand, we define the point-wise relative density $\rho_{\zmetric}$ of a physical element $\zphysical$ with respect to a reference element $\zmaster$. Specifically, for a $k$-dimensional physical element $\zphysical$ embedded in the Riemannian space $\left(\mathds{R}^n,\zmetric\right)$, the \textit{point-wise metric-aware density} of $\zphysical$ respect to a $k$-dimensional master element $\zmaster$ is given by $\sqrt{\det\ \left[\zJacobianphysical(\zxi)^\text{T}\ \zmetric\left(\zphysicalmap(\zxi)\right)\ \zJacobianphysical(\zxi)\right]}$.
	We also consider the \textit{point-wise metric-aware normalized density} as the quotient of the physical density by the ideal density
	\begin{equation}\label{eq:metricvolp}
		\rho_{\zmetric}(\zxi) := \sqrt{\frac{\det\ \left[\zJacobianphysical(\zxi)^\text{T}\ \zmetric\left(\zphysicalmap(\zxi)\right)\ \zJacobianphysical(\zxi)\right]}{\det\ \left[\zJacobianideal(\zxi)^\text{T}\ \zmetric\left(\zphysicalmap(\zxi)\right)\ \zJacobianideal(\zxi)\right]}},\quad \text{for}\quad \zxi\in\zmaster.
	\end{equation}
	Accordingly, we say that an element $\zphysical$ is \textit{unitary} if $\rho_{\zmetric} \equiv 1$ is satisfied for the element density and for the density of all its sub-entities.
	Considering the commutative diagram in Figure \ref{fig:xevisDiagram2}, we compute the metric-aware density as
	\begin{equation}\label{eq:metricvolpeq}
		\rho_{\zmetric}(\zxi) = \sqrt{\frac{\det\ \left[\zJacobianphysical(\zxi)^\text{T}\ \zmetric\left(\zphysicalmap(\zxi)\right)\ \zJacobianphysical(\zxi)\right]}{\det\ \left[\zJacobianequilater(\zxi)^\text{T}\ \zJacobianequilater(\zxi)\right]}},\quad \text{for}\quad \zxi\in\zmaster,
	\end{equation}
	%The unit element is the equilateral element of dimension $k$ with all the edges of unit length.
	where the unit element $\zequilater$ is a $k$-dimensional regular element with all the edges of unit length.
	Note that, any sub-entity of a regular element $\zequilater$ is also unitary because its density is one.
	%Note that, a unitary element is also edge-wise unitary \cite{loseille2011continuous}.
	%From Equation \eqref{eq:metricvolpeq}, we conclude that being $\zphysical$ unitary is equivalent to $\rho_{\zmetric} \equiv 1$.
	On the other hand, we define the \textit{element-wise metric-aware normalized measure} of $\zphysical$ according to the metric $\zmetric$ as
	\begin{equation}\label{eq:metricvol}
		V_{\zmetric}\left( \zphysical \right) := \frac{1}{V\left( \zmaster\right)}\int_{\zmaster} \rho_{\zmetric}(\zxi) \ \text{d}\zxi,
	\end{equation}
	where $V\left( \zmaster\right) = \int_{\zmaster} 1\ \text{d}\zxi$.
	%In this case, the measure does not depend on the reference element.
	%In contrast to the point-wise differential measure, being $\zphysical$ unitary is not equivalent to $V_{\zmetric}\left( \zphysical \right) = 1$.
	%This is because the element measure does not reflect the point-wise behavior of the element.
	%In particular, a non-unit element featuring low size at some regions while having great size at other regions can have unit element-wise measure.
	
	While it is common to consider only the element-wise length of the element edges, this does not illustrate if the element is unitary or not, specially for non-constant metrics or curved elements.
	In contrast, an element is unitary if the density of all its sub-entities is constant equal to one.
	For this reason, we measure how a mesh is unitary according to the metric by measuring all the mesh entities from Equation \eqref{eq:metricvol}.
	%In particular, we do this in terms of the measures of the mesh entities.
	That is, lengths of edges, areas of faces, and volumes of cells. For example, only when the mesh matches the metric, the lengths, areas, and volumes are unit and vice-versa. That is, they match the length, area, and volume of the equilateral element, respectively. In contrast, a higher stretching or non-unit volume of the intrinsic metric indicates that the mesh does not match the stretching or the volume of the metric, respectively. As a consequence, the lengths, areas, and volumes are non-unit and vice-versa.
	\section{Results}\label{sec:results}
	In this section, we apply the size-shape distortion minimization for curved $r$-adaptation.
	For this, we start comparing the behavior of the distortion minimization for the size-shape and the shape measures \cite{aparicio2018defining}.
	%analyze the behavior of the size-shape distortion measure.
	Then, we illustrate how the size-shape distortion minimization can be used for the improvement of the interpolation error of an input function.
	
	%\color{red}Paragraf que englobi totes les conclusions.
	%All examples show the capability of curved elements to capture sharp transition regions with curved alignment.
	%Specifically, the size-shape distortion minimization improves the matching between mesh and metric.
	%We illustrate this from the quality statistics, volumetric measure histograms, and interpolation, approximation, and numerical errors.
	%Finally, we illustrate how the distortion minimization improves the numerical error of a manufactured solution for the Poisson problem.
	%\color{black}
	
	First, in Section \ref{sec:implementation}, we describe the implementation details.
	Second, in Section \ref{sec:shape}, to illustrate the benefit and novelty of the size-shape minimization, we compare the shape and size-shape distortion measures for an analytic target metric.
	Finally, in Sections \ref{sec:min1}, \ref{sec:errormin}, \ref{sec:curving}, and \ref{sec:pde} we apply the size-shape distortion minimization for high-order function interpolation, approximation, or simulation.
	
%	how the metric-aware distortion minimization method allows curved elements to match the metric.
%	The experiments illustrate 
	
%	The presented examples show the capability of curved elements to capture sharp transition regions with curved alignment.
%	Specifically, the size-shape distortion minimization improves the matching between mesh and metric.

%	The purpose of the examples is to show the advantages of the size-shape distortion minimization approach. On the one hand, it illustrates the capability of curved elements to capture sharp transition regions with curved alignment. In particular, the success of this approach allows us to observe that curved elements approximate the metric better when compared to straight-sided elements. On the other hand, it shows the potential of the size-shape distortion minimization approach to enable optimized meshes featuring metrics, either manufactured or obtained from a finite element solution, defined over domains with flat or curved boundaries.
	
	We illustrate the advantages of the method from the quality statistics, volumetric measure histograms, and interpolation, approximation, and numerical errors.
	In particular, in Section \ref{sec:min1}, we consider a 2D case for degrees 1, 2, and 4, and a quadratic 3D example.
	Then, in Section \ref{sec:errormin}, we minimize the size-shape distortion for initial isotropic and initial adapted straight-edged quartic meshes.
	Next, in Section \ref{sec:curving}, we minimize the size-shape distortion for an initial adapted straight-edged cubic mesh according to a curved boundary.
	Finally, in Section \ref{sec:pde}, we minimize the size-shape distortion for
	a manufactured solution of a Poisson problem.
	
	%First, in Section \ref{sec:implementation}, we describe the implementation details.
	%Second, in Section \ref{sec:shape}, we compare the shape and size-shape distortion measures for an analytic target metric.
	%Third, in Sections \ref{sec:min1}, \ref{sec:errormin}, and  \ref{sec:curving}, we apply the size-shape distortion minimization for high-order interpolation.
	%Specifically, in Section \ref{sec:min1}, we consider a 2D case with varying degrees and a quadratic 3D example.
	%Fourth, in Section \ref{sec:errormin}, we minimize the size-shape distortion for initial isotropic and initial adapted straight-edged quartic meshes.
	%The results show that the size-shape distortion minimization improves the interpolation and approximation errors of the input function.
	%Fifth, in Section \ref{sec:curving}, we minimize the size-shape distortion for an initial adapted straight-edged cubic mesh according to a curved boundary.
	%The results show that the size-shape distortion minimization improves the interpolation and approximation errors of the input function while targeting a curved boundary.
	%Finally, in Section \ref{sec:pde}, we minimize the size-shape distortion for
	%a manufactured solution of a Poisson problem.
	%The results show that the potential of the size-shape distortion minimization to improve the numerical error of a finite element simulation.
	
	% Technicalities
	\subsection{Implementation}\label{sec:implementation}
	In all the examples, the mesh boundary nodes move tangentially to the boundary. If the domain boundary is a cartesian box, to slide the nodes on the boundary, we fix one coordinate, and we set the others as degrees of freedom. If the domain boundary is curved, to approximately move tangentially to the boundary, we consider all coordinates as degrees of freedom. Moreover, we minimize the geometric deviation between the mesh boundary and the CAD model with an implicit representation, see \cite{aparicio2023combining} for details.

	To evaluate the distortion measures and capture pronounced metric variations, we use up to $(3 p)^d$ quadrature points within an element, where $p$ is the degree and $d$ is the dimension. Nevertheless, for less pronounced metric variations, it is possible to use around $(2 p)^d$ quadrature points per element.
	
	Because our goal is to optimize the mesh distortion, instead of including mathematical proofs of mesh validity, we detail how we numerically enforce the positiveness of the element Jacobians.
	Specifically, we use a numerical valid-to-valid approach that uses four ingredients.
	First, because we want numerically valid results, we enforce mesh validity on the integration points.
	Second, to initialize the optimization, we start from a numerically valid mesh. Third, to penalize inverted elements, we modify the point-wise distortion to be infinity for non-positive Jacobians.
	Specifically, we regularize the element Jacobians to be zero for non-positive Jacobians, so their reciprocals are infinite, see Section \ref{sec:pointwise}.
	Note that these reciprocals appear in the distortion expression, and thus, they determine the infinite distortion value.
	Fourth, to enforce numerically valid mesh displacements, we equip Newton's method with a backtracking line-search.
	Specifically, if the mesh optimization update is invalid in any integration point, the objective function is infinite.
	In that case, the step is divided by two until it leads to a valid mesh update.
	Finally, to check that the mesh is numerically valid after the optimization, we evaluate the regularized distortion measure at a set of sample points. In particular, we choose a set of uniformly distributed sample points, being three times the mesh polynomial degree, $3p$, per edge.
	
%	Finally, to guarantee that the topology of the mesh is correct after the optimization we evaluate the regularized distortion measure at a set of sample points. In particular, we choose a set of uniformly distributed sample points, being three times the mesh polynomial degree, $3p$, per edge.
	
%	To evaluate the measures we use a quadrature distribution. Specifically, given a polynomial degree $p$ and a dimension $d$, we typically use around $d+1 \left(6*p\right)^d$ points. That is, around $12p$ for segments, $108p^2$ for triangles, and $864p^3$ for tetrahedra.
	
	Because we are detailing an approach to combine shape and size distortion, we do not detail how to obtain the input metric $M$. Nevertheless, it is possible to exploit existent high-order goal-oriented \cite{yano2012,fidkowski2011review} and interpolation-oriented \cite{loseille2011continuous,coulaud:VeryHighOrderAnisotropic} error estimators that provide a metric as an output. In practice, we interpolate these output metrics on a background mesh with the method detailed in \cite{aparicio2022metricinterpolation}.
	
	As a proof of concept, a mesh optimizer is developed in Julia 1.4.2 \cite{bezanson2017julia}. The Julia prototyping code is sequential, it corresponds to the implementation of the method presented in this work and the one presented in \cite{aparicio2018defining,aparicio2019imr,aparicio2020severoochoa}.
%	Unfortunately, the source code is not yet available to be published in an open repository.
	In all the examples, the optimization corresponds to finding a minimum of a nonlinear unconstrained multi-variable function $f$, see Equations \eqref{eq:minfun} and \eqref{eq:argmin}.
	In particular, the mesh optimizer uses an unconstrained line-search globalization with an iterative preconditioned conjugate gradients linear solver.
	The stopping condition is set to reach an absolute root mean square residual, defined as $\frac{\| \nabla f(x)\|_{\ell^2}}{\sqrt{n}}$ for $x\in\zR^n$, smaller than $10^{-4}$ or a length-step smaller than $10^{-4}$. 
	Each optimization process has been performed in a node featuring two Intel Xeon Platinum 8160 CPU with 24 cores, each at 2.10 GHz, and 96 GB of RAM memory.
	
%	The computational cost of the isotropic distortion minimization is carefully analyzed in \cite{ruiz2022automatic}. In addition, when the mesh is coupled with a metric, we should account for the computational cost of the metric derivatives evaluation \cite{aparicio2023combining}. However, this does not drawback nor limit the implementation of the distortion minimization.
	
%	Because the mesh and metric characteristics may stiffen the optimization problem, they challenge the global convergence of the non-linear solver and the solution of the corresponding linear systems. We alleviate the issues of standard solvers by considering a specific-purpose globalized and preconditioned Newton-CG solver.
	
%	Finally, to enforce convergence to a local minimum, we use a specific-purpose minimization method. Without this specific-purpose solver we cannot ensure convergence.
	
	Although we generate meshes adapted to a target metric with MMG \cite{dobrzynski2012mmg3d}, our goal is not to compare the distortion minimization with the MMG package. Actually, we acknowledge MMG because it generates an initial straight-sided mesh that matches the stretching and alignment of the target metric.
	
	\subsection{Shape versus size-shape distortion minimization: curved high-order mesh and analytic metric}\label{sec:shape}
	In what follows, to illustrate the potential for capturing not only the stretching and alignment but also the sizing of the target metric, we compare the shape and size-shape distortion measures presented in Section \ref{sec:pointwise}.
	Specifically, we do this for a curved high-order mesh and an analytic metric.
	For this, we first define the target metric.
	Then, we illustrate the initial and optimized meshes.
	Finally, we compare the distortion measures from the distribution and statistics of Riemannian length and area, see Section \ref{sec:intrinsicmetric} for the details.
	
	We consider the quadrilateral domain $\Omega=[-0.5,0.5]^2$ equipped with a metric matching a boundary layer.
	In particular, our target metric $\textbf{M}$ is characterized by a boundary layer metric with a diagonal matrix $\textbf{D}$, a deformation map $\varphi$, and the characteristic length $h_{\textrm{m}} := 0.25$ by the following expression
	\begin{equation}\label{eq:metricDeformation}
		\textbf{M} = \frac{1}{h_{\textrm{m}}^2}\nabla \varphi^{\text{T}}\ \textbf{D} \ \nabla \varphi.
	\end{equation}
	In what follows, we first detail the boundary layer metric $\textbf{D}$ and then the deformation map $\varphi$, see \cite{aparicio2018defining} for more details.
	
	On the one hand, the boundary layer aligns with the $x$-axis.
	It determines a constant unit element size along the $x$-direction, and a non-constant element size along the $y$-direction.
	This vertical element size grows linearly with the distance to the $x$-axis, with a factor $\alpha = 2$, and starts with the minimal value $h_{\min}=0.01$.
	Thus, the stretching ratio blends from $1:100$ to $1:1$ between $y=-0.5$ and $y=0.5$. 
	Specifically, we define the metric as:
	\begin{equation}\label{eq:test2D}
		\textbf{D} := \left(
		\begin{array}{cc}
			1 & 0\\
			0 & 1/h(y)^2
		\end{array}
		\right),
	\end{equation}
	where the function $h$ is defined by
	\begin{equation*}
		h(x):= h_{\min} + \alpha |x|.
	\end{equation*}
	
	On the other hand, the deformation map $\varphi$ in Equation \eqref{eq:metricDeformation} aligns the stretching of $\textbf{D}$ according to a given curve.
	In this case, we define the map $\varphi$ by
	\begin{equation*}
		\varphi(x,y) = \left(x,\frac{10y - \cos(2\pi x)}{\sqrt{100 + 4\pi^2}} \right).
	\end{equation*}
	Finally, the metric $\textbf{M}$ of Equation \eqref{eq:metricDeformation} attains the highest level of anisotropy close to the curve described by the points $(x,y)\in\Omega$ such that $\varphi(x,y)=(x,0)$.
	
	\begin{figure}[t!]
		\centering
		\hspace{-0.65cm}
		\setlength{\tabcolsep}{15pt}
		\begin{tabular}{ccc}
			\subfigure[]{\label{fig:sizeshape0}
				\includegraphics[width=0.25\textwidth]{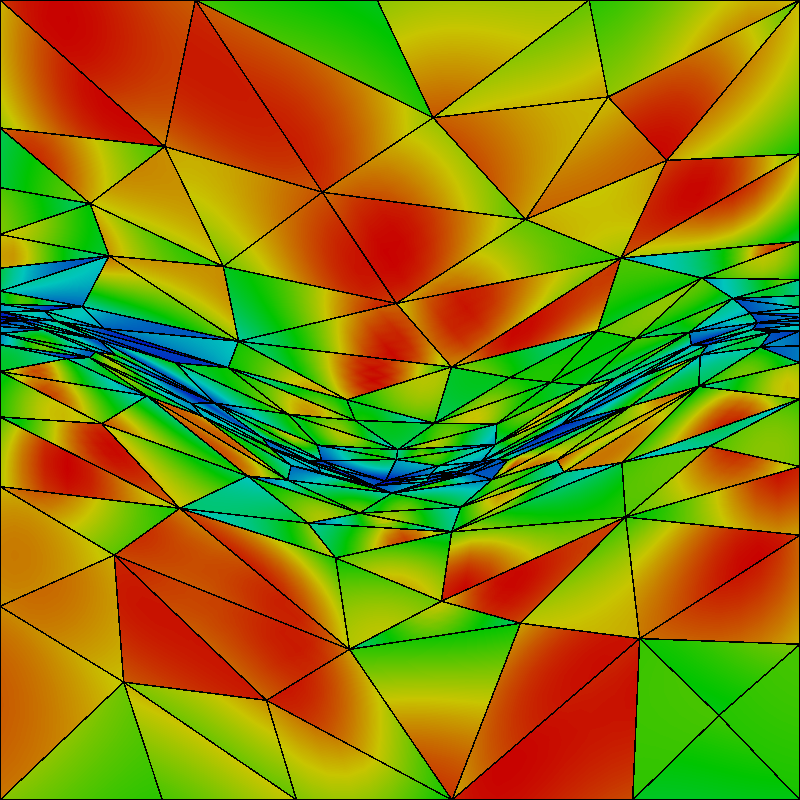}}
			&
			\subfigure[]{\label{fig:shape1}
				\includegraphics[width=0.25\textwidth]{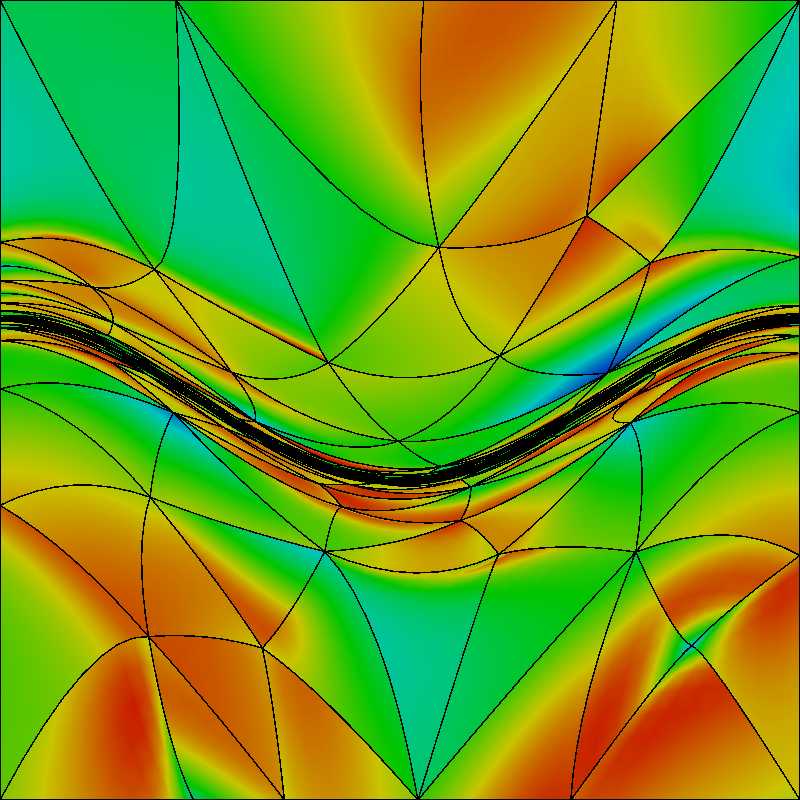}}
			&
			\subfigure[]{\label{fig:sizeshape1}
				\includegraphics[width=0.25\textwidth]{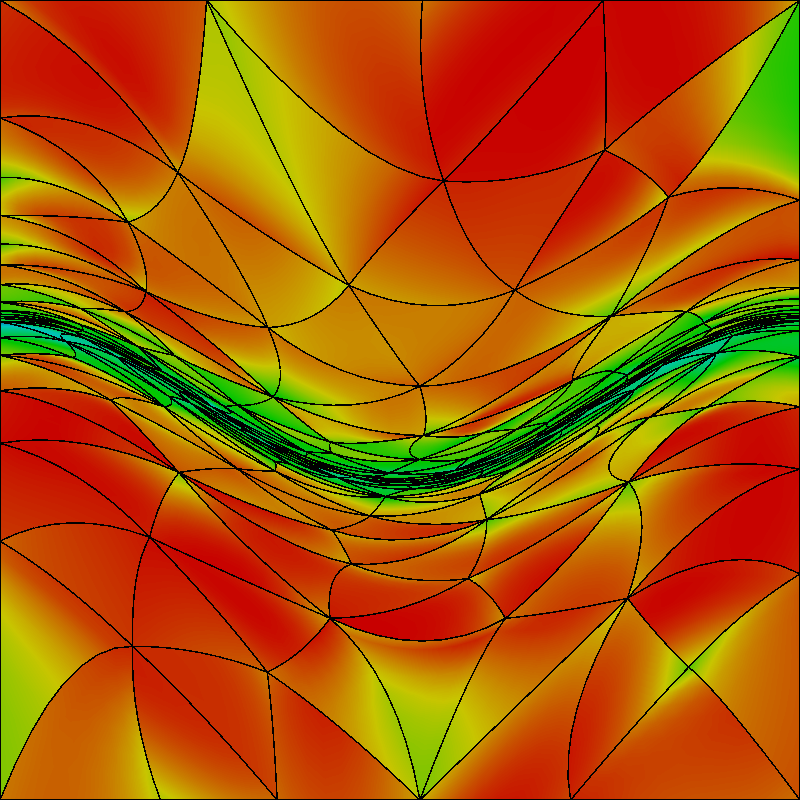}}
			\\
		\end{tabular}
		\\
		\includegraphics[width=0.25\textwidth]{./qualBarParaview_color}
		\caption{Point-wise size-shape quality measure for (a) initial and quadratic meshes optimized according to the (b) shape and (c) size-shape distortion measure, respectively.}
		\label{fig:sizevsshape}
	\end{figure}
	In Figure \ref{fig:sizevsshape}, we illustrate the initial and optimized quadratic meshes equipped with the input metric of Equation \eqref{eq:metricDeformation}. The meshes are colored according to the point-wise size-shape quality measure of Equation \eqref{eq:pointquality}.
	With MMG, we generate an initial anisotropic straight-edged mesh $\zmesh$ according to the target metric of Equation \eqref{eq:metricDeformation}. The obtained mesh is composed by 254 triangles and 553 nodes. From this initial mesh, we observe that the straight-edged elements are stretched, aligned, and scaled approximating the target metric. Then, we optimize the initial mesh $\zmesh$ according to the shape and size-shape distortion measures to obtain the corresponding optimized meshes $\zmesh^*_{\textrm{shape}}$ and $\zmesh^*$. Finally, we observe that the elements are curved according to the point-wise metric stretching and alignment for the mesh $\zmesh^*_{\textrm{shape}}$, and according to the point-wise metric stretching, alignment, and sizing for the mesh $\zmesh^*$.
	
	\begin{figure}[t!]
		\centering
		\hspace{-0.35cm}
		\setlength{\tabcolsep}{-2pt}
		\begin{tabular}{cc}
			\subfigure[]{\label{fig:lengths}
				\includegraphics[width=0.5\textwidth]{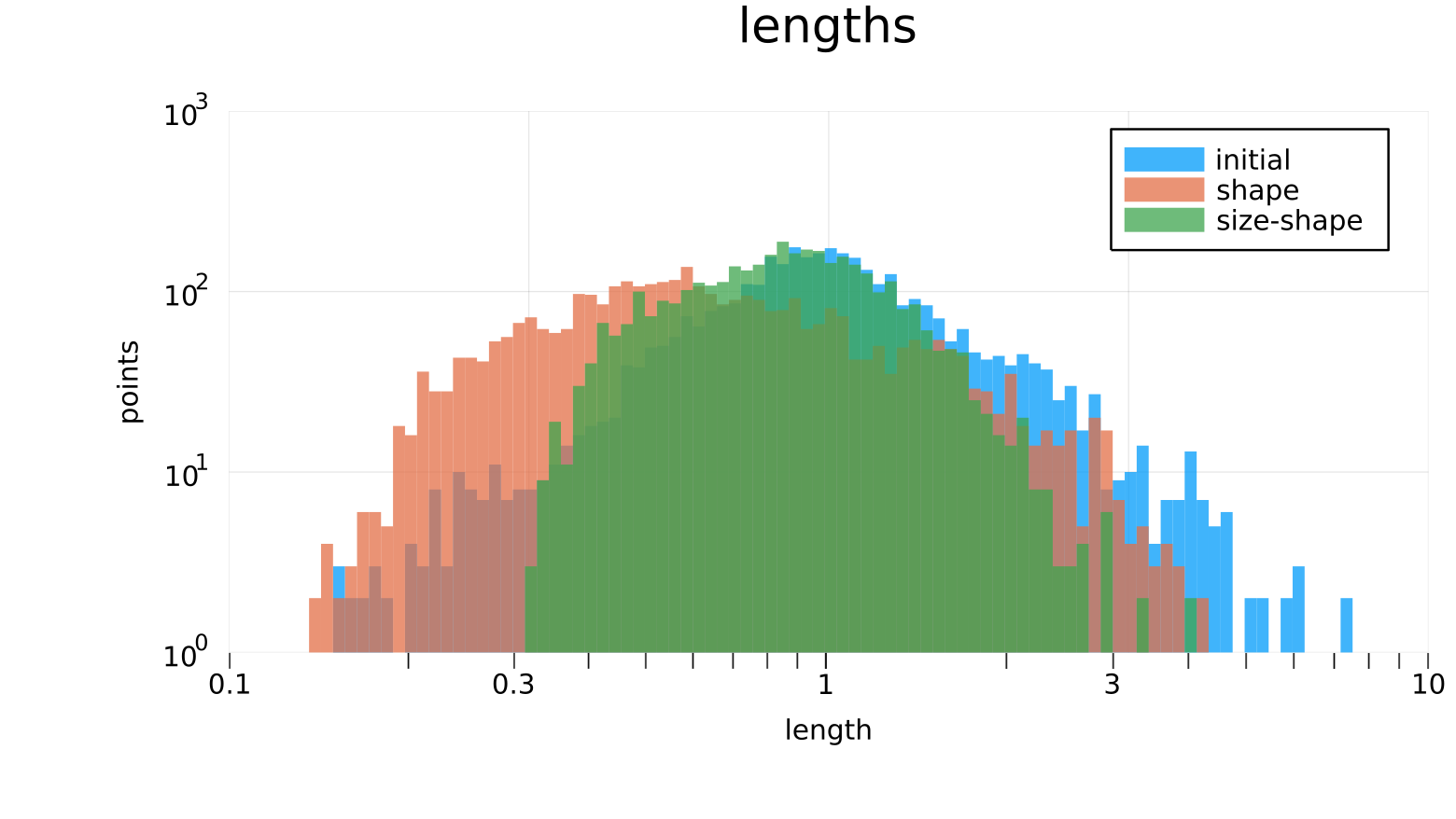}}
			&
			\subfigure[]{\label{fig:areas}
				\includegraphics[width=0.5\textwidth]{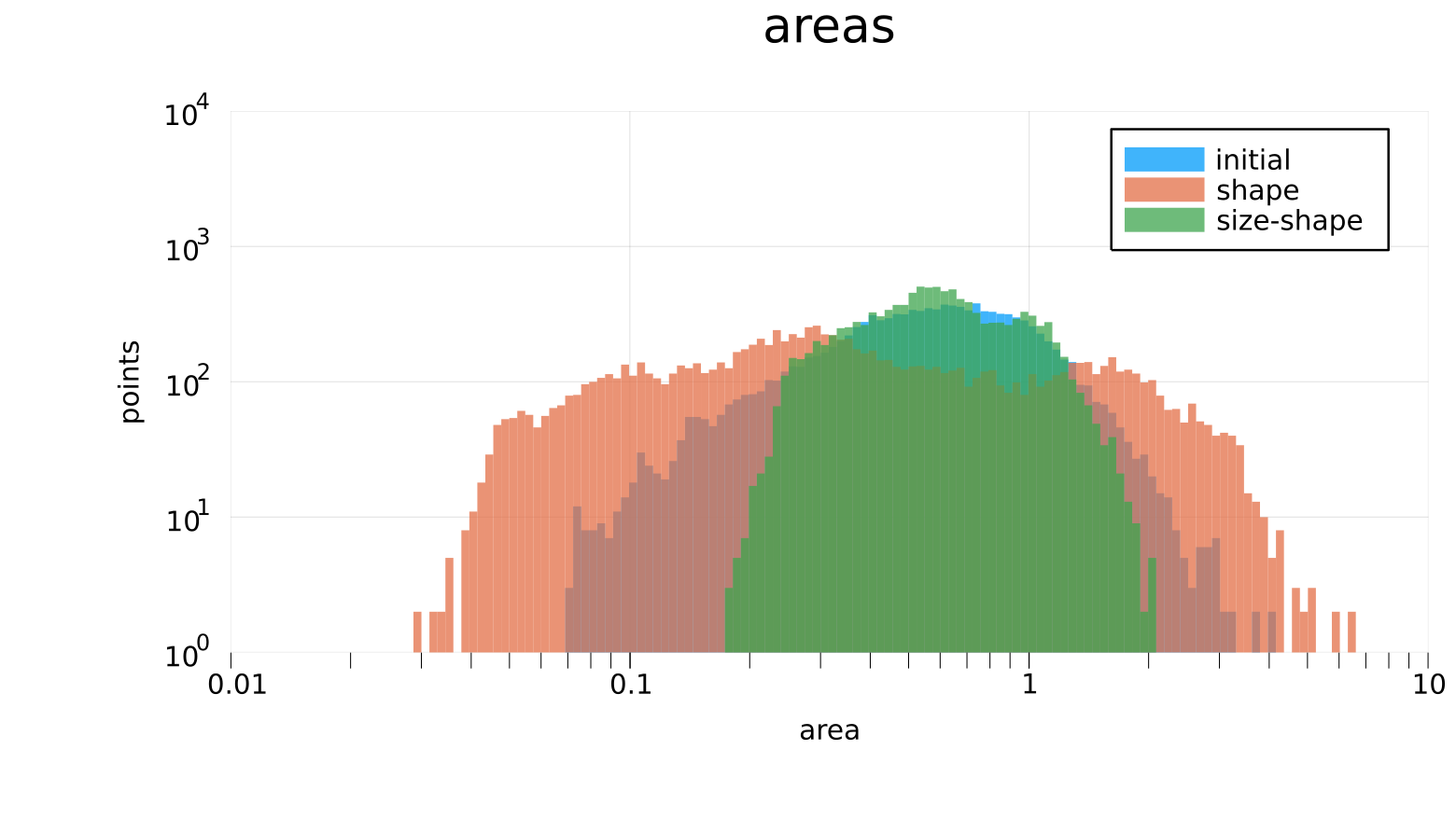}}
		\end{tabular}
		\caption{Logarithmic point-wise (first row) Riemannian length and (second row) area densities for (blue) initial and optimized quadratic meshes according to the (orange) shape and (green) size-shape distortion measures.}
		\label{fig:vs}
	\end{figure}
	From this example, we qualitatively compare the shape and size-shape distortion measures.
	For this, in Figure \ref{fig:vs}, we illustrate the logarithmic distributions of Riemannian length and area densities, see Equation \eqref{eq:metricvolpeq}.
	In particular, the shape minimization distorts the distribution of length and area.
	In contrast, when compared to the initial mesh and the shape minimization, the size-shape optimization concentrates more the distribution of length and area around unit values.
	From this, we conclude that the size-shape minimization matches more faithfully the target metric than the shape optimization.
	
	\begin{table}[t!]
		\caption{Size-shape quality and geometry statistics for the initial and optimized meshes according to the shape and size-shape distortion measures.}
		\label{table:vs}
		\centering
		\begin{tabular}{ c c c c c c}
			\hline\noalign{\smallskip}
			Measure&Mesh& Minimum&Maximum& Mean & Std dev.\\
			\noalign{\smallskip}\hline\noalign{\smallskip}
			&Initial& 0.0156& 0.9694& 0.4276& 0.2687\\
			Quality&Shape& 0.0950& 0.9104& 0.4927& 0.2207\\
			&Size-shape & 0.3136& 0.9882& 0.6337& 0.1976\\
			\noalign{\smallskip}\hline\noalign{\smallskip}
			&Initial & 0.2241& 3.8578& 1.1369& 0.5451\\
			Length&Shape & 0.1964& 3.1337& 0.7926& 0.5445\\
			&Size-shape & 0.3471& 2.3952& 0.9131& 0.3383\\
			\noalign{\smallskip}\hline\noalign{\smallskip}
			&Initial & 0.0724& 1.8048& 0.5593& 0.2931\\
			Area&Shape & 0.0417& 2.6045& 0.5594& 0.5843\\
			&Size-shape & 0.2148& 1.1885& 0.5593& 0.2310\\
			\noalign{\smallskip}\hline\noalign{\smallskip}
		\end{tabular}
	\end{table}
	We quantitatively compare the shape and size-shape distortion measures.
	For this, in Table \ref{table:vs}, we show the statistics of the elemental size-shape quality (Equation \eqref{eq:qualityreg}) and Riemannian length and area measures (Equation \eqref{eq:metricvol}). They allow us to compare the geometric quantities between the initial and optimized meshes, and between the shape and the size-shape quality measures.
	On the one hand, we observe that the shape minimization does not improve the length and area measure statistics from the initial mesh.
	This is because, the shape distortion does not take into account the local element size.
	In contrast, when compared from the initial mesh and the shape minimization, the size-shape optimization substantially improves the length and area measure statistics.
	This can be explained from the coupling between the size and shape distortion measures, which takes into account the local element size and shape deviation, see Section \ref{sec:pointwise}.
	From this, we conclude that the size-shape distortion minimization homogeneously matches more the geometric features of the input metric than the shape optimization.
	
%	For many results, we infer that for targets featuring curved features, curved meshes adapt to the target better than straight-edged meshes. In contrast, we should also infer that on those applications where the target metric presents ruled features, adapted straight-edged meshes might be more efficient than adapted curved meshes. Nevertheless, both for straight-edged and curved meshes, the optimization of the proposed size-shape distortion leads to meshes adapted better to the target metric.
	
%	When considering other metrics, we also expect that the size-shape distortion minimization allows an improved matching of the metric when compared to the shape distortion minimization.
	
%	We expect that the presented comparison remains valid for other metrics. Hence, we consider that, in general, the size-shape distortion minimization allows an improved matching of the metric when compared to the shape distortion minimization.
	
	\subsection{Size-shape distortion minimization for high-order interpolation: 2D different degrees and 3D quadratic}
	\label{sec:min1}
	
	Herein, we apply the size-shape distortion minimization for high-order interpolation.
	In particular, we consider a 2D case for degrees 1, 2, and 4, and a 3D quadratic example.
	For this, in Section \ref{sec:metric}, we determine our discrete metric for practical problems.
%	we set a discrete target metric from the higher-order derivatives of the input function.
	From this discrete metric, we minimize the size-shape distortion in Section \ref{sec:min}.
	To verify that the stretching, alignment, and sizing match the discrete metric, we measure the Riemannian lengths, areas, and volumes in Section \ref{sec:volumetric}.
	Then, to illustrate the potential of curved $r$-adaptation, we measure how the mesh represents the input function.
	In particular, we measure the interpolation and approximation $L^2$-errors in Section \ref{sec:error}.
	\subsubsection{Discrete high-order metric: high-order interpolation}\label{sec:metric}
	Next, to later check our distortion measure in a more realistic case, we compute a discrete metric from the input function as in \cite{coulaud:VeryHighOrderAnisotropic}.
	Specifically, for each polynomial interpolation degree, we obtain a discrete metric approximating the high-order derivatives of the function.
	
	\begin{figure}[t!]
		\centering
		\hspace{-0.35cm}
		\includegraphics[width=0.4\textwidth]{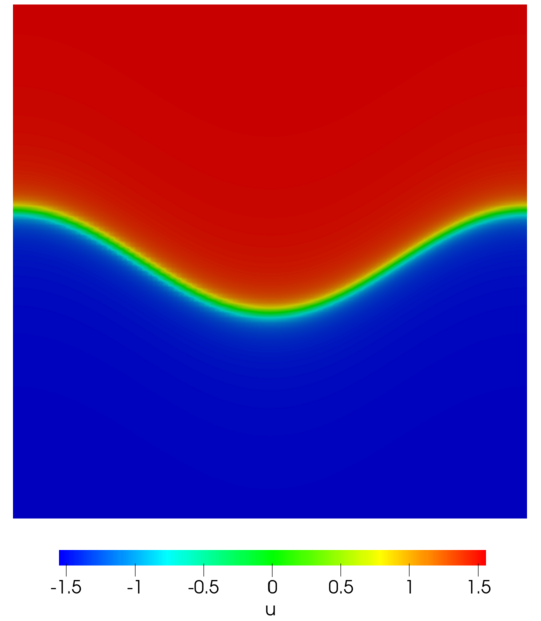}
		\caption{Values of the function $u$ for $\gamma = 10$.}
		\label{fig:u}
	\end{figure}
	In the 2D case, we consider a square domain $\Omega = \left[ -0.5,0.5 \right]^2$ and a function $u: \Omega \rightarrow \zR$ given by
	\begin{equation}\label{eq:u}
		u(x,y) := \arctan\left(\gamma\ \varphi(x,y) \right),\quad \varphi(x,y) := 10y + \cos(2\pi x).
	\end{equation}
	In Figure \ref{fig:u}, we show the values of $u$ for $\gamma = 10$.
	We observe that, near the curve $\varphi(x,y) = 0$ there is a sharp transition.
	Far away from such curve, the function is almost constant.
	
	In the 3D case, we consider a square domain $\Omega = \left[ -0.5,0.5 \right]^3$ and a function $u: \Omega \rightarrow \zR$ given by
	\begin{equation}\label{eq:u3D}
		u(x,y,z) := \arctan\left(\gamma\ \varphi(x,y,z) \right),\quad \varphi(x,y,z) := 10z + \cos(2\pi x)\cos(2\pi y).
	\end{equation}
	Analogously to the 2D case, near the surface $\varphi(x,y,z) = 0$ there is a sharp transition.
	Far away from such surface, the function is almost constant.
	
	We aim to approximate the function $u$. For this, in practical problems, the metric $\zbmetric$ is determined by an interpolation-oriented error estimator of the function $u$.
%	To approximate the function $u$ we consider an error indicator represented by a discrete target metric $\zbmetric$.
	We obtain the discrete metric $\zbmetric$ from the high-order derivatives of the function $u$ \cite{coulaud:VeryHighOrderAnisotropic}. In particular, for a mesh polynomial degree $q$, we consider the $(q+1)$th derivatives of $u$, $\nabla^{q+1} u$. Then, we obtain the discrete metric $\zbmetric$ in terms of $\nabla^{q+1} u$. To do this, we generate a background isotropic mesh $\zbmesh$ of polynomial degree $q$ and we evaluate the high-order derivatives, $\nabla^{q+1} u$, at the background mesh nodes. Finally, we obtain the values of an approximative discrete metric $\zbmetric$ at the background mesh nodes and we regularize this metric according to an $L^p$-norm and a fixed size $h$ \cite{loseille2011continuous}.
	
	\subsubsection{Size-shape distortion minimization: straight-sided anisotropic mesh adapted to the discrete metric}\label{sec:min}
	Herein, to check our distortion measure in a preliminary adaptation case, we minimize the size-shape distortion according to the discrete metric $\zbmetric$ of Section \ref{sec:metric}. To do this, we apply the methodology presented in \cite{aparicio2023combining}.
	The method considers two meshes: a background mesh $\zbmesh$ and a physical mesh $\zmesh$. First, we generate a background mesh $\zbmesh$ to interpolate the metric values $\zmetric$ in terms of the discrete metric $\zbmetric$. Then, we generate and optimize a physical mesh $\zmesh$ according to the interpolated metric $\zmetric$. This results in a triangular (tetrahedral) mesh $\zmesh^*$ with Riemannian lengths and areas (and volumes) closer to the metric unit, see Section \ref{sec:volumetric}. As a consequence, the interpolation and approximation errors are improved, see Section \ref{sec:error}.
	
	%\color{red}
	For this, we consider a background mesh $\zbmesh$ and a physical mesh $\zmesh$ of the same polynomial degree $q$, and the same characteristic size $h$. We first generate an isotropic background mesh $\zbmesh$ and we equip it with the discrete target metric $\zbmesh$. From this mesh, we generate an initial anisotropic physical mesh $\zmesh$ with the MMG mesh generator \cite{dobrzynski2012mmg3d}. In this situation, using a high-order background mesh is not possible because MMG does not allow high-order meshes as an input. Instead, we consider the linear metric interpolation in a uniformly subdivided linear background mesh $\zbmesh'$ from the generated one $\zbmesh$. We expect that both, the high-order $\zbmesh$ and the subdivided background $\zbmesh'$ meshes, represent faithfully the metric, even if their elemental node locations differ. Finally, we relocate the nodes of the initial physical mesh $\zmesh$ by minimizing the size-shape distortion measure, see Section \ref{sec:pointwise}. In this case, to obtain the point-wise varying metric $\zmetric$, we consider the high-order Log-Euclidean metric interpolation of the discrete target metric $\zbmetric$ at the high-order background mesh $\zbmesh$, see \cite{rochery2021p2,arsigny:Log-EuclideanMetrics} for the details.
	\color{black}
	
	\begin{figure}[t!]
		\centering
		\hspace{-0.35cm}
		\setlength{\tabcolsep}{15pt}
		\begin{tabular}{ccc}
			\subfigure[]{\label{fig:p1_0}
				\includegraphics[width=0.2\textwidth]{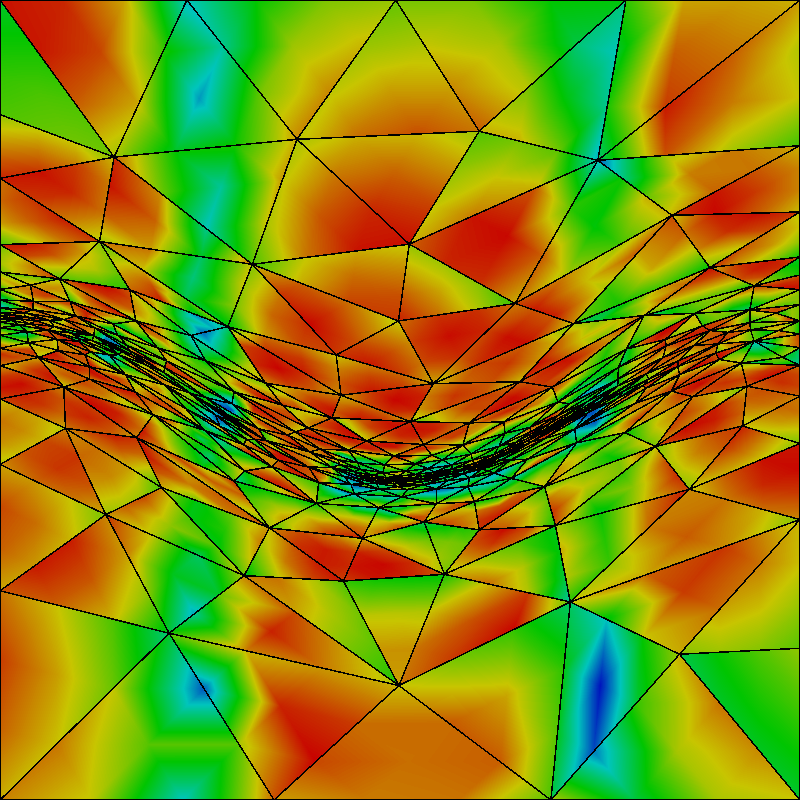}}
			&
			\subfigure[]{\label{fig:p2_0}
				\includegraphics[width=0.2\textwidth]{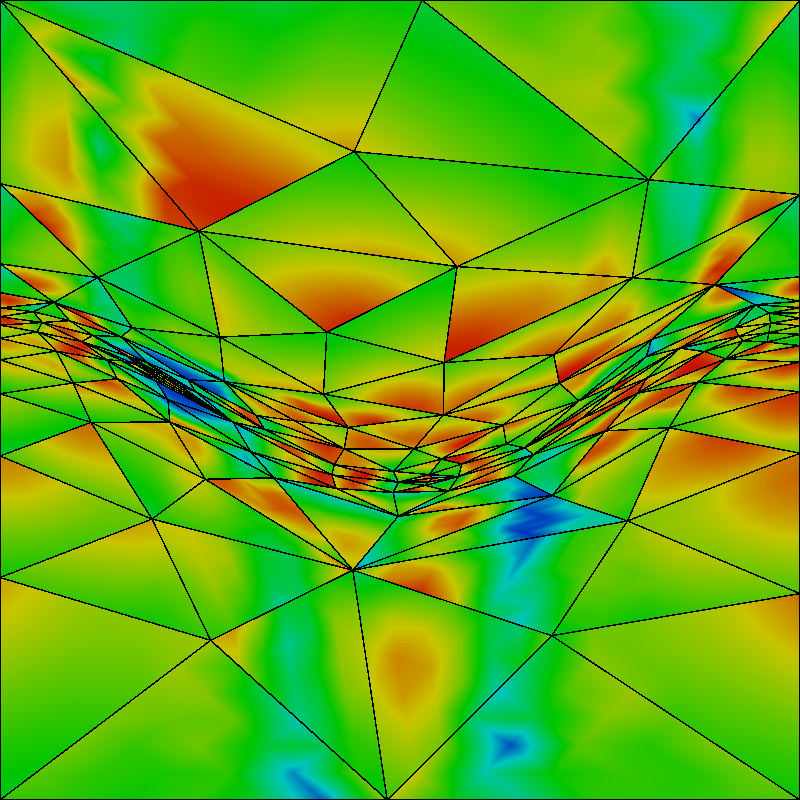}}
			&
			\subfigure[]{\label{fig:p4_0}
				\includegraphics[width=0.2\textwidth]{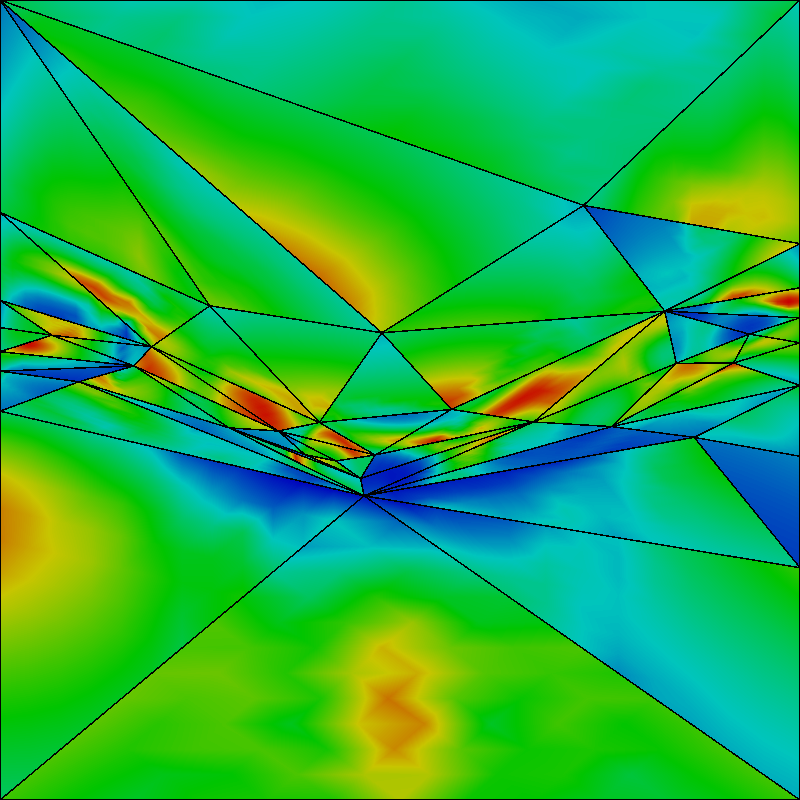}}
			\\
			\subfigure[]{\label{fig:p1_1}
				\includegraphics[width=0.2\textwidth]{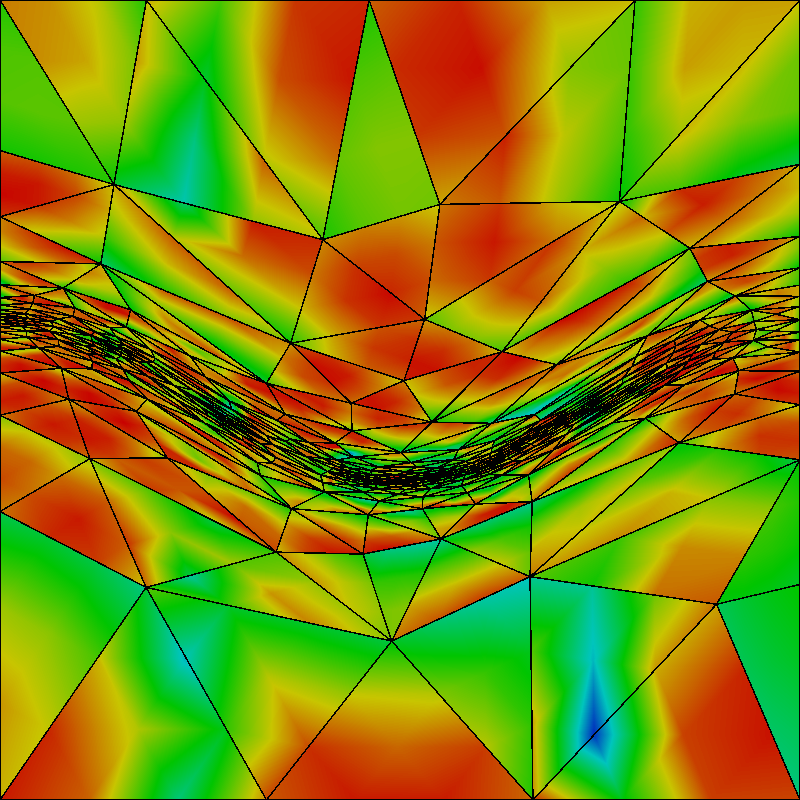}}
			&
			\subfigure[]{\label{fig:p2_1}
				\includegraphics[width=0.2\textwidth]{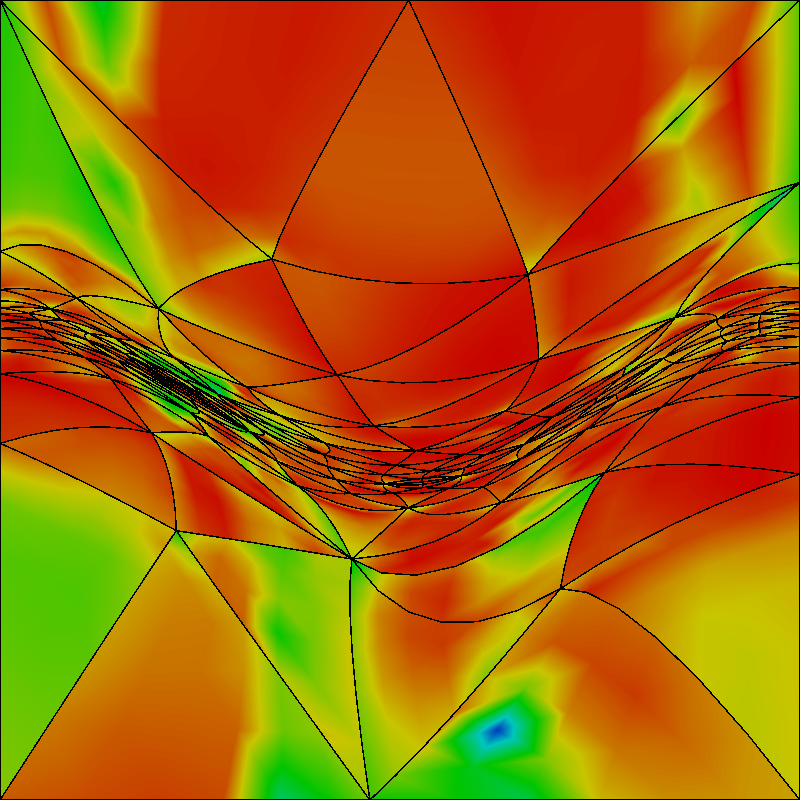}}
			&
			\subfigure[]{\label{fig:p4_1}
				\includegraphics[width=0.2\textwidth]{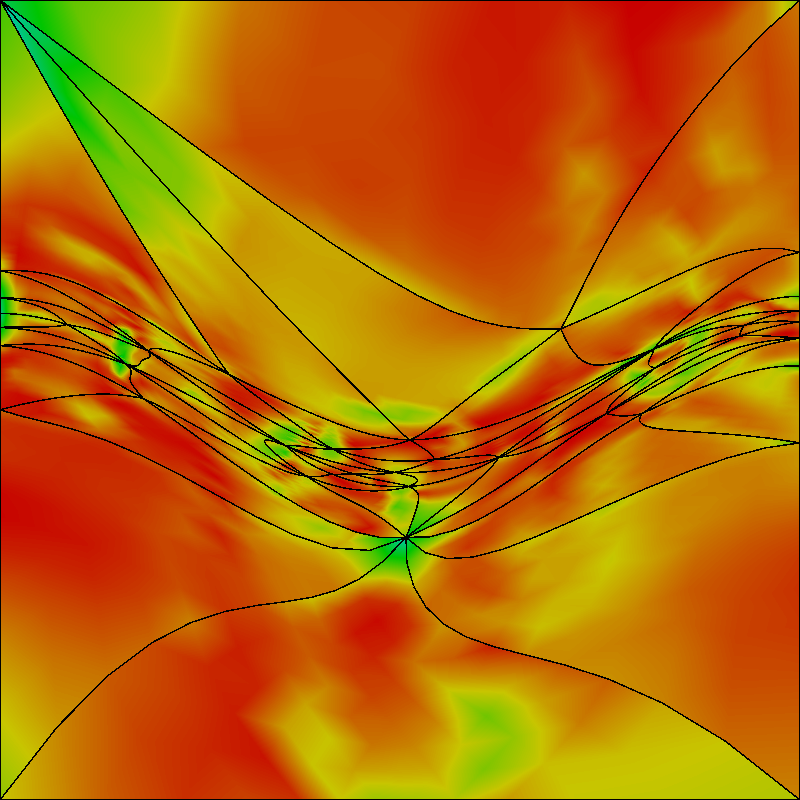}}
			\\
		\end{tabular}
		\\
		\includegraphics[width=0.25\textwidth]{./qualBarParaview_color}
		\caption{Point-wise size-shape quality measure for (rows) initial and optimized triangular meshes of (columns) polynomial degree 1, 2, and 4.}
		\label{fig:2dmeshes}
	\end{figure}
	%In Figure \ref{fig:2dmeshes}, we illustrate the initial and optimized triangular meshes equipped with the discrete metric. On the one hand, we obtain the metric from the function $u$ of Equation \eqref{eq:u} with $\gamma = 100$. In addition, we regularize the metric according to the $L^2(\Omega)$-norm, see Section \ref{sec:metric}. On the other hand, the meshes are colored according to the point-wise size-shape quality measure of Equation \eqref{eq:pointquality}. They are of polynomial degree $q=$ 1, 2, and 4, and of size $h = 0.05$. Each mesh is composed by 327, 491, and 523 nodes and 611, 230, and 61 triangles, respectively. From the initial meshes, we observe that, at the sharp transition region, the elements are stretched and aligned according to the metric. Moreover, when increasing the polynomial degree, the straight-sided elements cannot align with the curved transition region. In the optimized meshes, we observe that the elements are curved according to the point-wise stretching and alignment of the metric.
	%483, 548, and 611 nodes
	%nodes: 327, 491, 523
	%elements: 611, 230, 61
	
	\begin{figure}[t!]
		\centering
		\hspace{0.35cm}
		\setlength{\tabcolsep}{25pt}
		\begin{tabular}{cc}
			\subfigure[]{\label{fig:0}
				\includegraphics[width=0.25\textwidth]{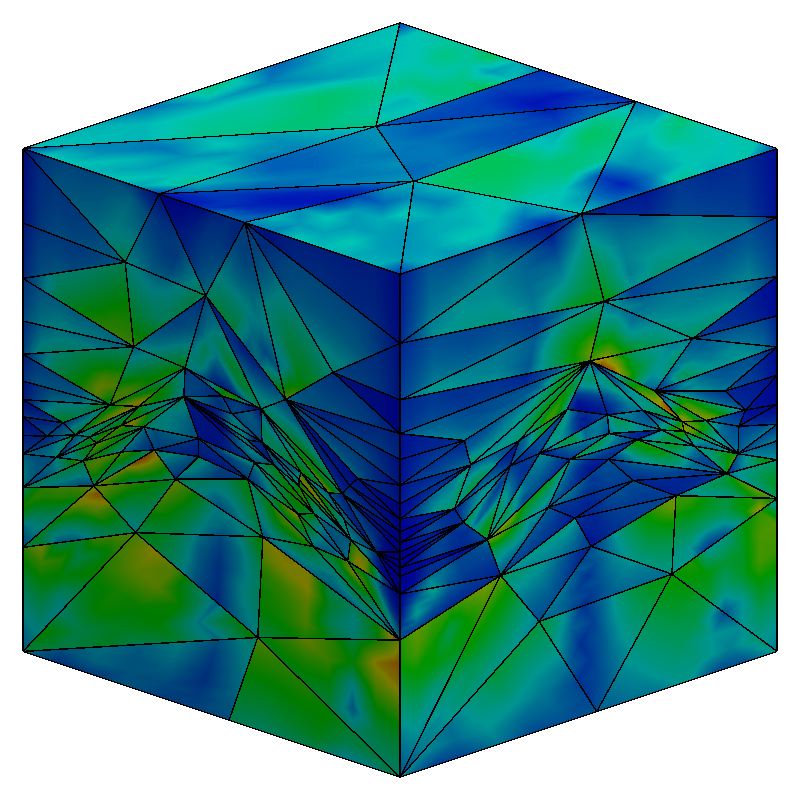}}
			&
			\subfigure[]{\label{fig:1}
				\includegraphics[width=0.25\textwidth]{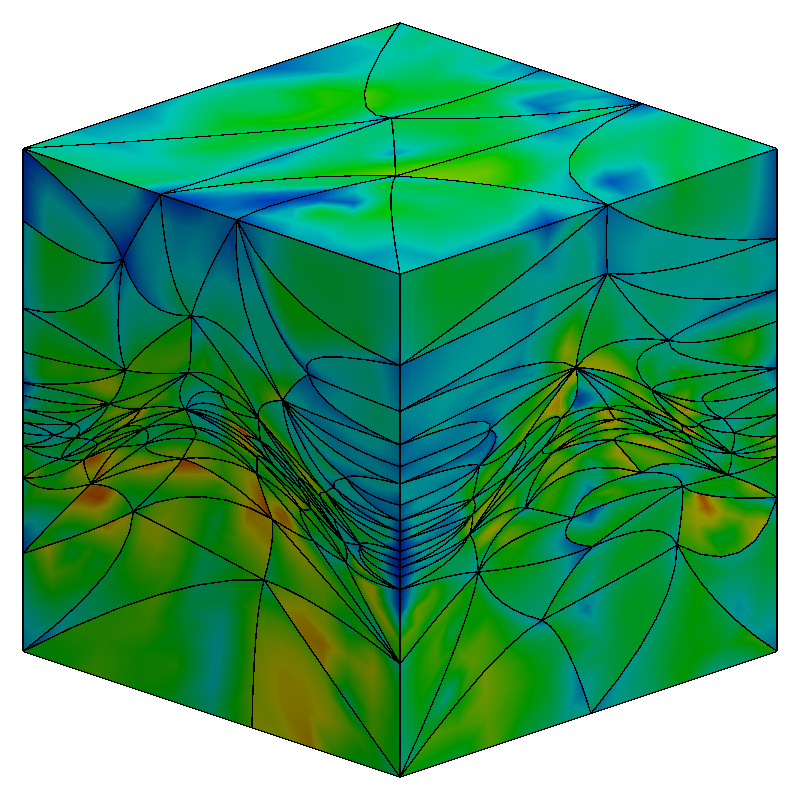}}
			\\
			\subfigure[]{\label{fig:0_clipped}
				\includegraphics[width=0.25\textwidth]{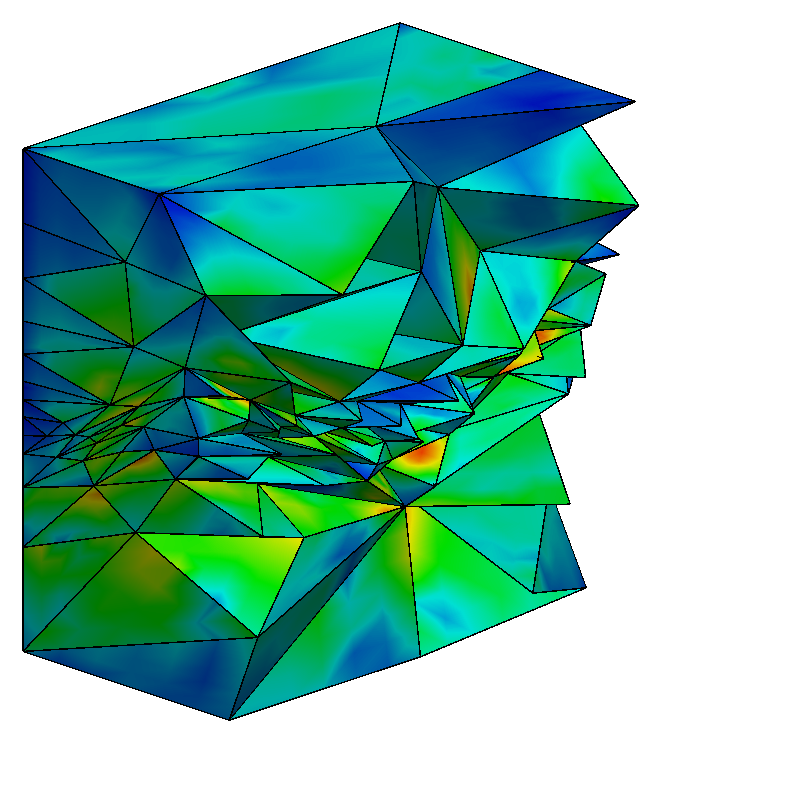}}
			&
			\subfigure[]{\label{fig:1_clipped}
				\includegraphics[width=0.25\textwidth]{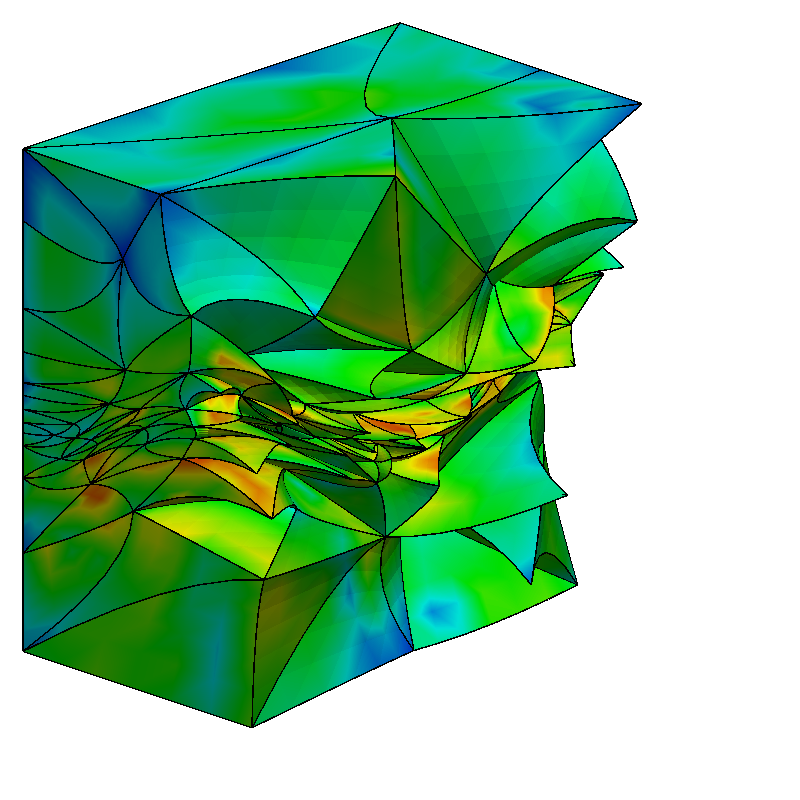}}
			\\
		\end{tabular}
		\\
		\includegraphics[width=0.25\textwidth]{./qualBarParaview_color}
		\caption{Point-wise size-shape quality measure for (columns) initial and optimized of (rows) full and clipped quadratic tetrahedral meshes.}
		\label{fig:3dmeshes}
	\end{figure}
	%\color{red}
	In Figures \ref{fig:2dmeshes} and \ref{fig:3dmeshes}, we illustrate the triangular and tetrahedral physical meshes, respectively. That is, the initial, $\zmesh$, and optimized, $\zmesh^*$, meshes equipped with the metric $\zmetric$. On the one hand, we consider the function $u$ of Equation \eqref{eq:u} with $\gamma = 100$, in 2D, and $\gamma = 10$, in 3D. Then, we obtain the metric $\zmetric$ by interpolating the discrete metric $\zbmetric$ at the background mesh $\zbmesh$. Note that, the metric scaling is imposed by regularizing the discrete metric $\zbmetric$ according to the $L^2(\Omega)$-norm, see Section \ref{sec:metric}. On the other hand, the 2D physical meshes are of polynomial degree $q=$ 1, 2, and 4, and of size $h = 0.05$. Each mesh is composed of 327, 491, and 523 nodes and of 611, 230, and 61 triangles, respectively. In the 3D case, the physical meshes are of polynomial degree $q=2$ and of size $h = 0.1$. They are composed of 3754 nodes and 2425 tetrahedra, respectively. From the initial mesh $\zmesh$, we observe that, at the sharp transition region, the elements are stretched, aligned, and scaled according to the metric. However, the straight-edged elements cannot align with the curved transition region. In contrast, for the optimized meshes $\zmesh^*$, we observe that the elements are curved according to the point-wise stretching, alignment, and sizing of the metric.
	%\color{black}
	%In Figure \ref{fig:3dmeshes}, we illustrate the initial and optimized tetrahedral meshes equipped with the discrete metric. On the one hand, we obtain the metric from the function $u$ of Equation \eqref{eq:u3D} with $\gamma = 10$. In addition, the metric is regularized according to the $L^2(\Omega)$-norm, see Section \ref{sec:metric}. On the other hand, the meshes are colored according to the point-wise size-shape quality measure of Equation \eqref{eq:pointquality}. They are of polynomial degree $q=2$ and of size $h = 0.1$. It is composed by 3754 nodes and 2425 tetrahedra, respectively. From the initial meshes, we observe that, at the sharp transition region, the elements are stretched and aligned according to the metric. However, the straight-sided elements cannot align with the curved transition region. In the optimized meshes, we observe that the elements are curved according to the point-wise stretching and alignment of the metric.
	%3754 nodes
	%2425 elements
	%background
	%2256 nodes
	%1400 tetrahedra
	\subsubsection{Verifying results: distributions for Riemannian measures of distortion and mesh entities}\label{sec:volumetric}
	Next, we illustrate how the size-shape distortion minimization enables an optimized mesh that approximates more faithfully the target metric than the initial one.
	For this, we measure the Riemannian length, area, and volume distributions of the mesh entities, see Equation \eqref{eq:metricvolpeq} and \eqref{eq:metricvol}. The results show that the size-shape distortion minimization enables an optimized mesh featuring an improved approximation of the target metric, when compared to the initial one.
	
	\begin{table}[t!]
		\caption{Size-shape quality, geometry, and error statistics of the initial meshes and the corresponding optimized triangular meshes.}
		\label{table:p124}
		\centering
		\tiny
		\setlength{\tabcolsep}{4pt}
		\begin{tabular}{ c c c c c c c c c c c}
			\hline\noalign{\smallskip}
			Measure &Mesh& \multicolumn{2}{c}{Minimum}&\multicolumn{2}{c}{Maximum}& \multicolumn{2}{c}{Mean} & \multicolumn{2}{c}{Standard deviation}\\
			&degree&Initial&Optimized&Initial&Optimized&Initial&Optimized&Initial&Optimized\\
			\noalign{\smallskip}\hline\noalign{\smallskip}
			&1 & 0.1019 & 0.2574 & 0.9779 & 0.9737 & 0.6710 & 0.7330 & 0.1898 & 0.1391\\
			Quality&2 & 0.0986 & 0.5229 & 0.9161 & 0.9812 & 0.6021 & 0.8538 & 0.1604 & 0.0883\\
			&4 & 0.0249 & 0.6881 & 0.7565 & 0.9275 & 0.3756 & 0.8307 & 0.1761 & 0.0523\\
			\noalign{\smallskip}\hline\noalign{\smallskip}
			&1 & 0.3641 & 0.4293 & 5.1197 & 3.2040 & 1.2916 & 1.2520 & 0.5376 & 0.3503 \\
			Length&2 & 0.4711 & 0.5726 & 4.6246 & 2.3879 & 1.1160 & 1.0269 & 0.5172 & 0.2619 \\
			&4 & 0.3109 & 0.3295 & 4.0158 & 1.7523 & 1.1334 & 0.9856 & 0.7727 & 0.2653 \\
			\noalign{\smallskip}\hline\noalign{\smallskip}
			&1 & 0.1998 & 0.3300 & 5.2956 & 3.5361 & 1.3135 & 1.3137 & 0.7967 & 0.6158 \\
			Area&2 & 0.2838 & 0.4376 & 5.1946 & 2.3034 & 0.8695 & 0.8696 & 0.6145 & 0.3219 \\
			&4 & 0.0911 & 0.4923 & 2.9104 & 1.5600 & 0.8195 & 0.8196 & 0.6847 & 0.2312 \\
			\noalign{\smallskip}\hline\noalign{\smallskip}
		\end{tabular}
	\end{table}
	%In Table \ref{table:p124}, we show the triangular mesh statistics for the logarithmic distributions of elemental qualities (Equation \eqref{eq:qualityreg}), lengths, and areas. They allow us to compare the geometric quantities between the initial and optimized triangular meshes in terms of the target metric. We observe that the maximum, minimum, mean, and standard deviation become closer to unit values in almost all cases. That is, in general, all statistics are improved.
	
	\begin{table}[t!]
		\caption{Size-shape quality, geometry, and error statistics of the initial and optimized quadratic tetrahedral meshes.}
		\label{table:3D}
		\centering
		\tiny
		\begin{tabular}{ c c c c c c c c c c}
			\hline\noalign{\smallskip}
			Measure & \multicolumn{2}{c}{Minimum}&\multicolumn{2}{c}{Maximum}& \multicolumn{2}{c}{Mean} & \multicolumn{2}{c}{Standard deviation}\\
			&Initial&Optimized&Initial&Optimized&Initial&Optimized&Initial&Optimized\\
			\noalign{\smallskip}\hline\noalign{\smallskip}
			Quality & 0.0021 & 0.2117 & 0.8351 & 0.8545 & 0.3414 & 0.5315 & 0.1580 & 0.1082\\
			Length& 0.1760 & 0.2388 & 3.4170 & 3.4307 & 0.9340 & 0.8728 & 0.3686 & 0.2560 \\
			Area& 0.0687 & 0.1673 & 3.6299 & 2.0579 & 0.5906 & 0.5557 & 0.3148 & 0.1797 \\
			Volume& 0.0229 & 0.1168 & 2.3427 & 0.9282 & 0.3097 & 0.3097 & 0.2179 & 0.0941 \\
			\noalign{\smallskip}\hline\noalign{\smallskip}
		\end{tabular}
	\end{table}
	%\color{green}
	In Tables \ref{table:p124} and \ref{table:3D}, we show the corresponding triangular and tetrahedral mesh statistics for the logarithmic distributions of elemental qualities (Equation \eqref{eq:qualityreg}) and Riemannian measures. That is, lengths and areas in 2D and lengths, areas, and volumes in 3D. They allow us to compare the geometric quantities between the initial and optimized physical meshes in terms of the target metric.
	%We observe that the minimum and standard deviation become closer to unit values in all cases.
	We observe that the maximum, minimum, mean, and standard deviation become closer to unit values in almost all cases. That is, in general, all statistics are improved.
	We also observe the mentioned behavior in Tables \ref{table:p3}, \ref{table:p4}, and \ref{table:poisson} corresponding to the following examples of Sections \ref{sec:errormin}, \ref{sec:curving}, and \ref{sec:pde}.
	\color{black}
	%Similarly, in Table \ref{table:3D}, we show the tetrahedral mesh statistics for the logarithmic distributions of elemental qualities (Equation \eqref{eq:qualityreg}), lengths, areas, and volumes. They allow us to compare the geometric quantities between the initial and optimized tetrahedral meshes in terms of the target metric. We observe that the minimum and standard deviation become closer to unit values in all cases.
	%In addition, the maximum become closer around unit values in almost all cases.
	% That is, in general, all statistics are improved.
	
	\begin{figure}[t!]
		\centering
		\hspace{-0.5cm}
		\setlength{\tabcolsep}{-2pt}
		\begin{tabular}{cc}
			\subfigure[]{\label{fig:qp1}
				\includegraphics[width=0.5\textwidth]{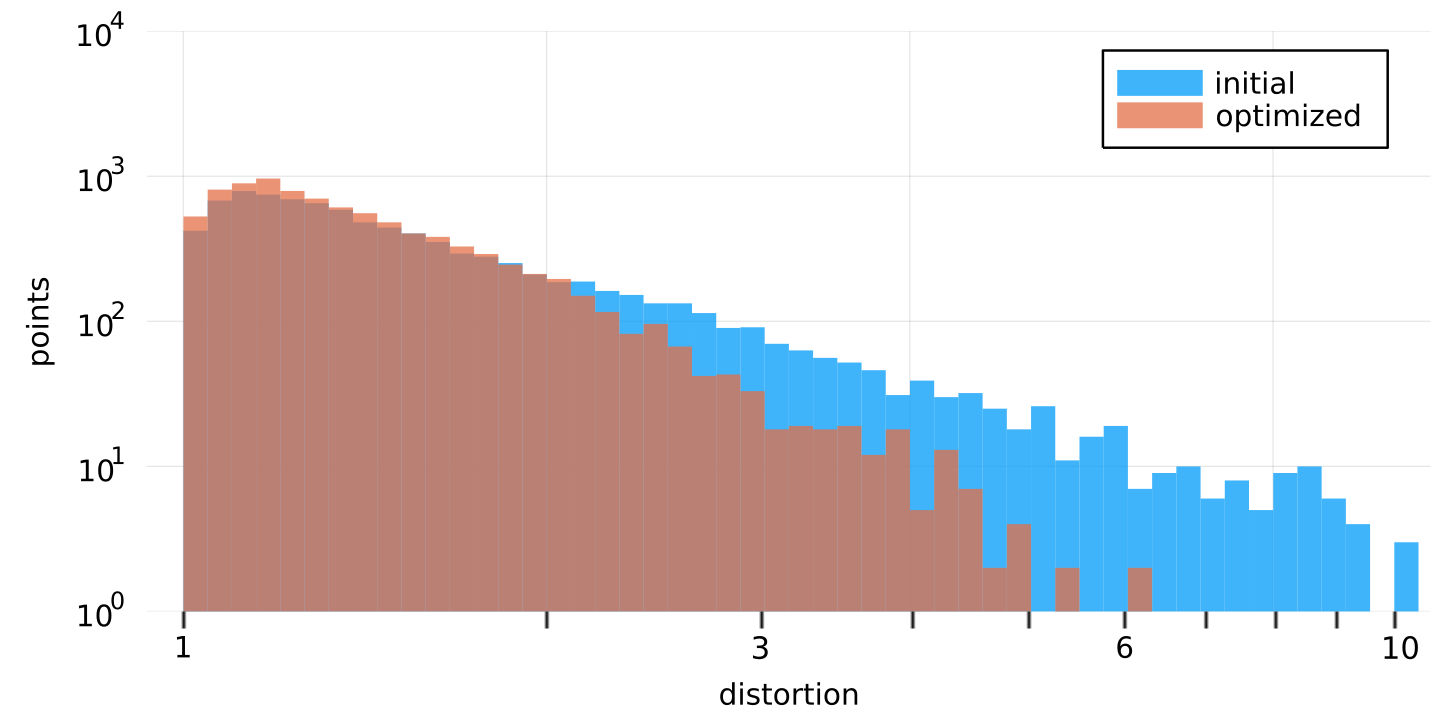}}
			&
			\subfigure[]{\label{fig:qp2}
				\includegraphics[width=0.5\textwidth]{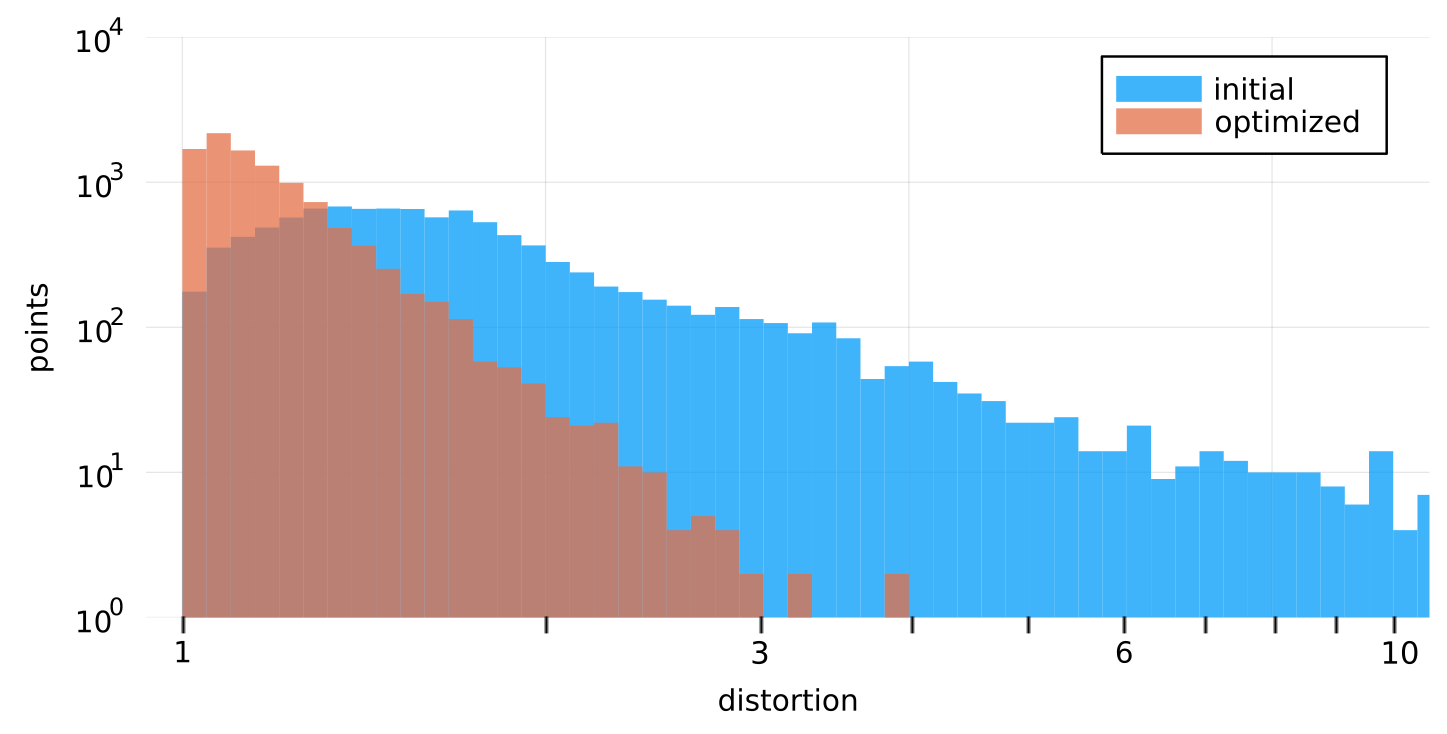}}
			\\
			\subfigure[]{\label{fig:qp4}
				\includegraphics[width=0.5\textwidth]{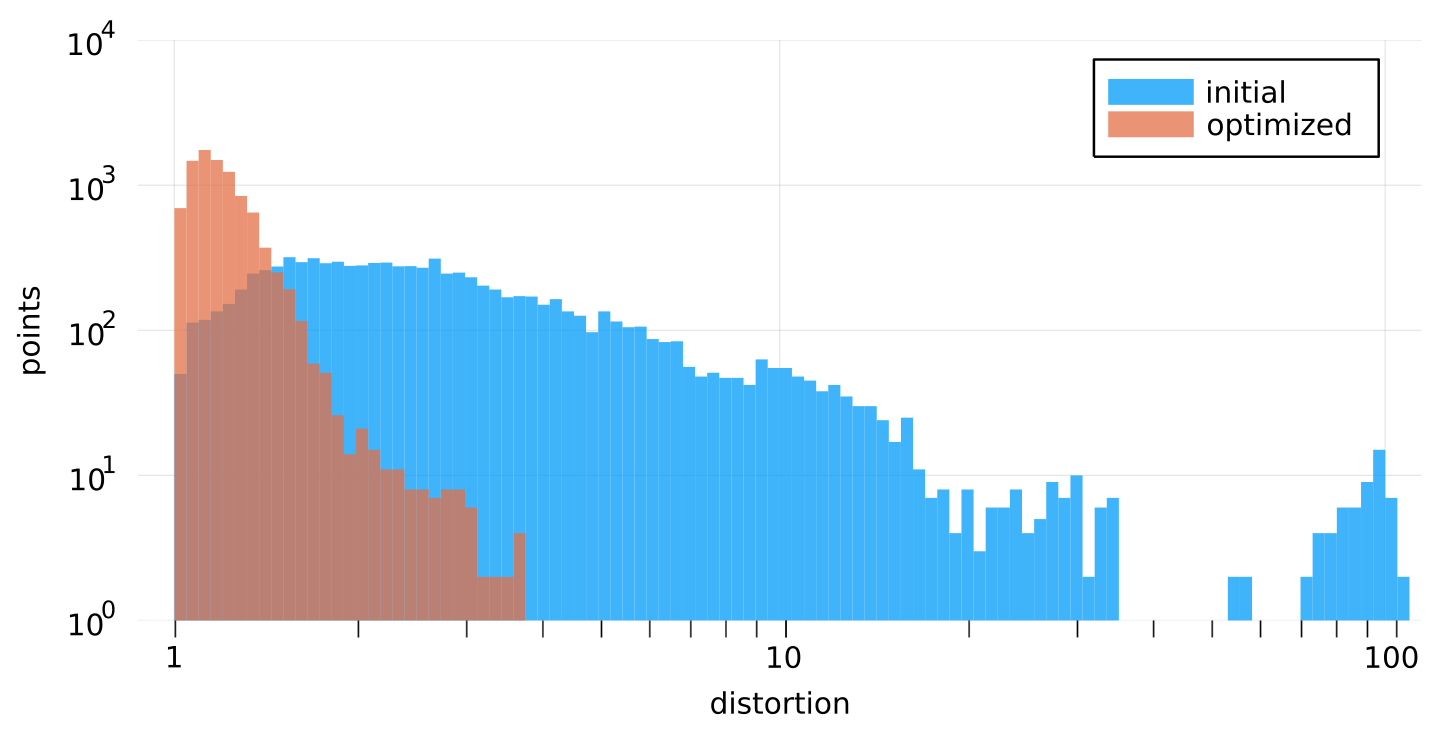}}
			\\
		\end{tabular}
		\caption{Logarithmic point-wise size-shape distortion histograms for (blue) initial and (orange) optimized meshes of polynomial degree 1, 2, and 4, respectively.}
		\label{fig:histquality}
	\end{figure}
	\begin{figure}[t!]
		\centering
		\hspace{-0.5cm}
		\setlength{\tabcolsep}{-2pt}
		\begin{tabular}{cc}
			\subfigure[]{\label{fig:lp1}
				\includegraphics[width=0.5\textwidth]{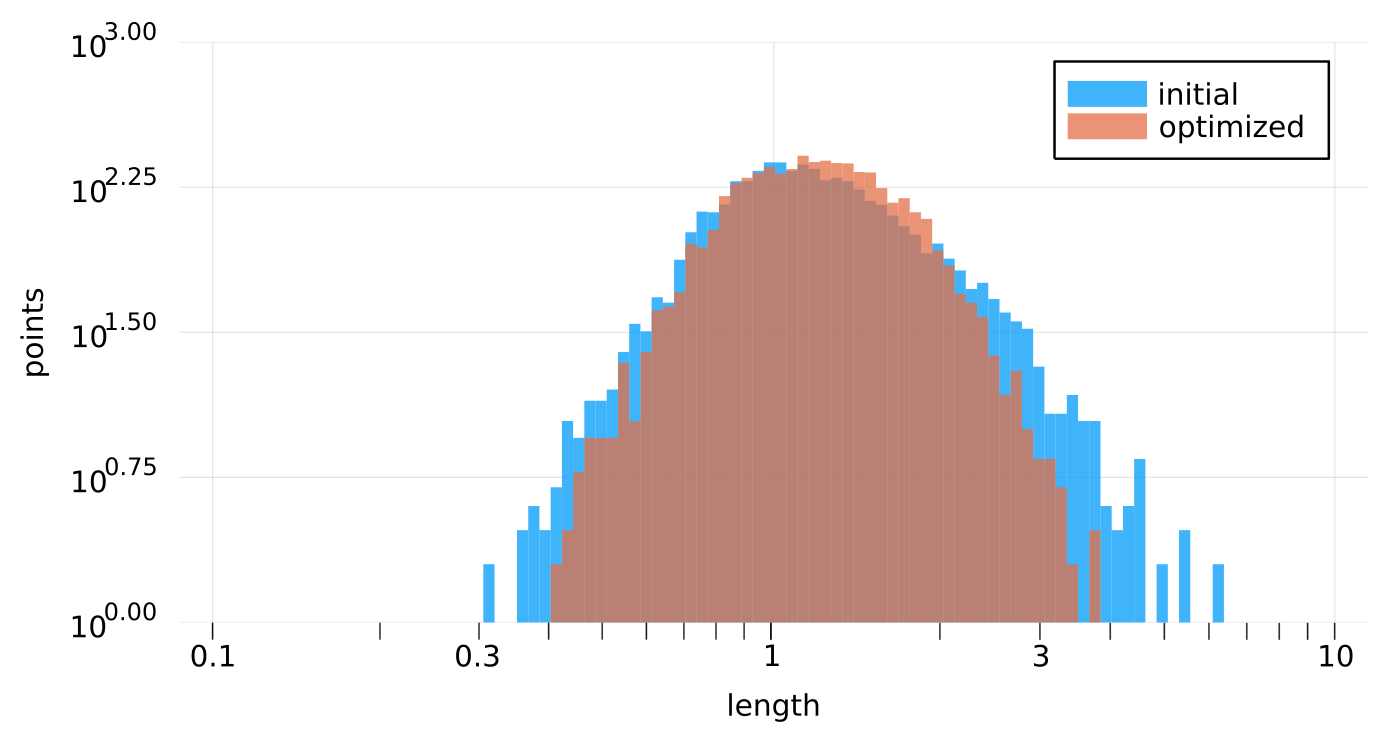}}
			&
			\subfigure[]{\label{fig:lp2}
				\includegraphics[width=0.5\textwidth]{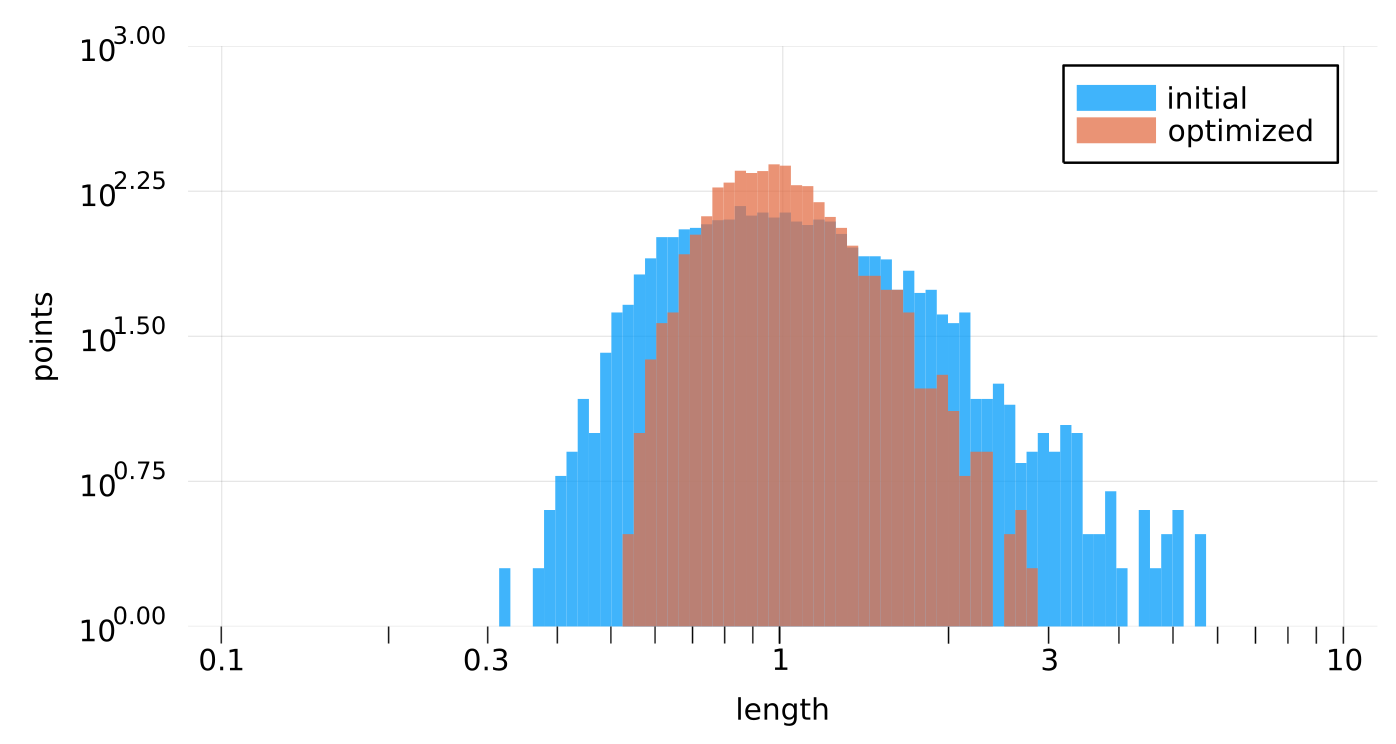}}
			\\
			\subfigure[]{\label{fig:lp4}
				\includegraphics[width=0.5\textwidth]{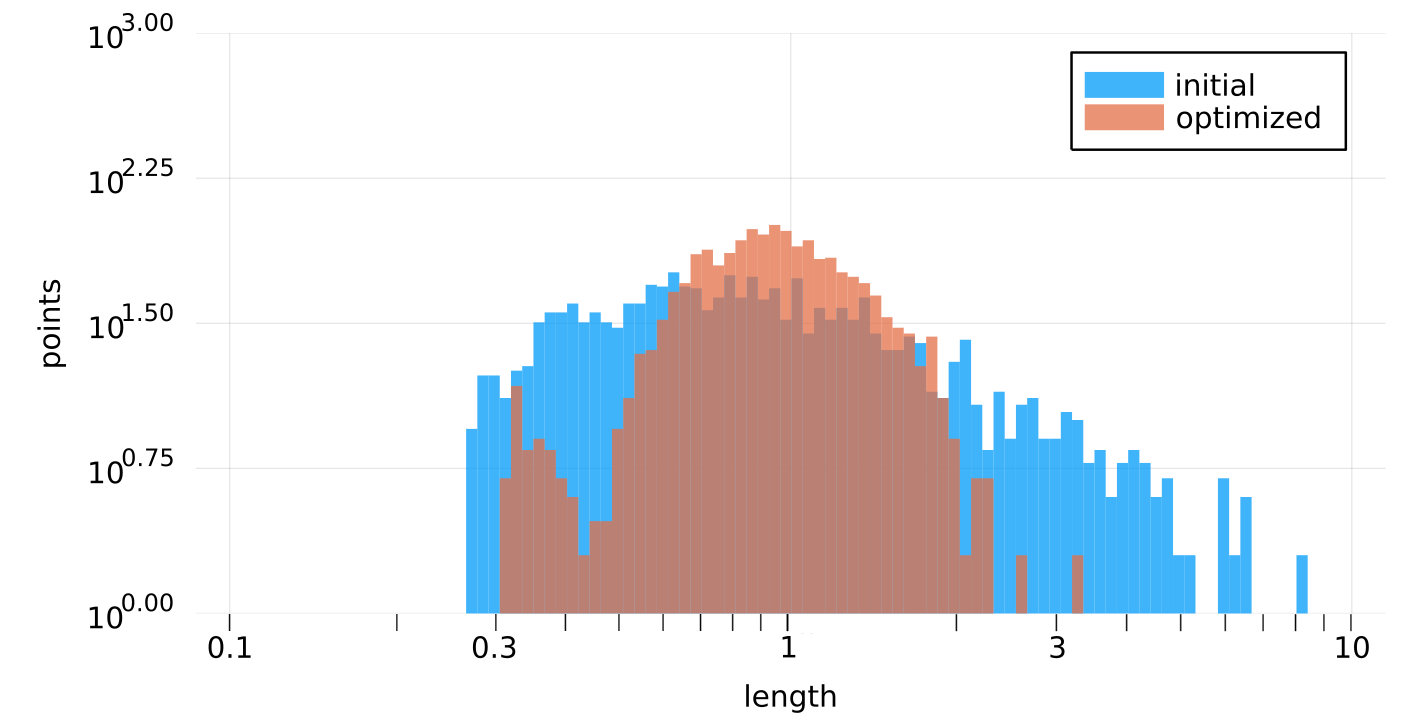}}
			\\
		\end{tabular}
		\caption{Logarithmic Riemannian length density histograms for (blue) initial and (orange) optimized meshes of polynomial degree 1, 2, and 4, respectively.}
		\label{fig:histlength}
	\end{figure}
	\begin{figure}[t!]
		\centering
		\hspace{-0.5cm}
		\setlength{\tabcolsep}{-2pt}
		\begin{tabular}{cc}
			\subfigure[]{\label{fig:ap1}
				\includegraphics[width=0.5\textwidth]{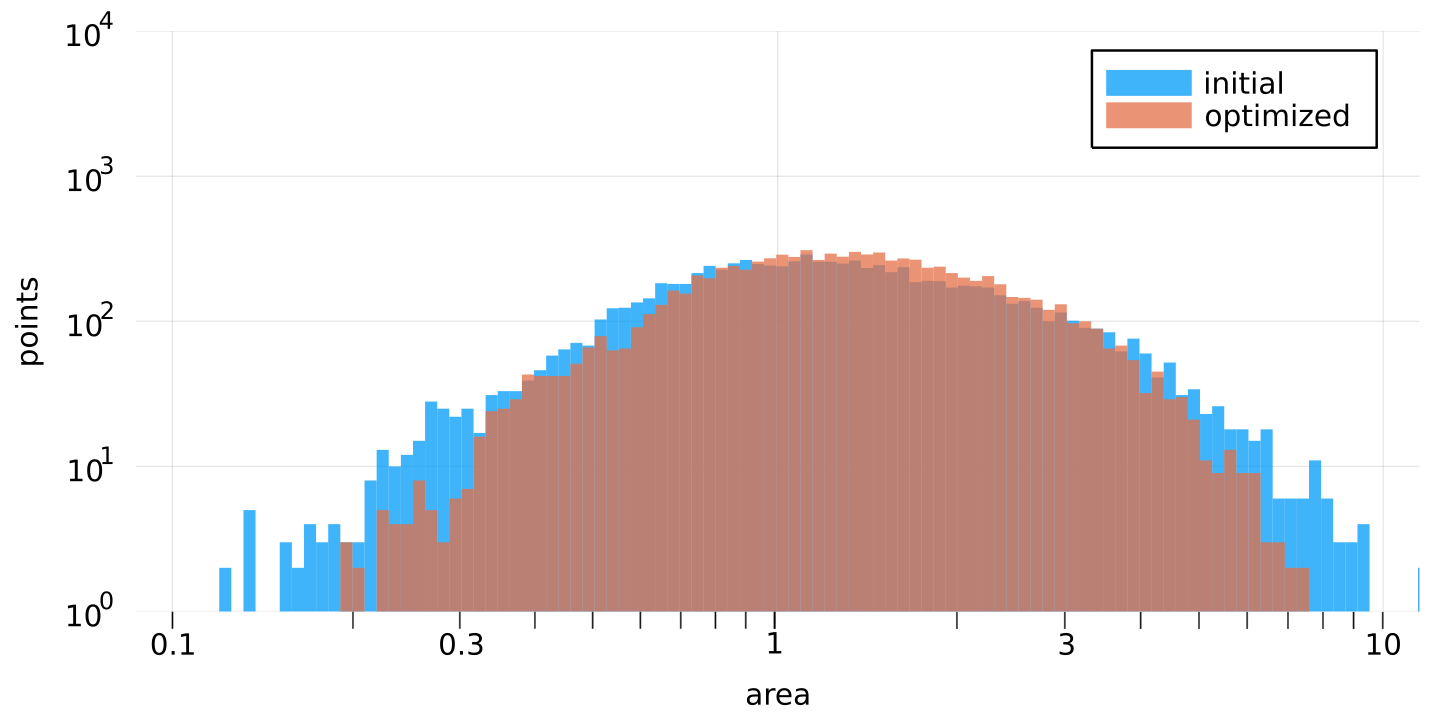}}
			&
			\subfigure[]{\label{fig:ap2}
				\includegraphics[width=0.5\textwidth]{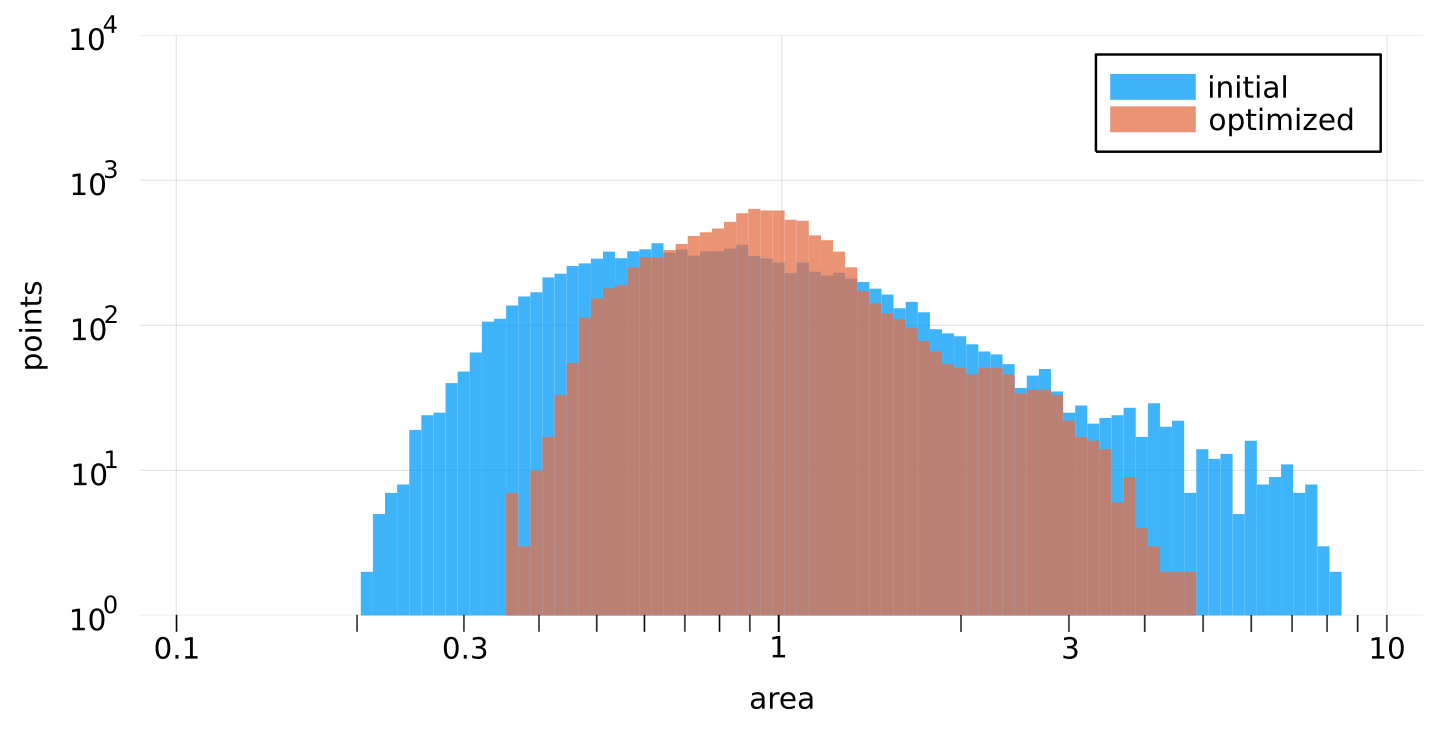}}
			\\
			\subfigure[]{\label{fig:ap4}
				\includegraphics[width=0.5\textwidth]{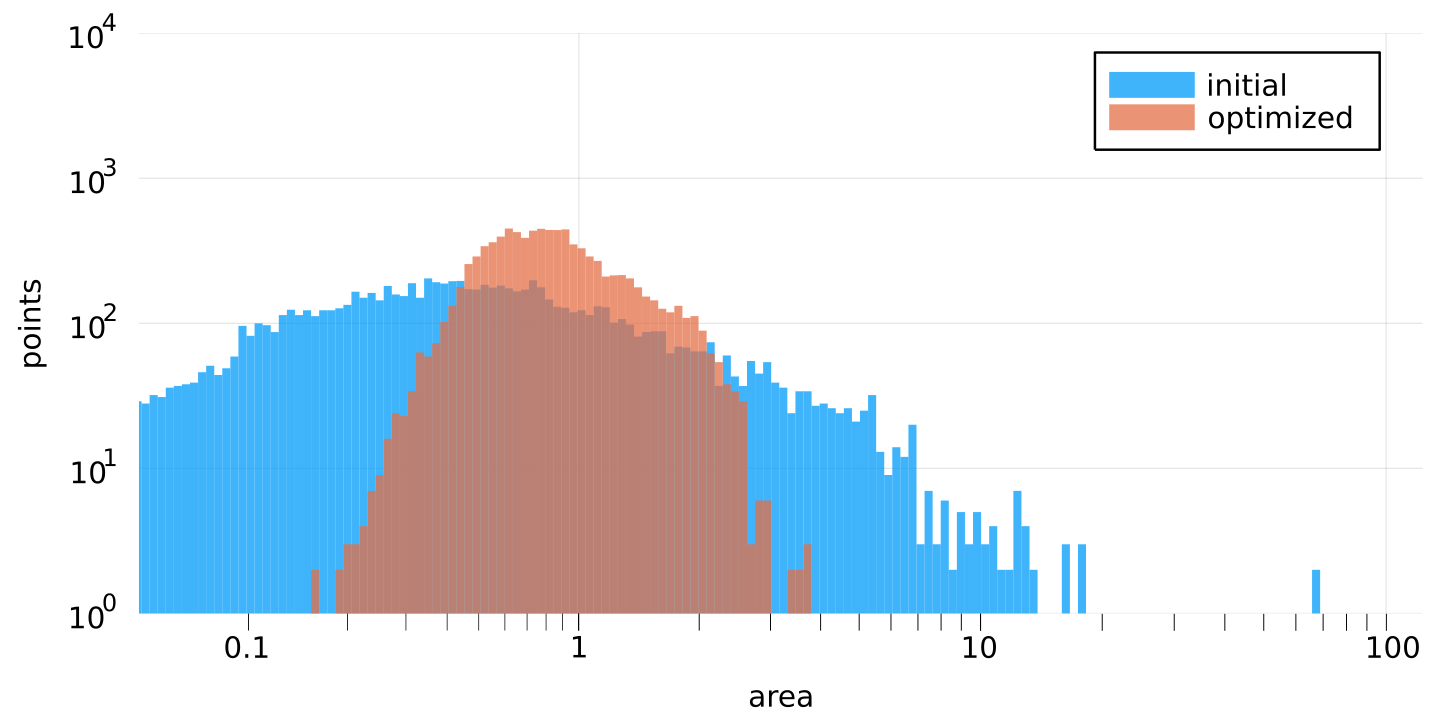}}
			\\
		\end{tabular}
		\caption{Logarithmic Riemannian area density histograms for (blue) initial and (orange) optimized meshes of polynomial degree 1, 2, and 4, respectively.}
		\label{fig:histarea}
	\end{figure}
	\begin{figure}[t!]
		\centering
		\hspace{-0.5cm}
		\setlength{\tabcolsep}{-2pt}
		\begin{tabular}{cc}
			%		\subfigure[]{\label{fig:qp3D}
				%			\includegraphics[width=0.5\textwidth]{./interpolation/caplan/3D/quality_pointwise.png}}
			%		&
			\subfigure[]{\label{fig:lp3D}
				\includegraphics[width=0.5\textwidth]{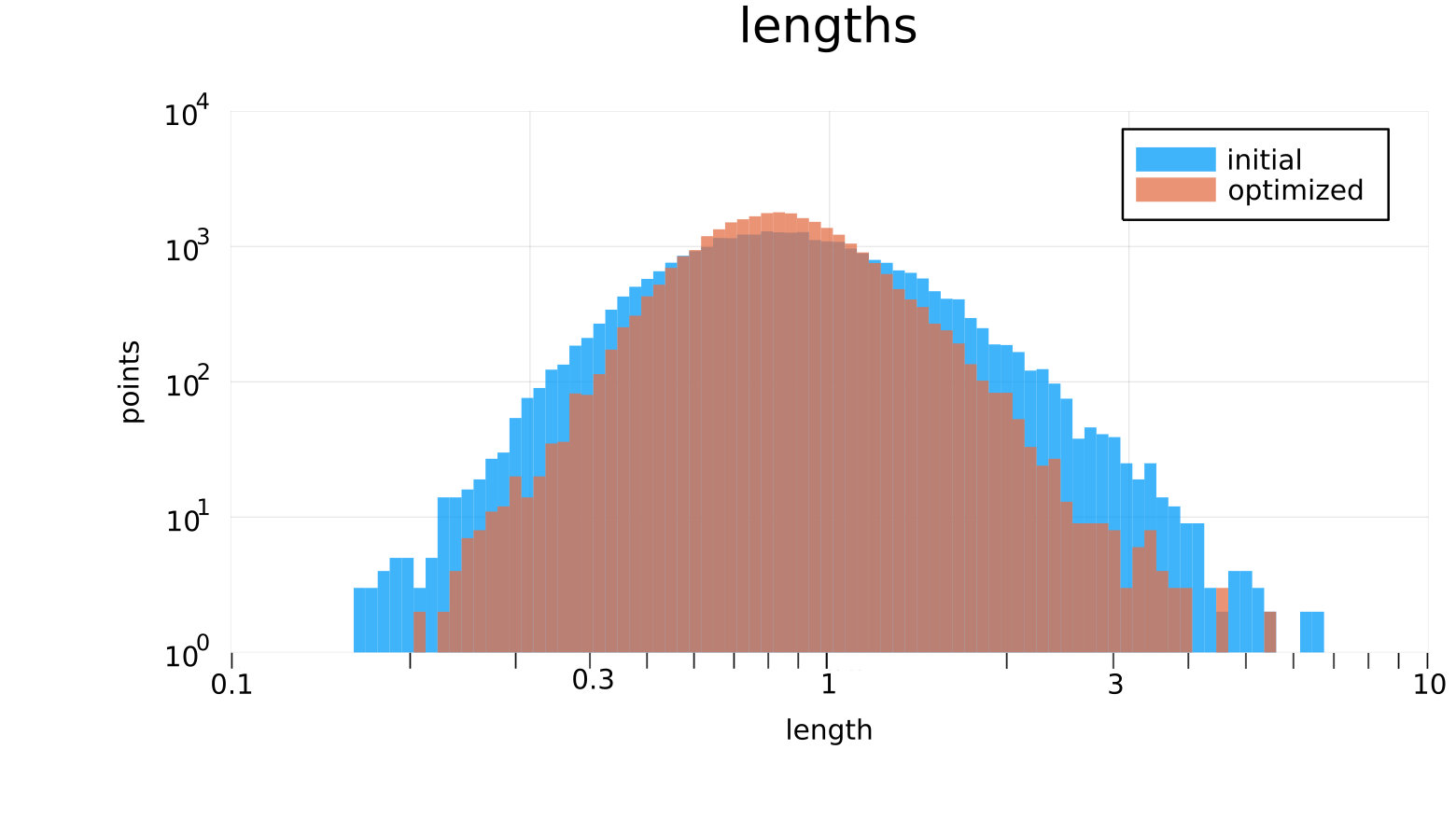}}
			&
			\subfigure[]{\label{fig:ap3D}
				\includegraphics[width=0.5\textwidth]{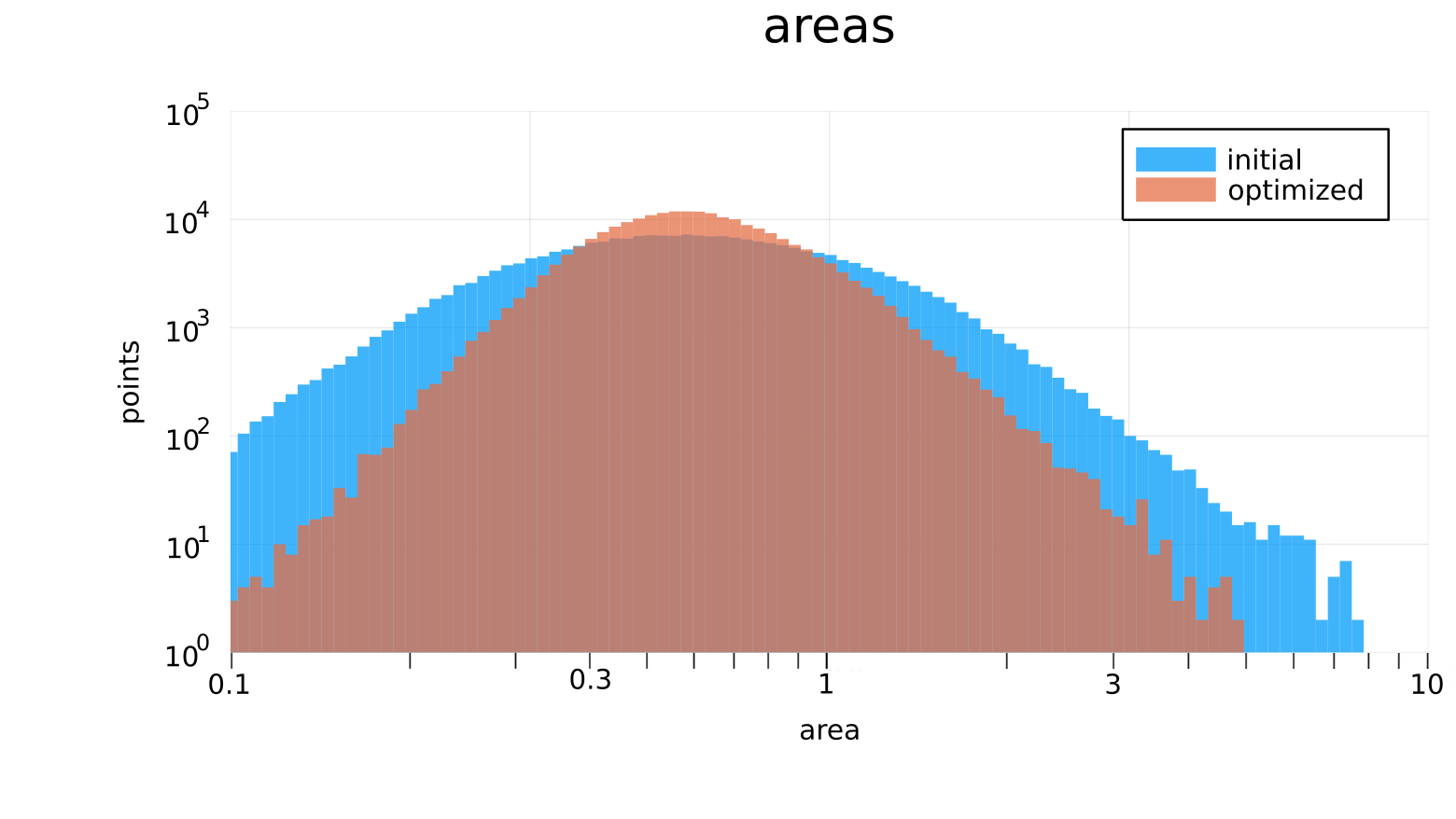}}
			\\
			\subfigure[]{\label{fig:vp3D}
				\includegraphics[width=0.5\textwidth]{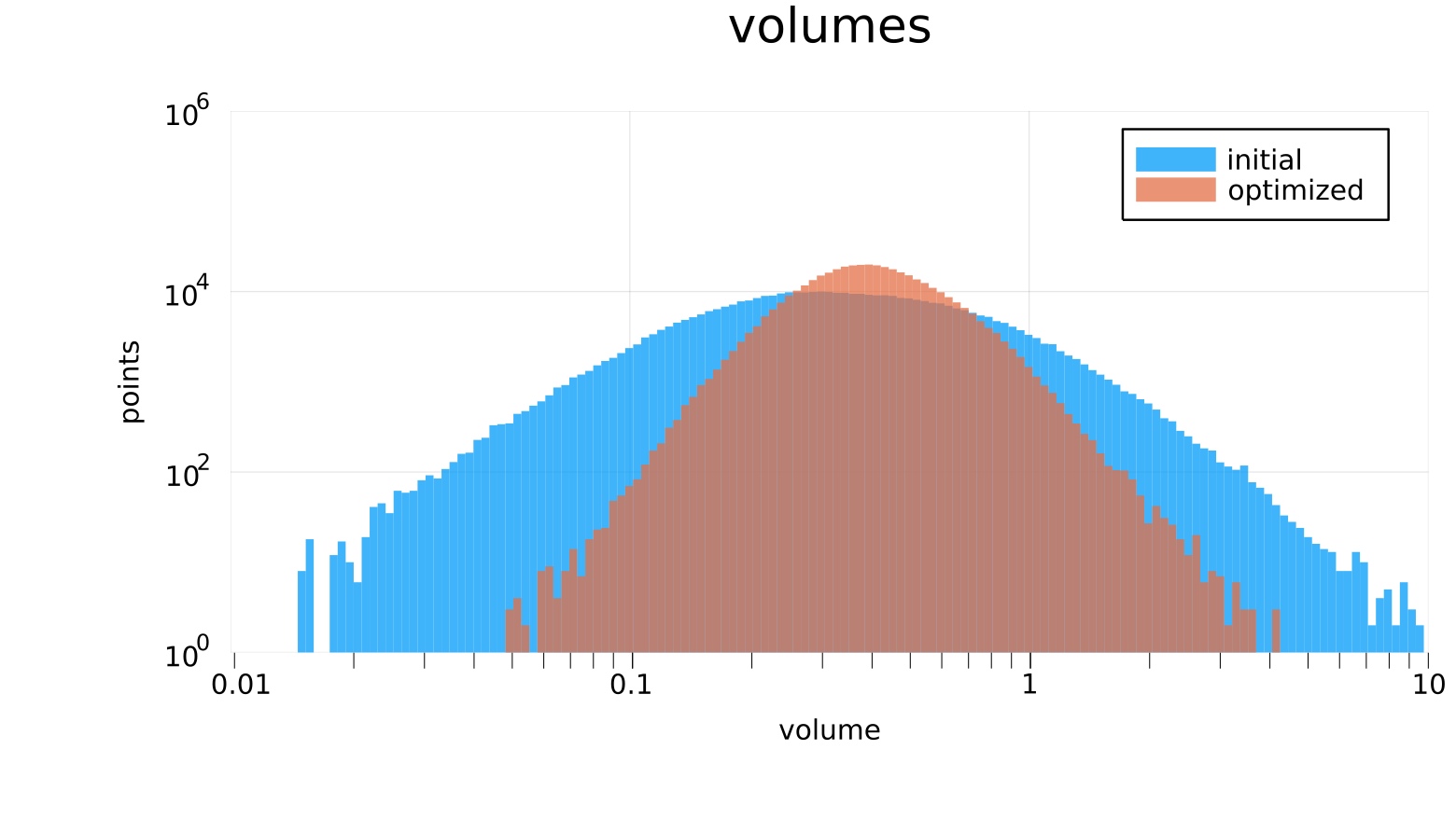}}
		\end{tabular}
		\caption{Logarithmic Riemannian length, area, and volume density histograms for (blue) initial and (orange) optimized quadratic tetrahedral meshes.}
		%	\caption{Logarithmic point-wise distortion, length, area, and volume histograms for (blue) initial and (orange) optimized quadratic tetrahedral meshes.}
		\label{fig:hist3D}
	\end{figure}
	%Similarly, in Figures, \ref{fig:qp3d}, \ref{fig:lp3D}, \ref{fig:ap3D}, and \ref{fig:vp3D} we respectively show the point-wise distortion, length, area, and volume of the initial and optimized tetrahedral meshes.
	%\color{orange}
	In Figures \ref{fig:histquality}, \ref{fig:histlength}, and \ref{fig:histarea}, we respectively show the point-wise distortion, length, and area of the initial and optimized triangular meshes. Similarly, in Figures \ref{fig:hist3D}\subref{fig:lp3D}, \ref{fig:hist3D}\subref{fig:ap3D}, and \ref{fig:hist3D}\subref{fig:vp3D}, we respectively show the point-wise Riemannian length, area, and volume densities of the initial and optimized quadratic tetrahedral meshes. Note that, the geometric quantities are typically compared in terms of ratios that is, in a multiplicative form. Accordingly, we use a logarithmic scale to illustrate the different scales of the corresponding ratios. The logarithmic representation illustrates the behavior near the minimum, maximum, and geometric mean of the distribution.
	
	From the reasoning presented in Section \ref{sec:intrinsicmetric}, we observe that almost all measure statistics are improved for the optimized meshes. On the one hand, for the geometric measures, the tails are reduced in measure (horizontal axis) and magnitude (vertical axis). This reduction is because the quality measure is sensitive to points with volume far from the unit. Hence, these regions gain priority during the optimization process. On the other hand, the distribution peak is increased. This increase is so because the global optimization of the squared quality measure tends to homogenize the points near a mean. Meanwhile, the measure and magnitude are almost preserved.
	We also observe the mentioned behavior in the histograms of Figures \ref{fig:histcurving} and \ref{fig:histpoisson} corresponding to the following examples of Sections \ref{sec:curving} and \ref{sec:pde}.
	
	\subsubsection{Interpolation and approximation error: curved high-order mesh matching the metric}\label{sec:error}
	%For this reason, we expect higher stretching values of the approximative metric only near the curve $\varphi(x,y) = 0$.
	To measure how a mesh $\zmesh$ supports the approximation of the function $u$, we consider two error indicators \cite{brenner2008mathematical}: the interpolation and the approximation errors.
	For briefness, we restrict to the $L^2(\Omega)$-norm error for a given domain $\Omega$.
	On the one hand, the interpolation error $e_I$ is defined by
	\begin{equation*}
		e_I = \| u - \Pi_{\zmesh} u \|_{L^2\left( \Omega \right)},
	\end{equation*}
	where $\Pi_{\zmesh}$ is the continuous mesh interpolation operator. It projects a function $u$ to an interpolative basis with the nodal distribution detailed in \cite{warburton2006explicit}.
	On the other hand, we consider the approximation error $e_A$ in the continuous Galerkin finite element space $V_{\zmesh}$ defined by
	\begin{equation*}
		e_A = \min\limits_{v\in V_{\zmesh}} \| u - v \|_{L^2\left( \Omega \right)}.
	\end{equation*}
	Note that, since the interpolated function belongs to the finite element space, that is $\Pi_{\zmesh} u$ in $V_{\zmesh}$, the approximation error is less or equal than the interpolation error, \textit{i.e.}, $e_A \leq e_I$.
	%In addition, we remark that the interpolation and approximation errors are defined for a global and element domains $\Omega$.
	
	\begin{table}[t!]
		\caption{Global interpolation and approximation $L^2$-error of the initial meshes and the corresponding optimized meshes.}
		\label{table:interpolation}
		\centering
		\begin{tabular}{c c c c c c c}
			\hline\noalign{\smallskip}
			Dimension&Mesh&Nodes&\multicolumn{2}{c}{Interpolation error}&\multicolumn{2}{c}{Approximation error}\\
			&degree&&Initial&Optimized&Initial&Optimized\\
			\noalign{\smallskip}\hline\noalign{\smallskip}
			2&1&327&0.0494&0.0382&0.0382&0.0302\\
			2&2&491&0.0404&0.0235&0.0314&0.0199\\
			2&4&523&0.0980&0.0336&0.0688&0.0251\\
			3&2&3754&0.0253&0.0121&0.0179&0.0089\\
			\noalign{\smallskip}\hline\noalign{\smallskip}
		\end{tabular}
	\end{table}
	\begin{table}[t!]
			\caption{Maximum interpolation and approximation elemental $L^2$-error of the initial meshes and the corresponding optimized meshes.}
			\label{table:interpolationmax}
			\centering
			\begin{tabular}{c c c c c c c}
				\hline\noalign{\smallskip}
				Dimension&Mesh&Nodes&\multicolumn{2}{c}{Interpolation error}&\multicolumn{2}{c}{Approximation error}\\
				&degree&&Initial&Optimized&Initial&Optimized\\
				\noalign{\smallskip}\hline\noalign{\smallskip}
				2&1&327&0.0151&0.0138&0.0382&0.0302\\
				2&2&491&0.0200&0.0071&0.0314&0.0199\\
				2&4&523&0.0524&0.0140&0.0688&0.0251\\
				3&2&3754&0.0059&0.0018&0.0179&0.0089\\
				\noalign{\smallskip}\hline\noalign{\smallskip}
			\end{tabular}
		\end{table}
	%\color{purple}
	The presented example shows how our method can be used to improve the error of a straight-edged mesh.
	In Tables \ref{table:interpolation} and \ref{table:interpolationmax}, we respectively show the global and maximum interpolation and approximation errors of the initial and optimized triangular and tetrahedral meshes of Section \ref{sec:min}.
%	In Table \ref{table:interpolation}, we show the global interpolation and approximation error of the initial and optimized triangular and tetrahedral meshes of Section \ref{sec:min}.
	We observe that all quantities are improved.
	Note that the global approximation error is less than the global interpolation one. This is so because the best approximation approximates better the analytic function than the interpolated one.
	We also observe this behavior in Tables \ref{table:interpolation3}, \ref{table:interpolation4}, and \ref{table:interpolationpoisson} corresponding to the following examples of Sections \ref{sec:errormin}, \ref{sec:curving}, and \ref{sec:pde}.
	
	%\color{blue}
	In addition, the presented example and those of Sections \ref{sec:errormin}, \ref{sec:curving}, and \ref{sec:pde}, show the capability of curved elements to capture sharp curved transition regions with straight-edged (Sections \ref{sec:errormin} and \ref{sec:pde}) and curved boundaries (Section \ref{sec:curving}). We observe that, even if the straight-edged elements approximate the curved transition region, this is not sufficient. Only when we curve them, we reduce the interpolation and approximation errors. Furthermore, when considering a manufactured solution, also the numerical error is reduced, see Section \ref{sec:pde}.
	
	%\color{teal}
	Note that, similarly to the quality and geometry measures, a greater improvement is achieved for the high-order cases. In particular, we observe that the quartic triangular mesh is the one featuring the worst interpolation and approximation error. This is because the initial quartic mesh has low quality elements. Accordingly, the approximation of the function for the optimized mesh is limited by the initial mesh quality.
	
	\begin{figure}[t!]
		\centering
		\hspace{-0.5cm}
		\setlength{\tabcolsep}{-2pt}
		\begin{tabular}{cc}
			\subfigure[]{\label{fig:ierror3D}
				\includegraphics[width=0.5\textwidth]{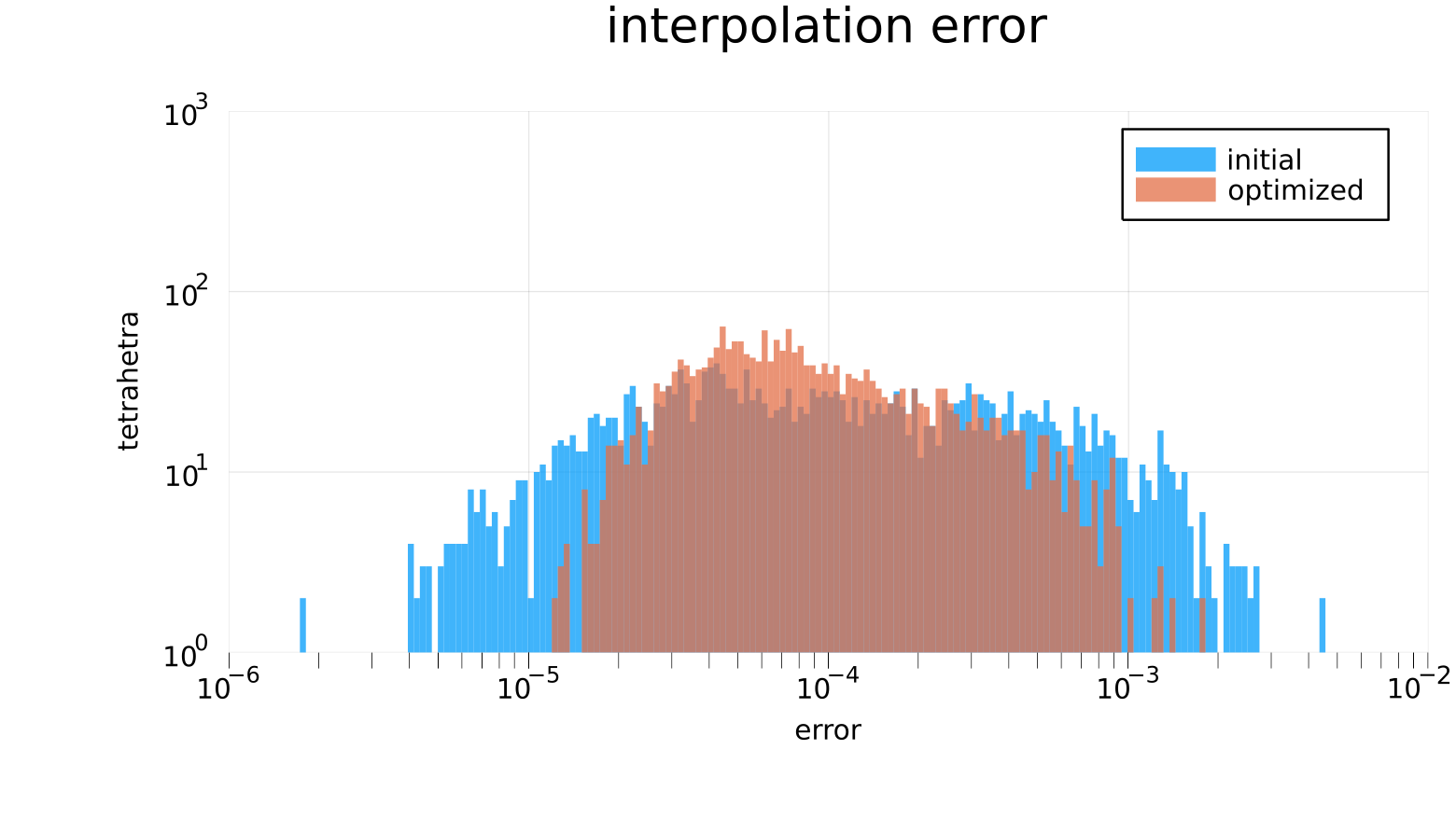}}
			&
			\subfigure[]{\label{fig:aerror3D}
				\includegraphics[width=0.5\textwidth]{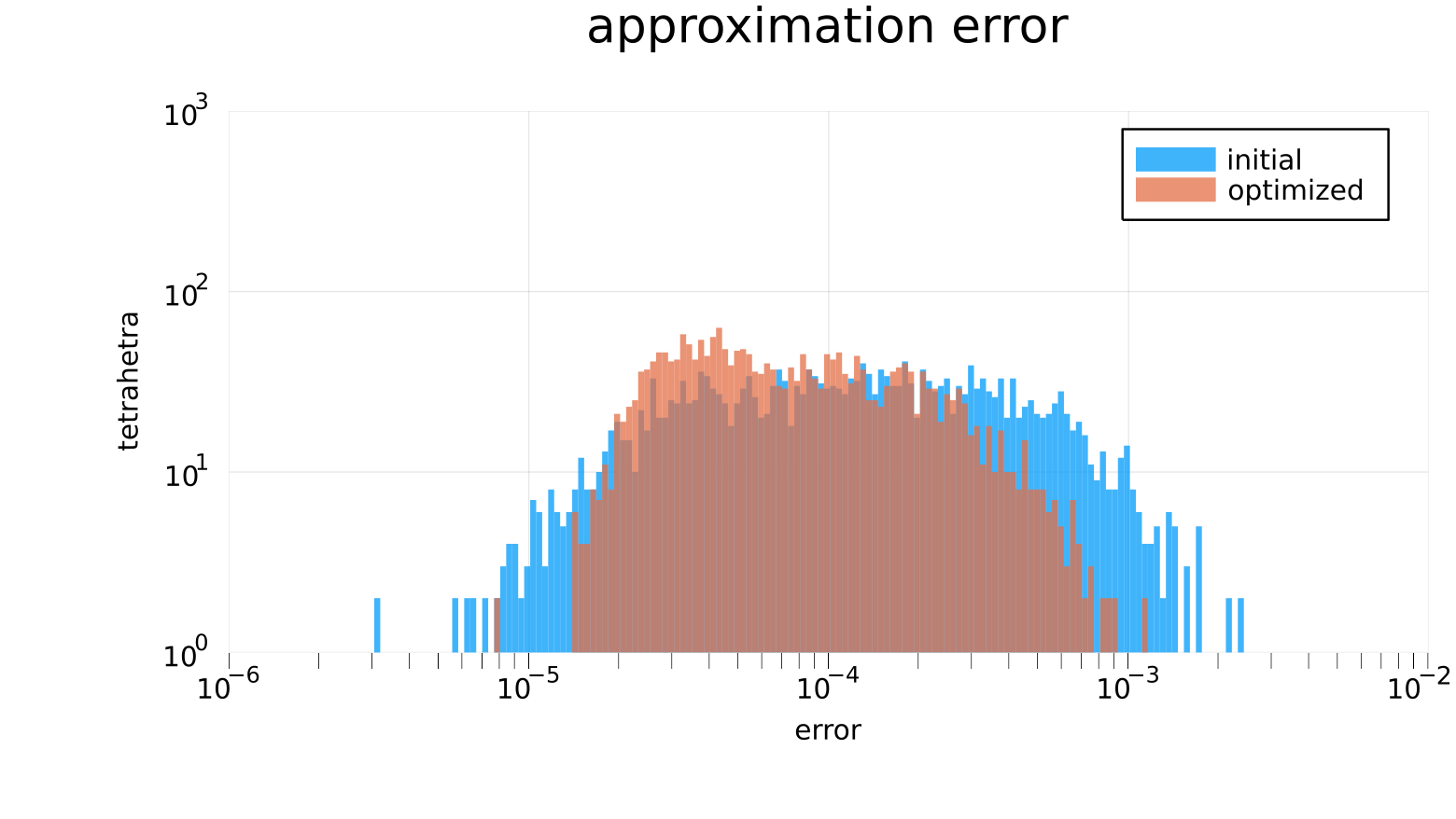}}\\
		\end{tabular}
		\caption{Logarithmic distribution for the elemental interpolation and approximation error histograms for (blue) initial and (orange) optimized quadratic tetrahedral meshes.}
		\label{fig:error3D}
	\end{figure}
	%\color{teal}
	In Figure \ref{fig:error3D}, we illustrate the distribution of the elemental interpolation and approximation error for the initial and optimized quadratic tetrahedral meshes.
	On the one hand, the tails are reduced in measure (horizontal axis) and magnitude (vertical axis). This reduction shows that the maximum and minimum elemental error become closer in the optimized mesh than in the initial one. This also illustrates a reduced standard deviation for the optimized mesh. On the other hand, the distribution peak is increased. Moreover, this distribution peak is slightly translated to the left. This illustrates that the optimized mesh enables a more concentrated and reduced mean error than the initial mesh. From these observations, we conclude that the optimized mesh enables improved error statistics when compared to the initial one.
	%Meanwhile, the measure and magnitude are almost preserved.
	
	%\color{black}
	
	%\color{blue}In addition, the presented example shows the capability of curved elements to capture sharp curved transition regions. For the initial anisotropic mesh, we observe that, even if the straight-edged elements approximate the curved transition region, this is not sufficient. Only when we curve them, we gain one order of magnitude for the interpolation and approximation error.
	
	%\color{blue}In addition, the presented example shows the capability of curved elements to capture sharp curved transition regions with curved boundaries. We observe that, even if the straight-edged elements approximate the curved transition region, this is not sufficient. Only when we curve them, we reduce the interpolation and approximation error.
	
	%\color{blue}In addition, the presented example shows the capability of curved elements to capture sharp curved transition regions. We observe that, even if the straight-edged elements approximate the curved transition region, this is not sufficient. Only when we curve them, we reduce the interpolation, approximation, and numerical error.
	
	\subsection{Size-shape distortion minimization for quartic interpolation: isotropic and anisotropic initial straight-edged meshes}\label{sec:errormin}
	\begin{figure}[t!]
		\centering
		%	\hspace{0.35cm}
		\setlength{\tabcolsep}{25pt}
		\begin{tabular}{cc}
			\subfigure[]{\label{fig:iso40}
				\includegraphics[width=0.3\textwidth]{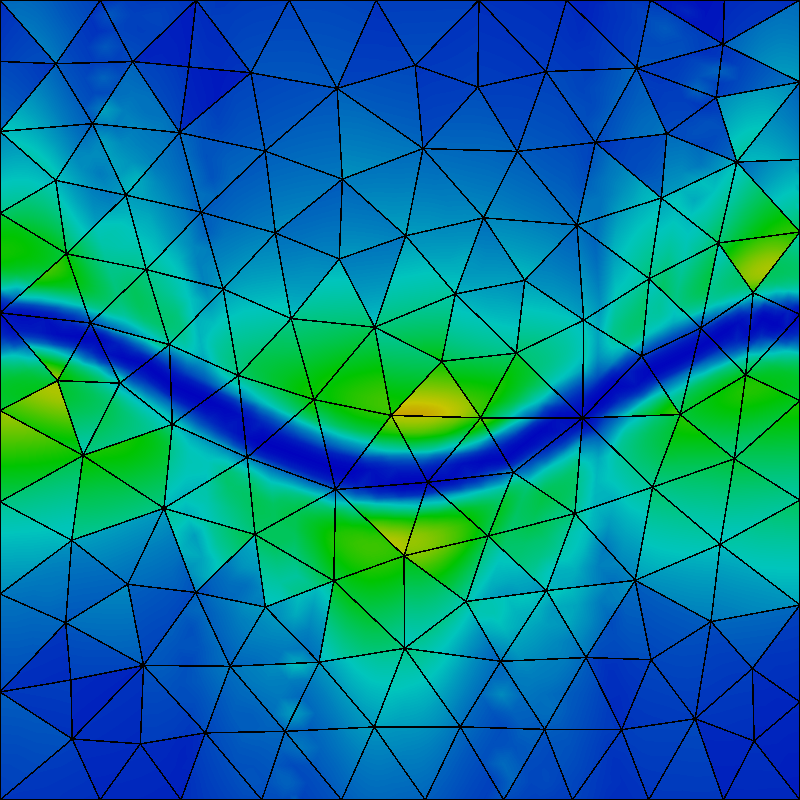}}
			&
			\subfigure[]{\label{fig:40}
				\includegraphics[width=0.3\textwidth]{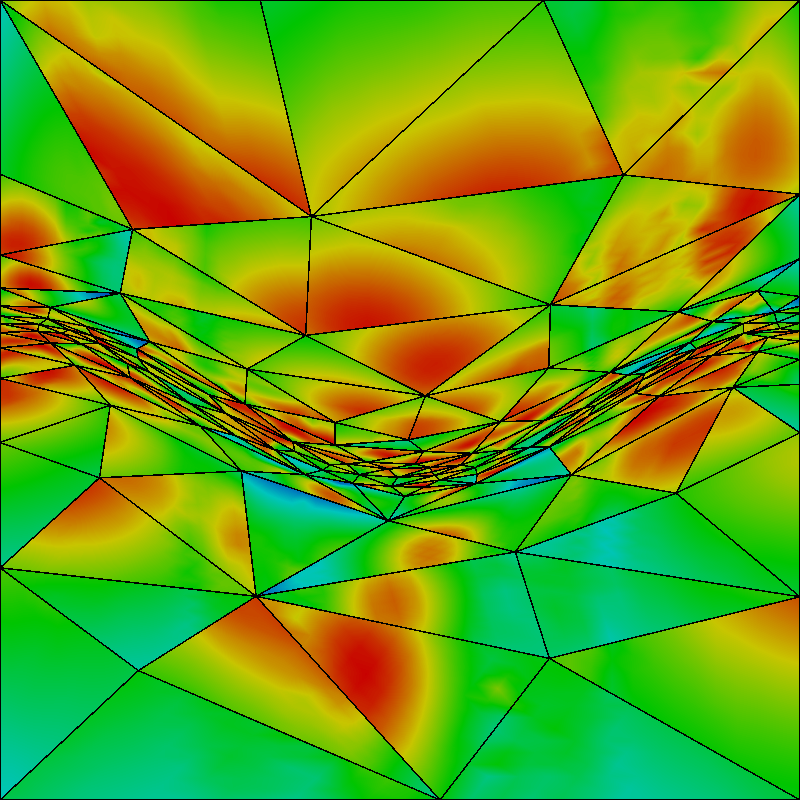}}
			\\
			\subfigure[]{\label{fig:iso41}
				\includegraphics[width=0.3\textwidth]{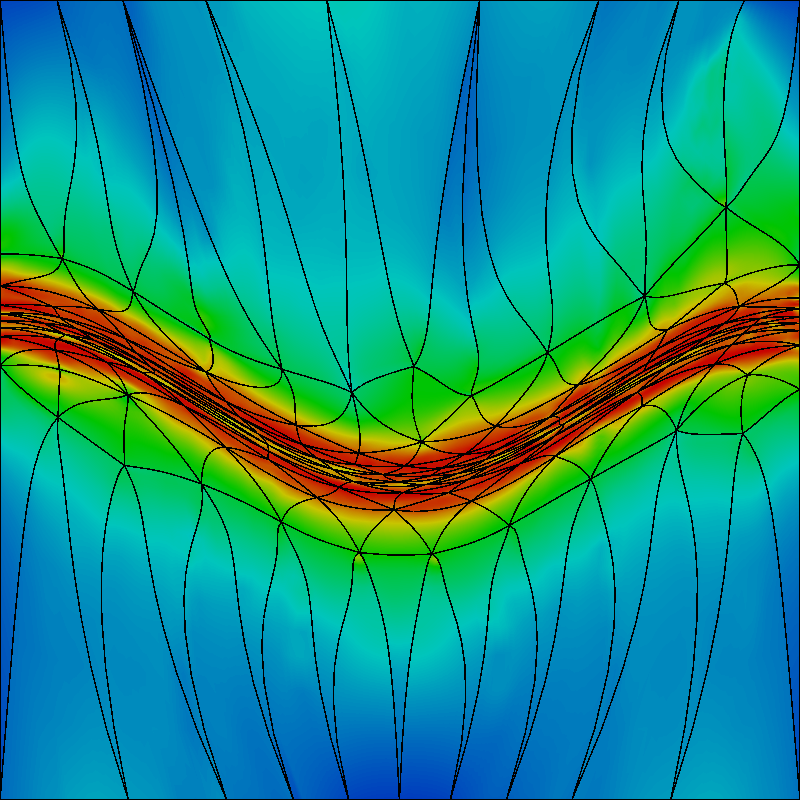}}
			&
			\subfigure[]{\label{fig:41}
				\includegraphics[width=0.3\textwidth]{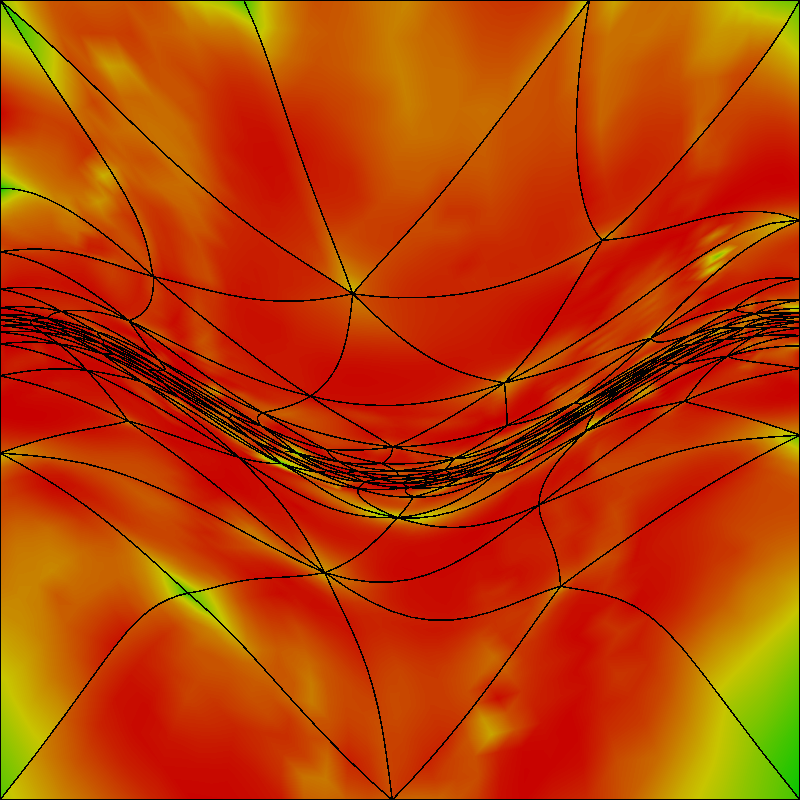}}
		\end{tabular}
		\\	
		\includegraphics[width=0.3\textwidth]{./qualBarParaview_color}
		\caption{Point-wise size-shape quality for (columns) initial isotropic and anisotropic straight-edged meshes, and (rows) initial and optimized quartic meshes.}
		\label{fig:p4}
	\end{figure}
	In the following example, to illustrate the potential adaptation advantages of the proposed distortion for any type of initial mesh, we apply the size-shape distortion minimization for quartic interpolation of a function to isotropic and anisotropic initial straight-edged meshes.
	For this, we consider the function $u$ of Equation \eqref{eq:u} with $\gamma = 100$.
	We generate a background mesh $\zbmesh$ and two initial physical meshes $\zmesh$ of polynomial degree $q = 4$, and size $h = 0.1$.
	Specifically, the isotropic and anisotropic physical meshes are composed of 1923 and 1917 nodes and of 231 and 257 elements, respectively.
	We show the physical meshes in Figure \ref{fig:p4}, where they are colored according to the point-wise size-shape quality measure of Equation \eqref{eq:pointquality}.
	On the one hand, we generate an initial isotropic mesh, see Figure \ref{fig:p4}\subref{fig:iso40}. In this case, the initial physical mesh $\zmesh$ and the background mesh $\zbmesh$ coincide. We observe that almost all elements are of low size-shape quality. This is because the element stretching, alignment, or sizing does not match with the metric. As expected, the lowest quality elements lie in the sharp transition region.
	On the other hand, we generate an initial anisotropic mesh according to the discrete metric $\zbmetric$ of the input function $u$, see Figure \ref{fig:p4}\subref{fig:40}. We observe that almost all elements are of medium quality. In addition, the straight-edged elements approximate the curved transition region. Finally, the corresponding optimized meshes $\zmesh^*$ are shown in Figures \ref{fig:p4}\subref{fig:iso41} and \ref{fig:p4}\subref{fig:41}. We observe that, in both cases, the elements are accumulated and match the metric stretching, alignment, and sizing at the sharp transition region.
	
	\begin{table}[t!]
		\caption{Size-shape quality and geometry statistics of the initial isotropic and optimized quartic meshes.}
		\label{table:isop4}
		\centering
		\tiny
		\begin{tabular}{ c c c c c c c c c c}
			\hline\noalign{\smallskip}
			Measure & \multicolumn{2}{c}{Minimum}&\multicolumn{2}{c}{Maximum}& \multicolumn{2}{c}{Mean} & \multicolumn{2}{c}{Standard deviation}\\
			&Initial&Optimized&Initial&Optimized&Initial&Optimized&Initial&Optimized\\
			\noalign{\smallskip}\hline\noalign{\smallskip}
			Quality& 0.0189 & 0.1765 & 0.6413 & 0.9819 & 0.1613 & 0.7328 & 0.1149 & 0.2504\\
			Length& 0.1232 & 0.1462 & 9.1069 & 2.9094 & 0.9304 & 1.0989 & 1.5793 & 0.4197 \\
			Area& 0.0219 & 0.3318 & 15.7890 & 1.8669 & 0.8658 & 0.8657 & 2.3650 & 0.3310 \\
			%		Log-error&2.4230&2.7244&8.1721&6.7385&5.5217&5.4263&1.3147&1.0173\\
			\noalign{\smallskip}\hline\noalign{\smallskip}
		\end{tabular}
	\end{table}
	\begin{table}[t!]
		\caption{Size-shape quality and geometry statistics of the initial anisotropic and optimized quartic meshes.}
		\label{table:p4}
		\centering
		\tiny
		\begin{tabular}{ c c c c c c c c c c}
			\hline\noalign{\smallskip}
			Measure & \multicolumn{2}{c}{Minimum}&\multicolumn{2}{c}{Maximum}& \multicolumn{2}{c}{Mean} & \multicolumn{2}{c}{Standard deviation}\\
			&Initial&Optimized&Initial&Optimized&Initial&Optimized&Initial&Optimized\\
			\noalign{\smallskip}\hline\noalign{\smallskip}
			Quality& 0.1959 & 0.8063 & 0.9270 & 0.9955 & 0.6454 & 0.9288 & 0.1578 & 0.0384\\
			Length& 0.4171 & 0.5999 & 2.5246 & 1.6854 & 1.0921 & 1.0101 & 0.3823 & 0.1914 \\
			Area& 0.2493 & 0.5571 & 2.7515 & 1.4499 & 0.8620 &  0.8620 & 0.4149 & 0.1775 \\
			%		Log-error&2.4230&2.7244&8.1721&6.7385&5.5217&5.4263&1.3147&1.0173\\
			\noalign{\smallskip}\hline\noalign{\smallskip}
		\end{tabular}
	\end{table}
	In Tables \ref{table:isop4} and \ref{table:p4}, we illustrate the size-shape quality, and Riemannian length and area measure statistics.
	The statistics and histograms illustrate an improved metric matching of the optimized meshes versus the initial ones, see the reasoning in Section \ref{sec:volumetric}.
	As expected, we observe a greater improvement for the case with initial isotropic mesh.
	%the geometric quantities
	%\color{green}In Tables \ref{table:isop4} and \ref{table:p4}, we show the statistics for elemental qualities (Equation \eqref{eq:qualityreg}) and Riemannian lengths and areas. They allow us to compare between the initial and optimized meshes in terms of the target metric. We observe that the maximum, minimum, mean, and standard deviation become closer to unit values in almost all cases. That is, in general, all statistics are improved. As expected, we observe a greater improvement for the initial isotropic mesh.
	
	\begin{table}[t!]
		\caption{Interpolation and approximation $L^2$ error of the initial isotropic and anisotropic quartic meshes and the corresponding optimized meshes.}
		\label{table:interpolation4}
		\centering
		\begin{tabular}{c c c c c c}
			\hline\noalign{\smallskip}
			%		Mesh&\multicolumn{2}{c}{$-\log_{10}\| u - \Pi u \|$}&\multicolumn{2}{c}{$-\log_{10} \min\limits_{v} \| u - v \|$}\\
			Initial&Nodes&\multicolumn{2}{c}{Interpolation error}&\multicolumn{2}{c}{Approximation error}\\
			Mesh&&Initial&Optimized&Initial&Optimized\\
			\noalign{\smallskip}\hline\noalign{\smallskip}
			%		Isotropic&1923&0.8032&2.3702&0.9544&2.5009\\
			%		Anisotropic&2115&1.9891&2.4274&2.1312&2.5550\\
			Isotropic&1923&0.1573&0.0029&0.1111&0.0022\\
			Anisotropic&1917&0.0138&0.0031&0.0095&0.0024\\
			\noalign{\smallskip}\hline\noalign{\smallskip}
		\end{tabular}
		%elements:231, 257
	\end{table}
	%\color{purple}The presented example shows how our method can be used to improve the error of a straight-edged mesh. In Table \ref{table:interpolation4}, we present the global interpolation and approximation error of the initial and optimized meshes. First, we observe that the approximation error is less than the interpolation one. This is because the approximation error compares the analytic function with its best approximation in the continuous finite element space, see Section \ref{sec:error}. Since the best approximation approximates better the analytic function than the interpolated one, the approximation error is less than the interpolation one.
	
	%\color{teal}
	The presented example shows how our method can be used to improve the error of a straight-edged mesh. 
	In Table \ref{table:interpolation4}, we present the global interpolation and approximation error of the initial and optimized meshes.
	We observe that all quantities are improved for the optimized meshes. In particular, they are improved by almost two orders of magnitude for the initial isotropic mesh and by almost one order of magnitude for the initial anisotropic mesh. This is because the initial anisotropic mesh approximates the metric better than the initial isotropic one. Finally, the errors of the optimized meshes corresponding to the initial isotropic mesh and the initial anisotropic one are of the same order of magnitude. This phenomenon illustrates the potential of curved \textit{r}-adaptation.
	
	%\color{blue}In addition, the presented example shows the capability of curved elements to capture sharp curved transition regions. For the initial anisotropic mesh, we observe that, even if the straight-edged elements approximate the curved transition region, this is not sufficient. Only when we curve them, we gain one order of magnitude for the interpolation and approximation error.
	
	\begin{figure}[t!]
		\centering
		\hspace{-0.35cm}
		\setlength{\tabcolsep}{25pt}
		\begin{tabular}{cc}
			\subfigure[]{\label{fig:isoerror0}
				\includegraphics[width=0.3\textwidth]{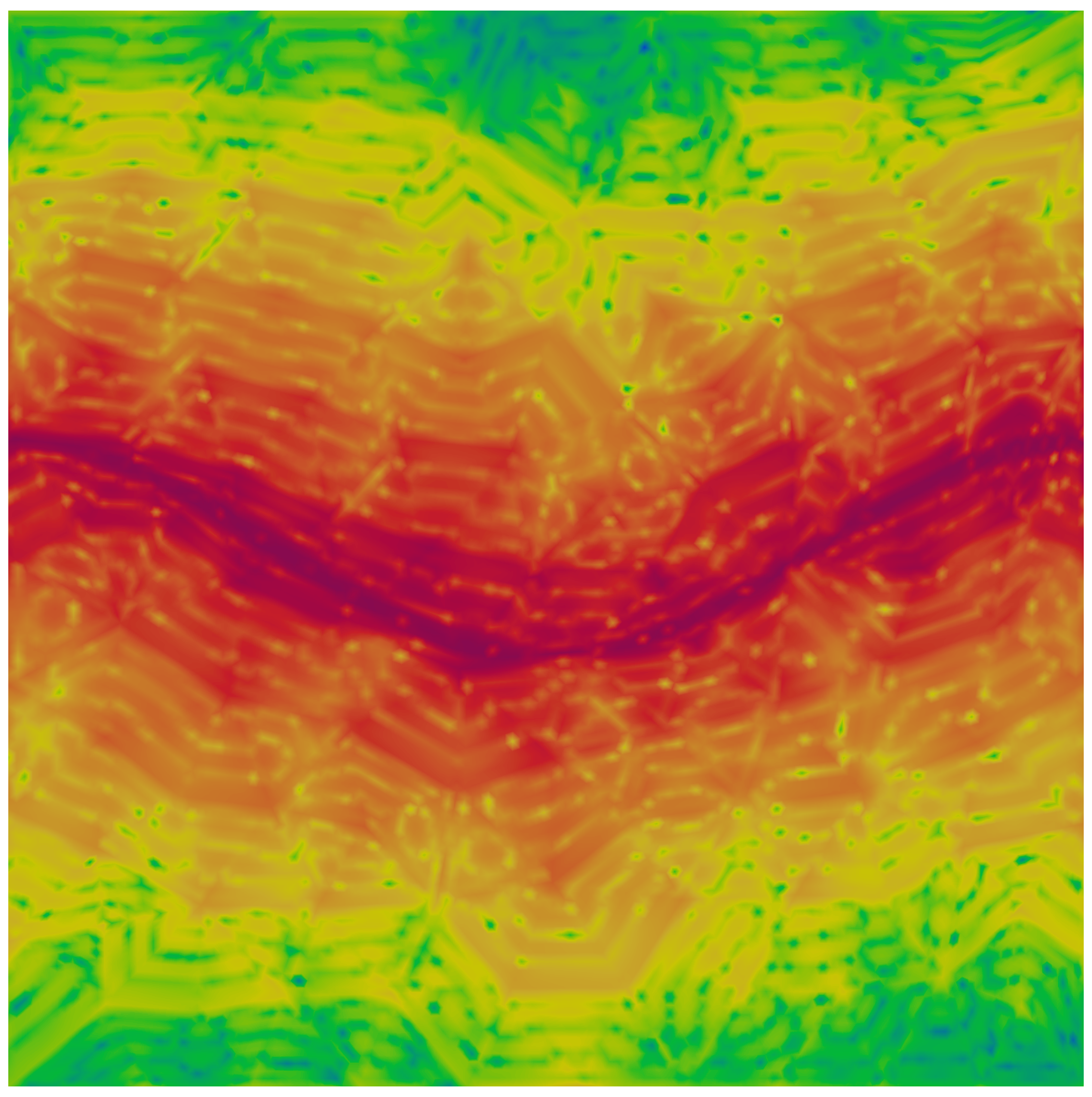}}
			&
			\subfigure[]{\label{fig:anisoerror0}
				\includegraphics[width=0.3\textwidth]{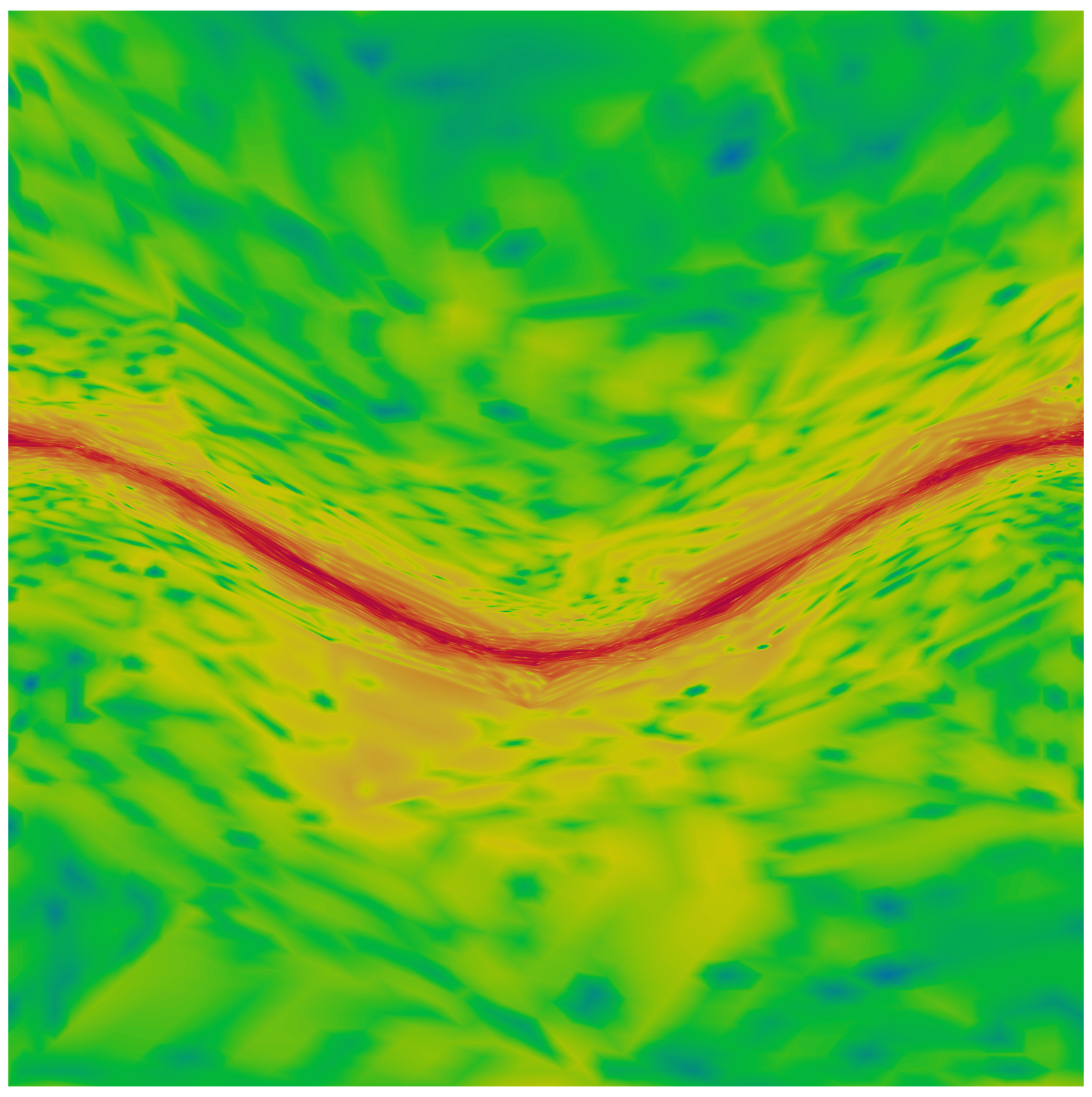}}
			\\
			\subfigure[]{\label{fig:isoerror1}
				\includegraphics[width=0.3\textwidth]{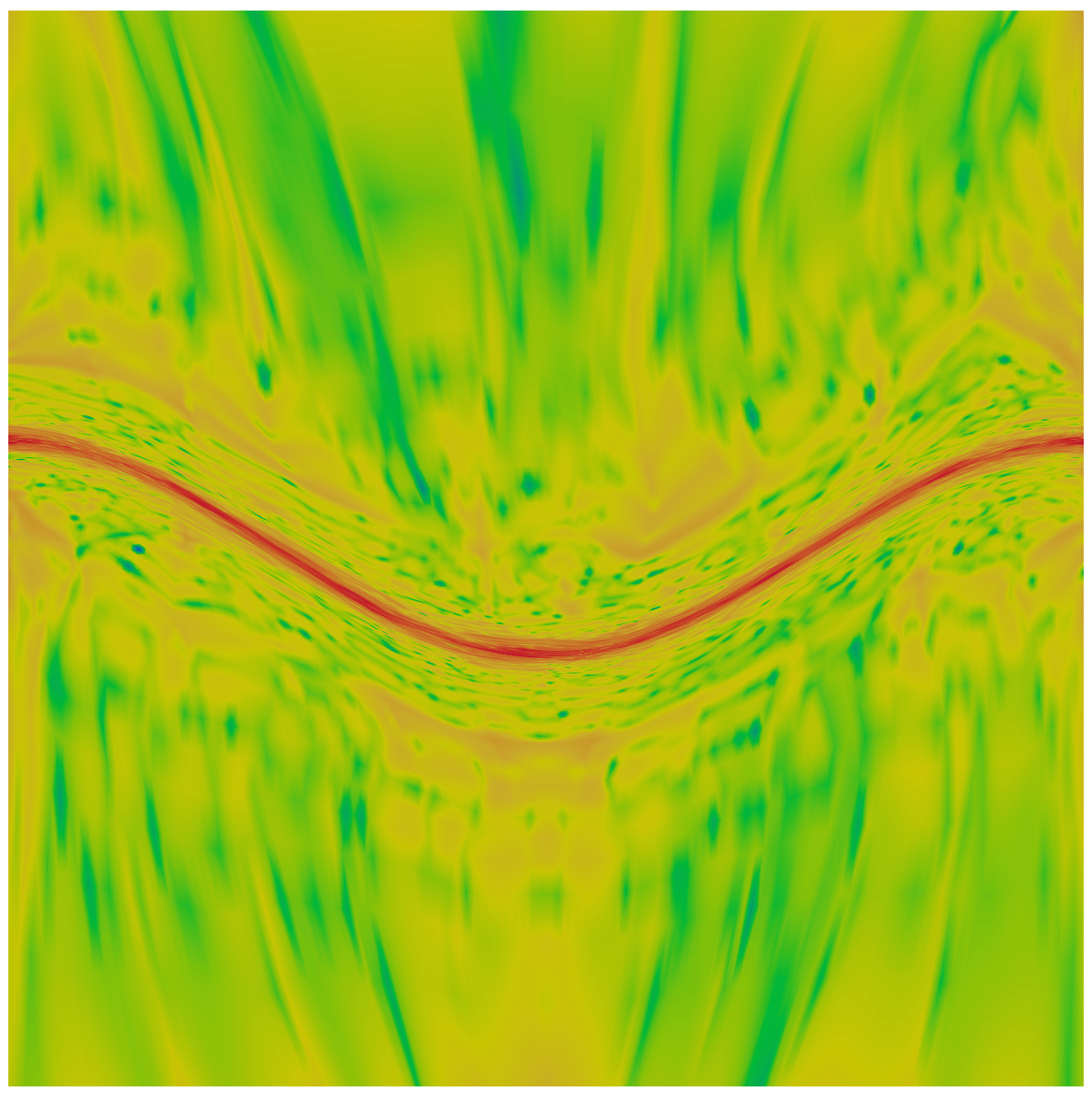}}
			&
			\subfigure[]{\label{fig:anisoerror1}
				\includegraphics[width=0.3\textwidth]{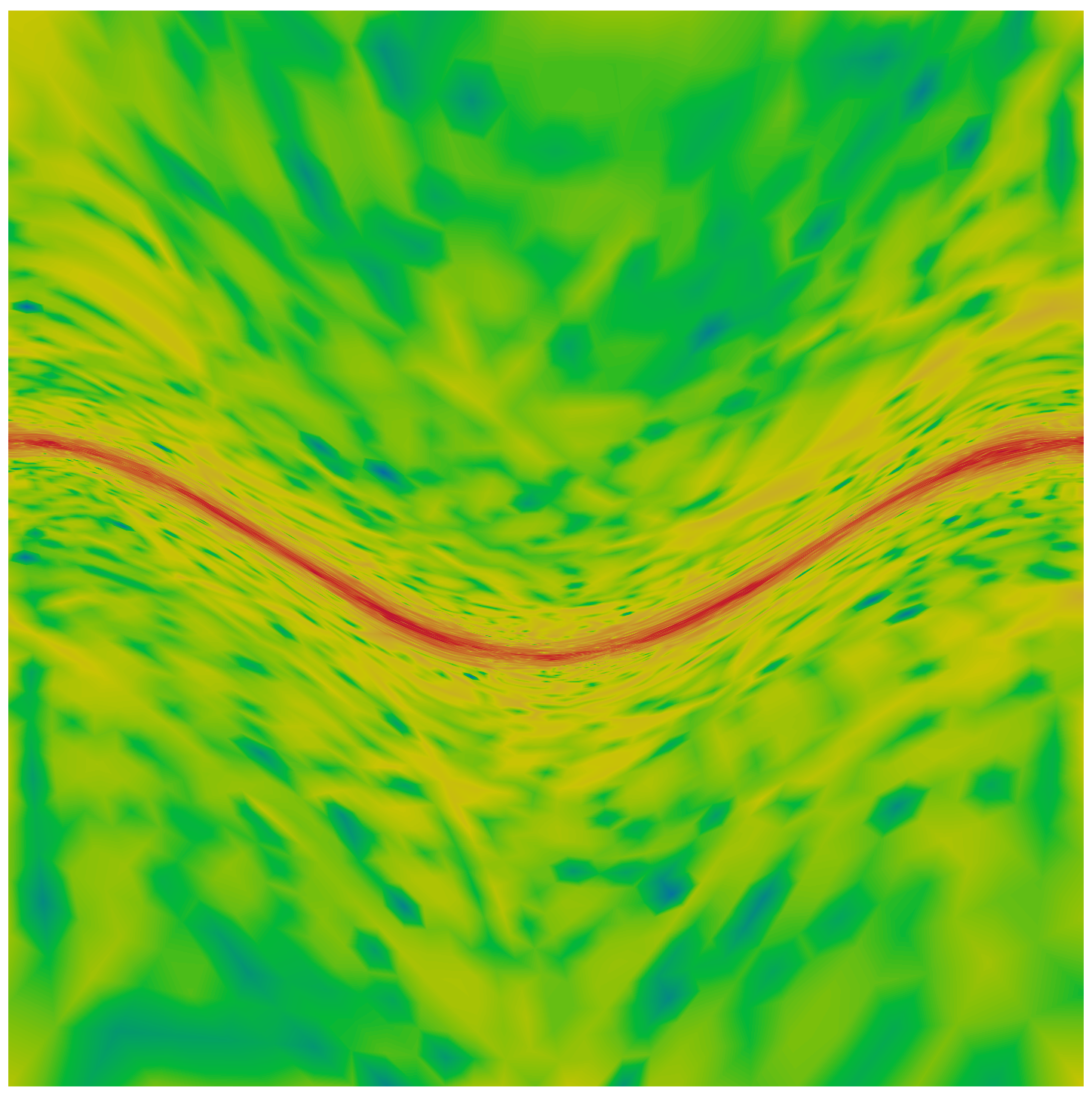}}
		\end{tabular}
		\\
		\includegraphics[width=0.5\textwidth]{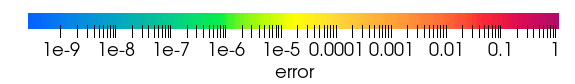}
		\caption{Point-wise error between the function $u$ and its best $L^2(\Omega)$ approximation $u_{\zmesh}$ for (columns) initial isotropic and anisotropic straight-edged, and (rows) initial and optimized quartic meshes.}
		\label{fig:p4error}
	\end{figure}
	% 7 colors 7 ordres de magnitud
	%\color{teal}
	In Figure \ref{fig:p4error}, we show the point-wise $L^2$ approximation error. For the initial isotropic mesh, we observe that the error increases as we approximate to the sharp transition region. This is because the isotropic elements cannot represent the sharp transition of the function. Then, in the optimized mesh, we observe that the error is localized at the sharp transition region in a smaller magnitude compared to the initial mesh. This is because the elements are stretched and aligned to match the sharp curved transition region. For the initial anisotropic mesh, we observe that the error is localized at the sharp transition region only. This is because the mesh has been previously adapted to match, with straight-edged elements, the sharp transition region. In the optimized mesh, we observe that this error fits the curved sharp transition region in a slightly smaller magnitude. This shows the potential of reducing function errors by approximating a metric-based error estimator with curved elements.
	%\color{black}
	\subsection{Size-shape distortion minimization with curved boundary for cubic interpolation: anisotropic initial straight-sided mesh}\label{sec:curving}
	% Background
	%	linear: 87 nodes and 133 triangles 
	%	subdivided: 660 nodes and 1197 triangles
	%	high-order: 660 nodes and 133 triangles
	%Physical: 951 nodes and 191 triangles
	In the following example, to illustrate preliminary adaptation results of the proposed distortion when the domain has curved boundaries, we apply the size-shape distortion minimization with curved boundary for cubic interpolation to an anisotropic initial straight-edged mesh.
	For this, we consider the function $u$ of Equation \eqref{eq:u} with $\gamma = 100$ over the square domain with a circular hole $\Omega$.
	%over the square domain $K_1$ with a circular hole $C_1$, $\Omega_1 = $.
	Specifically, we denote the domain by $\Omega = K \backslash C$, where $K = [-0.5,0.5]^2$ is a square, and where $C$ is the circle with radius equal to $0.18$ and centered at the origin.
	The domain $\Omega$ has two boundaries, the one of the square $K$ and the one of the circle $C$.
	%We illustrate in Figure \ref{fig:2Dmodels}\subref{fig:circle} a global implicit representation of the boundary $\zmodel_1:=\partial\Omega_1$, using the method presented in Section \ref{sec:cadimplicit}.
	Although the inner boundary is smooth, the outer boundary contains sharp features such as corners.
	
	\begin{figure}[t!]
		\centering
		%	\hspace{-1cm}
		%	\setlength{\tabcolsep}{10pt}
		\begin{tabular}{ccc}
			\subfigure[]{\label{fig:bmesh}
				\includegraphics[width=0.28\textwidth]{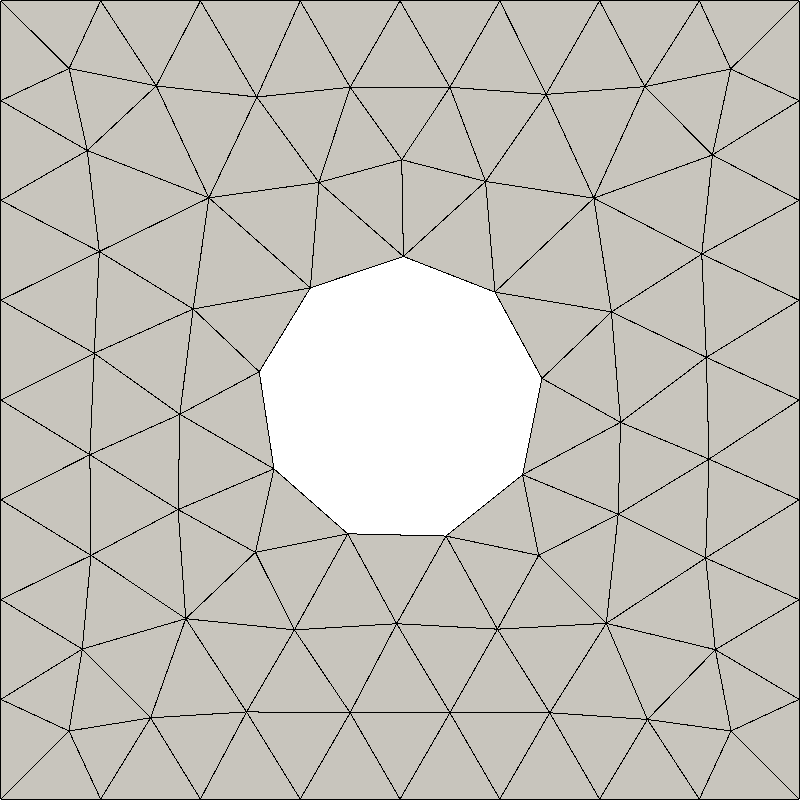}}
			&
			\subfigure[]{\label{fig:mesh0}
				\includegraphics[width=0.28\textwidth]{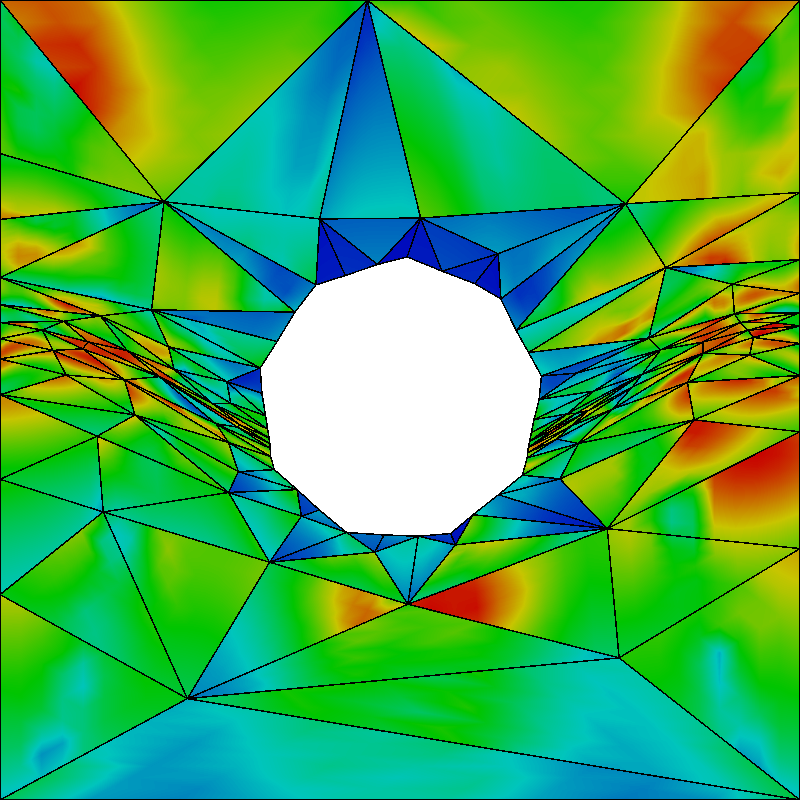}}
			&
			\subfigure[]{\label{fig:mesh1}
				\includegraphics[width=0.28\textwidth]{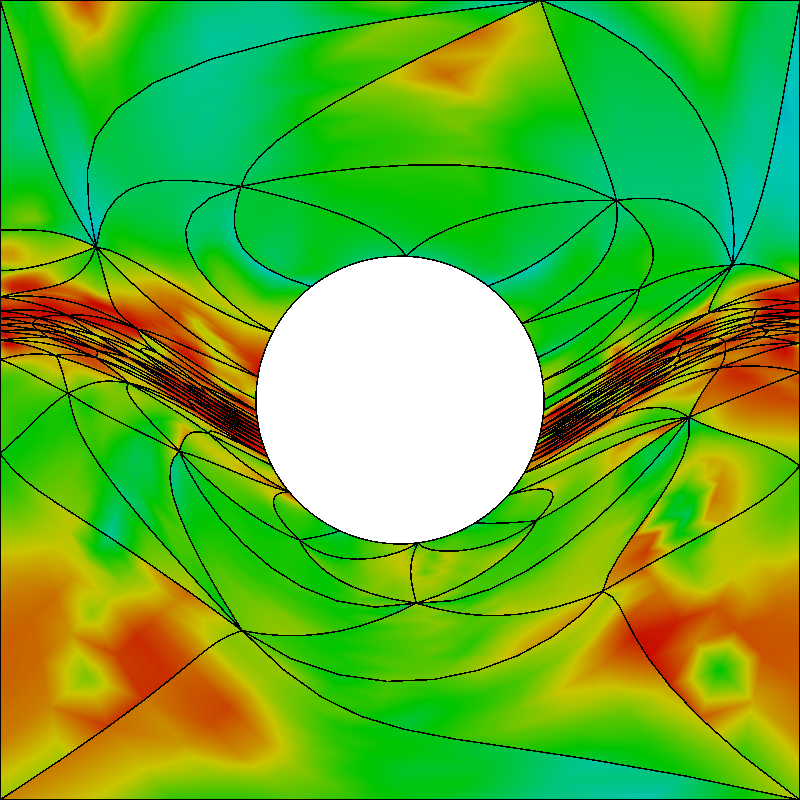}}
		\end{tabular}
		\\	
		\includegraphics[width=0.3\textwidth]{./qualBarParaview_color}
		\caption{Background, initial straight-edged mesh (already adapted to the metric), and optimized cubic meshes. Initial and optimized meshes are colored with the point-wise size-shape quality measure.}
		\label{fig:p3}
	\end{figure}
	%\color{red}
	In Figure \ref{fig:p3}, we show the background $\left(\zbmesh\right)$ and physical $\left(\zmesh,\ \zmesh^* \right)$ meshes, where the physical meshes are colored according to the point-wise size-shape quality measure of Equation \eqref{eq:pointquality}.
	The background mesh $\zbmesh$ is of polynomial degree $q = 3$, and size $h \approx 0.042$, see Figure \ref{fig:p3}\subref{fig:bmesh}.
	In particular, it is composed of 87 vertices, 660 nodes, and 133 triangles.
	From this background mesh $\zbmesh$, we generate a physical cubic mesh $\zmesh$ according to the input discrete metric $\zbmetric$ and preserving the background mesh boundary $\partial \zbmesh$.
	Specifically, in order to obtain an output MMG mesh, we uniformly subdivide the background mesh and we evaluate the fourth derivatives of $u$, $\nabla^4 u$, at the subdivided background mesh vertices.
	In this case, the input MMG linear mesh $\zbmesh'$, which is different from the high-order background mesh $\zbmesh$, is composed of 660 vertices-nodes and 1197 triangles.
	Then, we obtain the discrete metric $\zbmetric$ from the derivatives $\nabla^4 u$ by applying the log-simplex algorithm \cite{coulaud:VeryHighOrderAnisotropic}.
	The output MMG mesh is an adapted straight-edged physical mesh $\zmesh$ composed of 951 nodes and 191 triangles, see Figure \ref{fig:p3}\subref{fig:mesh0}.
	%The initial adapted straight-edged mesh is generated according to the discrete metric of the input function, see Figure \ref{fig:p3}\subref{fig:mesh0}.
	We observe that almost all elements are of medium quality. In addition, the straight-edged elements approximate the curved transition region. Finally, we show the corresponding optimized mesh $\zmesh^*$ in Figure \ref{fig:p3}\subref{fig:mesh1}. We observe that the elements are accumulated and match the metric stretching, alignment, and sizing at the sharp transition region.
	
\begin{figure}[t!]
		\centering
		\hspace{-0.5cm}
		\setlength{\tabcolsep}{-2pt}
		\begin{tabular}{cc}
			\subfigure[]{\label{fig:lpc}
				\includegraphics[width=0.5\textwidth]{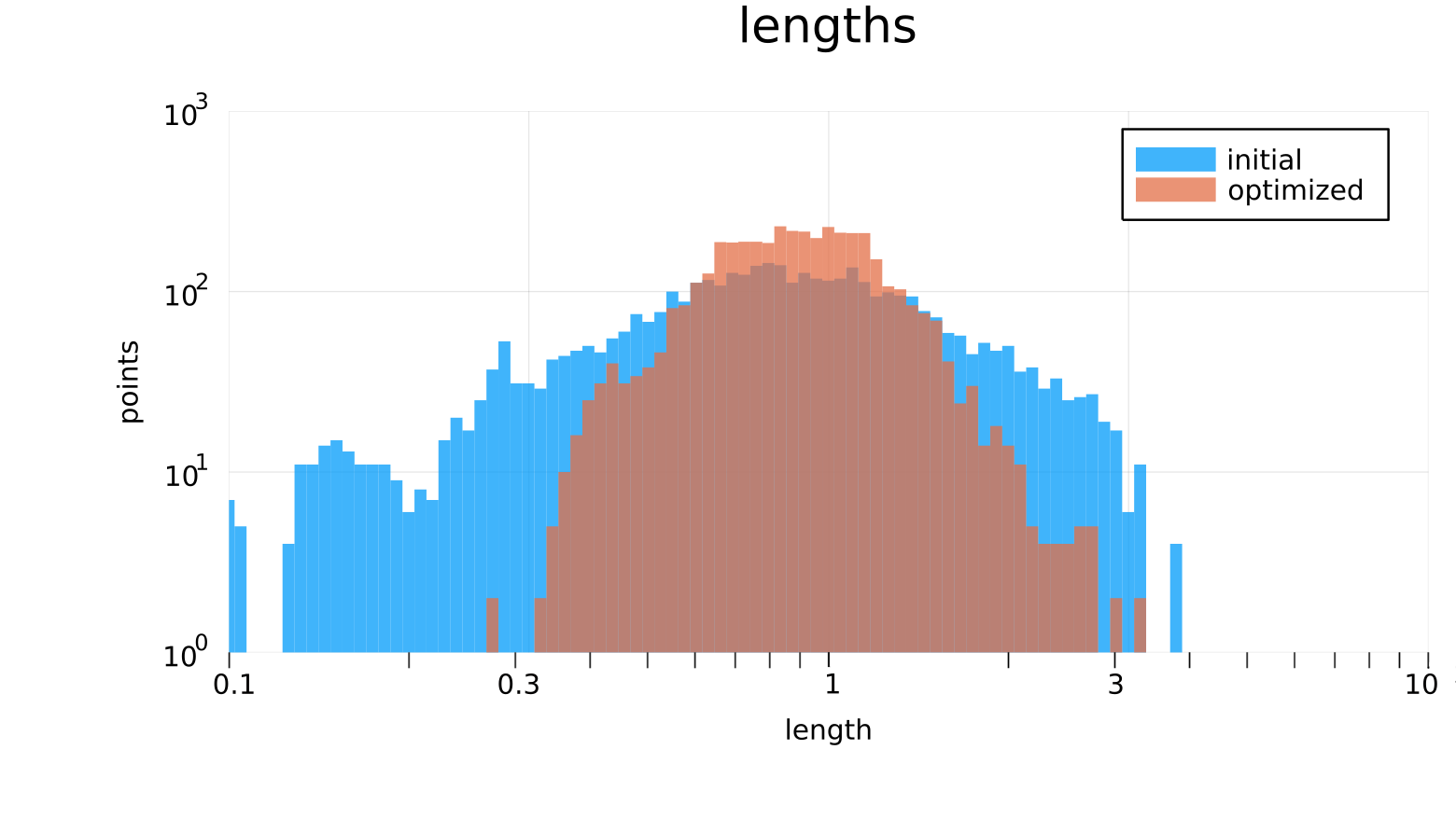}}
			&
			\subfigure[]{\label{fig:apc}
				\includegraphics[width=0.5\textwidth]{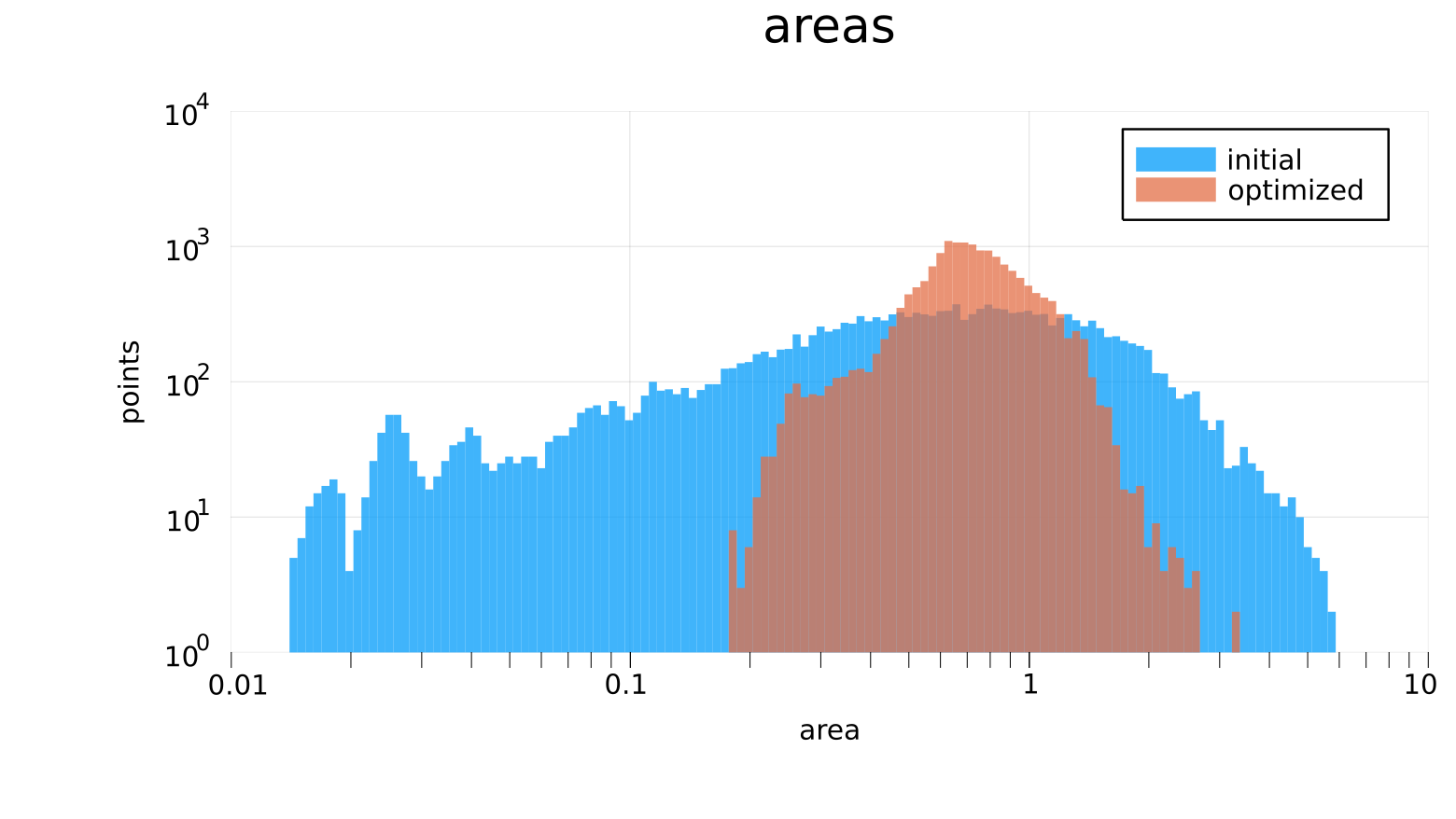}}
		\end{tabular}
		\caption{Logarithmic Riemannian length and area density histograms for (blue) initial and (orange) optimized cubic triangular meshes.}
		%	\caption{Logarithmic point-wise distortion, length, area, and volume histograms for (blue) initial and (orange) optimized quadratic tetrahedral meshes.}
		\label{fig:histcurving}
	\end{figure}

	\begin{table}[t!]
		\caption{Size-shape quality and geometry statistics of the initial adapted straight-edged and optimized cubic meshes.}
		\label{table:p3}
		\centering
		\tiny
		\begin{tabular}{ c c c c c c c c c c}
			\hline\noalign{\smallskip}
			Measure & \multicolumn{2}{c}{Minimum}&\multicolumn{2}{c}{Maximum}& \multicolumn{2}{c}{Mean} & \multicolumn{2}{c}{Standard deviation}\\
			&Initial&Optimized&Initial&Optimized&Initial&Optimized&Initial&Optimized\\
			\noalign{\smallskip}\hline\noalign{\smallskip}
			Quality& 0.0161 & 0.3549 & 0.9292 & 0.9864 & 0.4667 & 0.7804 & 0.2312 & 0.1554\\
			Length& 0.1022 & 0.3714 & 2.7706 & 2.1434 & 0.9635 & 0.9310 & 0.5344 & 0.2760 \\
			Area& 0.0148 & 0.2235 & 3.0501 & 1.2100 & 0.6696 & 0.6428 & 0.5215 & 0.1758 \\
			\noalign{\smallskip}\hline\noalign{\smallskip}
		\end{tabular}
	\end{table}
	In Table \ref{table:p3}, we illustrate the size-shape quality, and Riemannian length and area measure statistics.
	In addition, Figure \ref{fig:histcurving} shows the point-wise Riemannian length and area density histograms for the initial and optimized triangular cubic meshes.
	The statistics and histograms illustrate an improved metric matching of the optimized mesh versus the initial one, see the reasoning in Section \ref{sec:volumetric}.
	%\color{green}In Table \ref{table:p3}, we show the statistics for elemental qualities (Equation \eqref{eq:qualityreg}) and Riemannian lengths and areas. The table allow us to compare between the initial and optimized meshes in terms of the target metric. We observe that the maximum, minimum, mean, and standard deviation become closer to unit values in almost all cases. That is, in general, all statistics are improved.
	
	%\color{orange}In Figure \ref{fig:histcurving}, we show the point-wise Riemannian length and area of the initial and optimized triangular cubic meshes. As in Section \ref{sec:volumetric}, we use a logarithmic scale to illustrate the different scales of the corresponding ratios. The logarithmic representation illustrates the behavior near the minimum, maximum, and geometric mean of the distribution.
	
	%\color{brown}From the reasoning presented in Section \ref{sec:intrinsicmetric}, we observe that almost all measure statistics are improved for the optimized meshes. On the one hand, for the geometric measures, the tails are reduced in measure (horizontal axis) and magnitude (vertical axis). This reduction is because the quality measure is sensitive to points with volume far from the unit. Hence, these regions gain priority during the optimization process. On the other hand, the distribution peak is increased. This increase is so because the global optimization of the squared quality measure tends to homogenize the points near a mean. Meanwhile, the measure and magnitude are almost preserved.
	
	%Comment small elements near the boundary are compensated, see also histograms.
	%\color{teal}
	From the results, we observe that, when compared with straight-edged elements, curved elements approximate more faithfully the metric while preserving the curved features of the boundary.
	In this case, the stretching direction is almost aligned according to the tangent of the geometry.
	When considering straight-edged elements, in Figure \ref{fig:p3}\subref{fig:mesh0}, accumulating more degrees of freedom in the stretched regions may worsen the boundary representation at non-stretched regions.
	Moreover, this accumulation leads to triangles with small area near the boundary, see the area density histogram in Figure \ref{fig:histcurving}\subref{fig:apc}.
	In contrast, when considering curved elements, in Figure \ref{fig:p3}\subref{fig:mesh1}, we observe that a single curved element represents the boundary more faithfully than several straight-edged elements.
	This flexibility of curved elements allows the degrees of freedom to slide and accumulate, from non-stretched regions to the stretched regions, featuring high-quality elements.
	In addition, those small elements initially generated near the boundary are enlarged according to the metric size and to the domain boundary, see the area density histogram in Figure \ref{fig:histcurving}\subref{fig:apc}.
	For that reason, we observe how the elements are stretched, aligned, sized, and curved according to the stretching, alignment, and sizing of the metric.
	Hence, curved elements allow an improved representation of the metric while preserving the curved features of the boundary.
	
	\begin{table}[t!]
		\caption{Interpolation and approximation $L^2$ error of the initial adapted straight-edged and optimized cubic meshes.}
		\label{table:interpolation3}
		\centering
		\begin{tabular}{c c c}
			\hline\noalign{\smallskip}
			Mesh&Interpolation error&Approximation error\\
			\noalign{\smallskip}\hline\noalign{\smallskip}
			Initial&0.0197&0.0144\\
			Optimized&0.0058&0.0046\\
			\noalign{\smallskip}\hline\noalign{\smallskip}
			%		\multicolumn{2}{c}{Interpolation error}&\multicolumn{2}{c}{Approximation error}\\
			%		Initial&Optimized&Initial&Optimized\\
			%		\noalign{\smallskip}\hline\noalign{\smallskip}
			%		0.0197&0.0144&0.0058&0.0046\\
			%		\noalign{\smallskip}\hline\noalign{\smallskip}
		\end{tabular}
	\end{table}
	%\color{purple}The presented example shows how our method can be used to improve the error of a straight-edged mesh according to a curved boundary $\partial \Omega$. In Table \ref{table:interpolation3}, we present the global interpolation and approximation error of the initial and optimized mesh. As before, we observe that the approximation error is less than the interpolation one. This is because the approximation error compares the analytic function with its best approximation in the continuous finite element space, see Section \ref{sec:error}. Since the best approximation approximates better the analytic function than the interpolated one, the approximation error is less than the interpolation one.
	
	%\color{teal}
	The presented example shows how our method can be used to improve the error of a straight-edged mesh according to a curved boundary $\partial \Omega$. In Table \ref{table:interpolation3}, we present the global interpolation and approximation error of the initial and optimized mesh.
	We observe that the errors are improved three times for the optimized mesh. This is because the optimized mesh approximates better the metric of the function than the initial one, reducing the interpolation and approximation errors, see the reasoning of Section \ref{sec:error}.
	
	%\color{blue}In addition, the presented example shows the capability of curved elements to capture sharp curved transition regions with curved boundaries. We observe that, even if the straight-edged elements approximate the curved transition region, this is not sufficient. Only when we curve them, we reduce the interpolation and approximation error.
	\color{black}
	
	\subsection{Size-shape distortion minimization for quadratic approximation of the Poisson problem: anisotropic initial straight-sided mesh}\label{sec:pde}
	
	In the following example we apply the size-shape distortion minimization to the numerical approximation of a Poisson problem with an anisotropic curved quadratic mesh. Our purpose is to illustrate the potential advantages of the distortion minimization procedure to anisotropic curved mesh adaptation. It is out of the scope of this example to compare the presented procedure with existing adaptation strategies or error estimators.
	
	We consider the Poisson problem with Neumann boundary conditions corresponding to the function $u$ with $\gamma := 20$, see Equation \eqref{eq:u}. Specifically, we consider the problem
	\begin{equation}\label{eq:Poisson}
		\left\lbrace 
		\begin{array}{cc}
			-\Delta u = f & \textrm{in}\quad \Omega\\
			\frac{\partial u}{\partial n} = g & \textrm{on}\quad \partial\Omega\\
		\end{array}\right.,
		\quad \textrm{with} \quad u\in V:= \left\lbrace v \in H^1(\Omega):\ \int_{\Omega} v = 0 \right\rbrace.
	\end{equation}
	%where the functions $f$ and $g$ correspond to $-\Delta u$ and $\frac{\partial u}{\partial n}$, respectively.
	To enforce the coercivity of the operator $-\Delta$ in $V$, we impose the orthogonality constraint against constant functions, that is,
	\begin{equation*}
		\int_{\Omega} v = 0\quad \textrm{for}\quad v\in V.
	\end{equation*}
	Consequently, there exists a unique solution $u$ for Equation \eqref{eq:Poisson}.
	Finally, given a numerical approximation $u_{\zmesh}$, we consider the numerical error $e_{\zmesh}$, against the exact solution $u$, as $e_{\zmesh} := \|u - u_{\zmesh}\|_{L^2(\Omega)}$.
	%u has zero integral because of symmetry.$
	%over the function space $V:= \left\lbrace u \in H^1(\Omega):\ \int_{\Omega} u = 0 \right\rbrace$.
	%Here, we respectively compute the functions $f$ and $g$ from $-\Delta u$ and $\frac{\partial u}{\partial n}$.
	%Note that, to ensure the coercivity of $-\Delta$ in $V$, the function space $V$ must be orthogonal to constant functions.
	
	%\color{red}
	For an input mesh $\zmesh$, we detail the adaptation procedure.
	First, we solve the Poisson problem of Equation \eqref{eq:Poisson}.
	In particular, we compute a numerical solution $u_{\zmesh}$ belonging to the finite element space $V_{\zmesh}$, a continuous piece-wise quadratic interpolative Galerkin approximation of $V$.
	Second, from this solution $u_{\zmesh}$, we compute an \emph{a posteriori} error estimator.
	Specifically, we estimate the third-order derivatives $\nabla^3 u_{\zmesh}$ by embedding the function $u_{\zmesh}$ into a piece-wise cubic polynomial space $\tilde{V}_{\zmesh}$ \cite{bank1993posteriori}.
	We denote the embedded function in $\tilde{V}_{\zmesh}$ as $\tilde{u}_{\zmesh}$.
	%Specifically, we compute the third-order derivatives of the shape functions from [?].
	%Note that, for a straight-edged mesh we do not need the higher-order derivatives of the mesh mappings $\nabla^n \phi$ ($n \geq 2$).
	Third, from the error estimator $\nabla^3 \tilde{u}_{\zmesh}$, we obtain a discrete metric $\zbmetric$, see \cite{coulaud:VeryHighOrderAnisotropic}.
	This provides the metric values at the element nodes.
	To assemble the values at the mesh nodes, we average the values at a single node from a log-Euclidean mean of adjacent elements.
	Finally, from the input mesh $\zmesh$ and the metric $\zbmetric$, we obtain an output mesh $\zmesh^*$.
	In our case, this can be an adapted straight-edged mesh or an optimized curved mesh.
	
	%Solve finite element problem with quadratic elements in a finite element space with zero mean $V$, $u_h$ numerical solution, compute \emph{a posteriori} third-order error estimator $\nabla^3 u_h$ \cite{bank1993posteriori}, obtain its third order derivatives from the shape-functions' derivatives (since the initial mesh is straight-edged, no mapping derivatives $> 2$ are required), minimize log-simplex problem from the third-order tensor to obtain a discrete metric, optimize mesh.
	
	%We apply the method presented above.
	%First, we generate an isotropic quadratic mesh of size $h = 0.075$ and composed of 917 nodes and 428 elements.
	%Second, after 3 iterations of the adaptation procedure presented before, we generate an adapted straight-edged mesh of size $h$ and composed of 839 nodes and 400 elements.
	%Third, we consider this adapted mesh as the initial mesh $\zmesh$ and as the background mesh $\zbmesh = \zmesh$.
	%In addition, we compute the discrete metric $\zbmetric$ corresponding to the \emph{a posteriori} error estimator of the Poisson problem.
	%Finally, we optimize the straight-edged mesh $\zmesh$ to obtain a curved mesh $\zmesh^*$.
	
	We apply the method presented above.
	First, we consider an initial adapted straight-edged mesh $\zmesh$ of size $h = 0.075$ and composed of 863 nodes and 414 elements.
	Second, we consider a linear background mesh $\zbmesh$, obtained from a midpoint subdivision of the physical mesh $\zmesh$.
	Third, we compute the discrete metric $\zbmetric$, corresponding to the \emph{a posteriori} error estimator of the Poisson problem.
	Finally, we optimize the straight-edged mesh $\zmesh$, according to the discrete metric $\zbmetric$, to obtain a curved mesh $\zmesh^*$.
	% linear interpolation of subdivided mesh
	\begin{figure}[t!]
		\centering
		%	\hspace{0.35cm}
		\setlength{\tabcolsep}{25pt}
		\begin{tabular}{cc}
			\subfigure[]{\label{fig:poisson0}
				\includegraphics[width=0.35\textwidth]{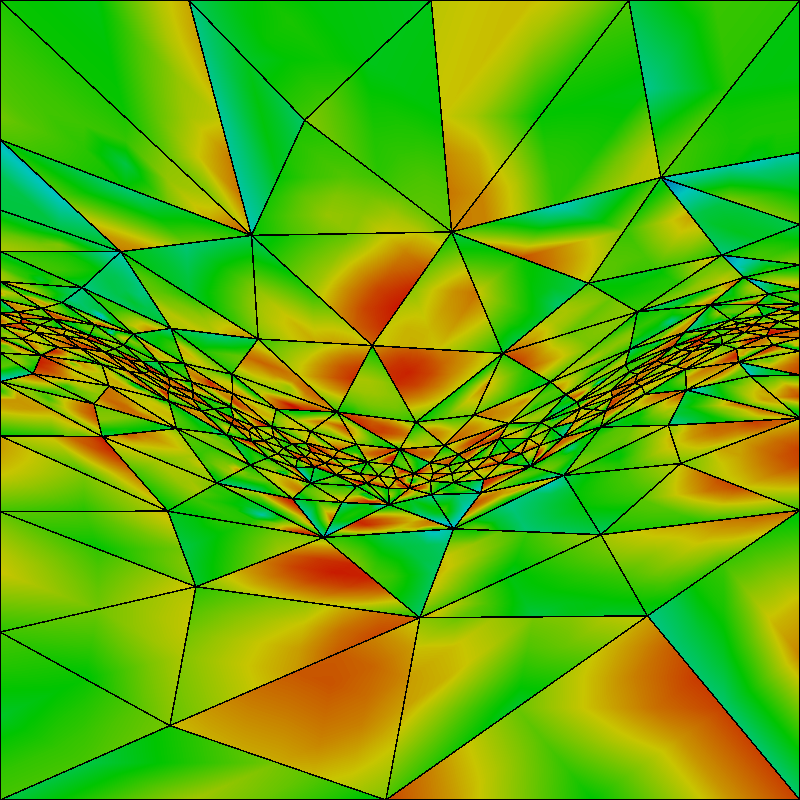}}
			&
			\subfigure[]{\label{fig:poisson1}
				\includegraphics[width=0.35\textwidth]{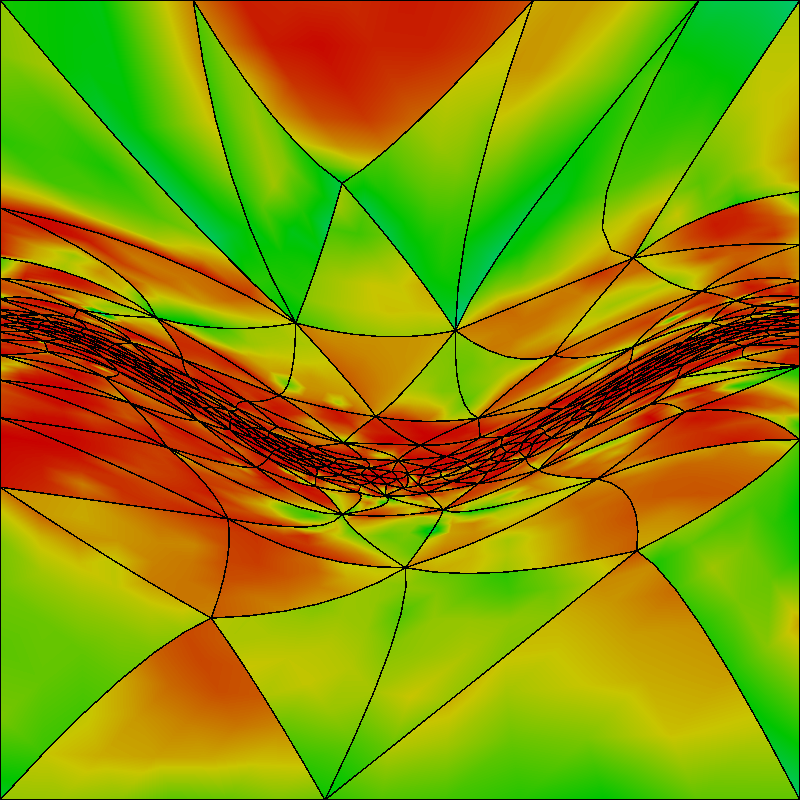}}
		\end{tabular}
		\\	
		\includegraphics[width=0.3\textwidth]{./qualBarParaview_color}
		\\
		%	\vspace{1cm}
		\begin{tabular}{cc}
			\subfigure[]{\label{fig:epoisson0}
				\includegraphics[width=0.35\textwidth]{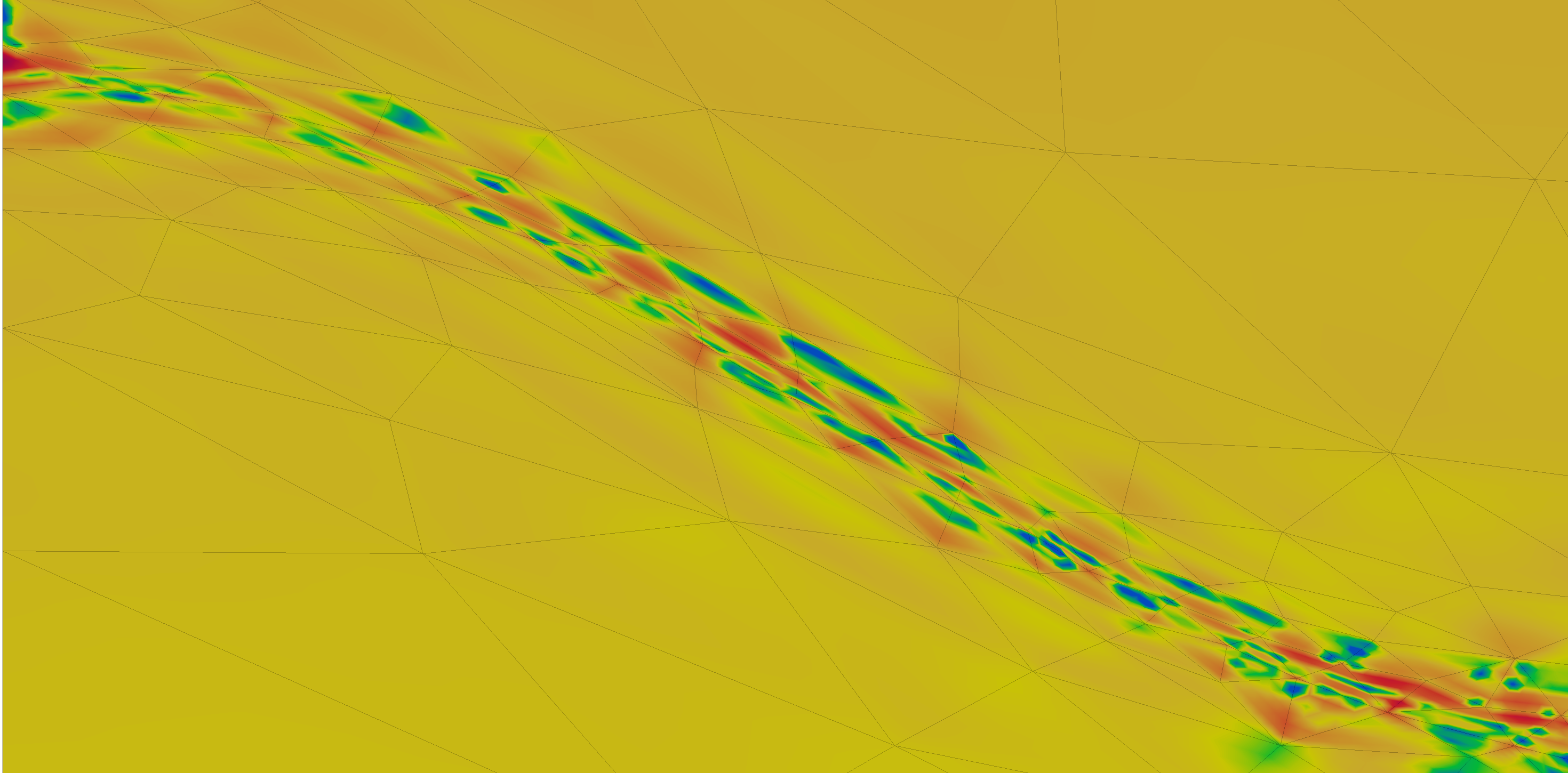}}
			&
			\subfigure[]{\label{fig:epoisson1}
				\includegraphics[width=0.35\textwidth]{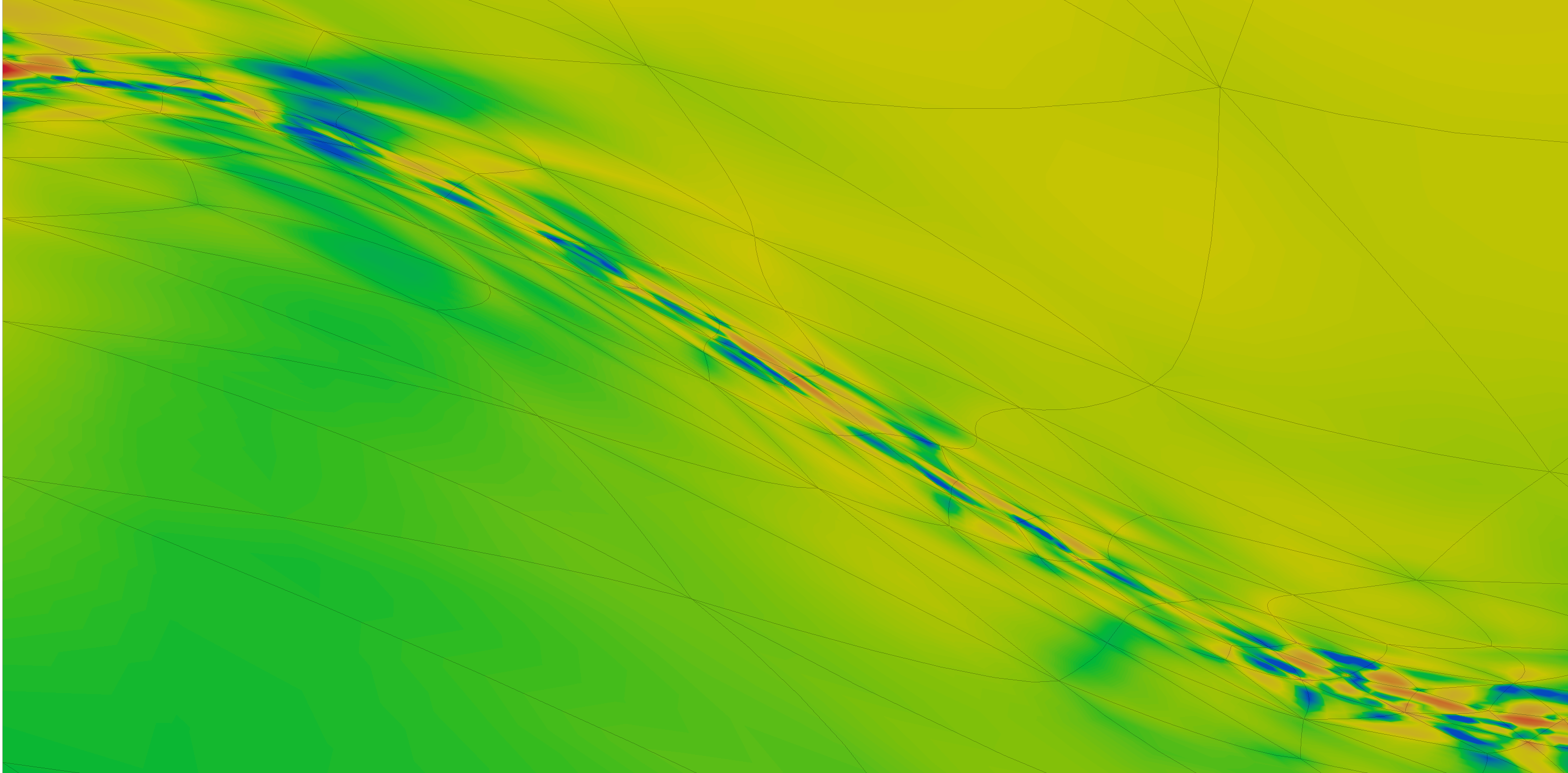}}
		\end{tabular}
		\\
		\includegraphics[width=0.7\textwidth]{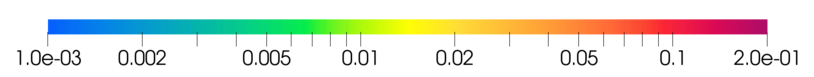}
		\caption{(Top) point-wise size-shape quality measure and (bottom) logarithmic absolute error between analytic and numerical solutions of the Poisson problem for initial adapted straight-edged and optimized quadratic meshes.}
		%	\caption{Initial adapted straight-edged and optimized quadratic meshes colored with the point-wise size-shape quality measure.}
		\label{fig:poisson}
	\end{figure}
	
	\begin{table}[t!]
		\caption{Size-shape quality and geometry statistics of the initial adapted straight-edged and optimized quadratic meshes.}
		\label{table:poisson}
		\centering
		\tiny
		\begin{tabular}{ c c c c c c c c c c}
			\hline\noalign{\smallskip}
			Measure & \multicolumn{2}{c}{Minimum}&\multicolumn{2}{c}{Maximum}& \multicolumn{2}{c}{Mean} & \multicolumn{2}{c}{Standard deviation}\\
			&Initial&Optimized&Initial&Optimized&Initial&Optimized&Initial&Optimized\\
			\noalign{\smallskip}\hline\noalign{\smallskip}
			Quality&0.3409&0.5399&0.9274&0.9900&0.6727&0.9145&0.1266&0.0773\\
			Length&0.3927&0.4047&2.2340&2.0123&1.0819&1.0248&0.3268&0.1885\\
			Area&0.2548&0.5591&2.1157&1.5185&0.8594&0.8595&0.3303&0.1385\\
			\noalign{\smallskip}\hline\noalign{\smallskip}
		\end{tabular}
	\end{table}
	\begin{figure}[t!]
		\centering
%		\hspace{-2.cm}
		\setlength{\tabcolsep}{-5pt}
		\begin{tabular}{cc}
			\subfigure[]{\label{fig:lpcp}
				\includegraphics[width=0.5\textwidth]{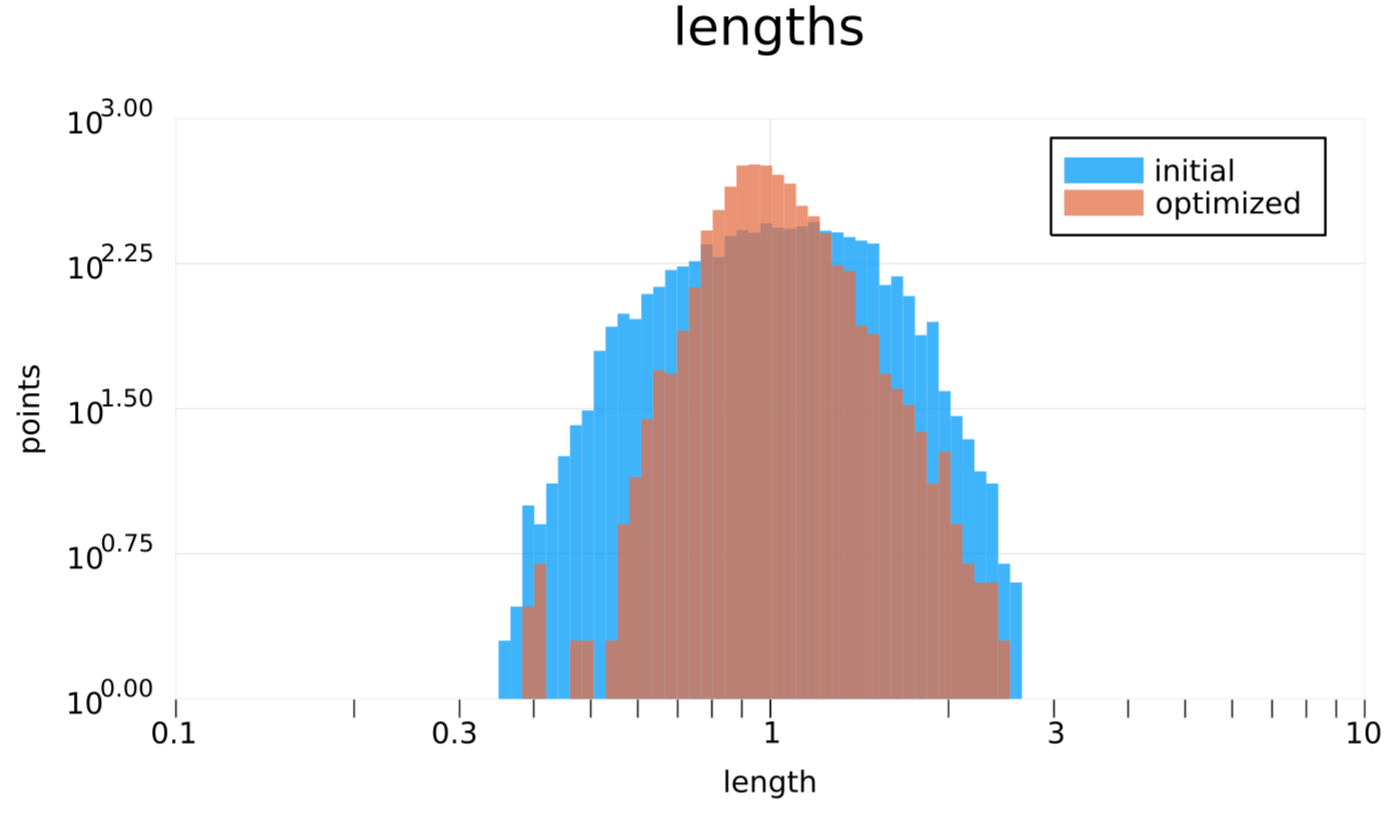}}
			&
			\subfigure[]{\label{fig:apcp}
				\includegraphics[width=0.5\textwidth]{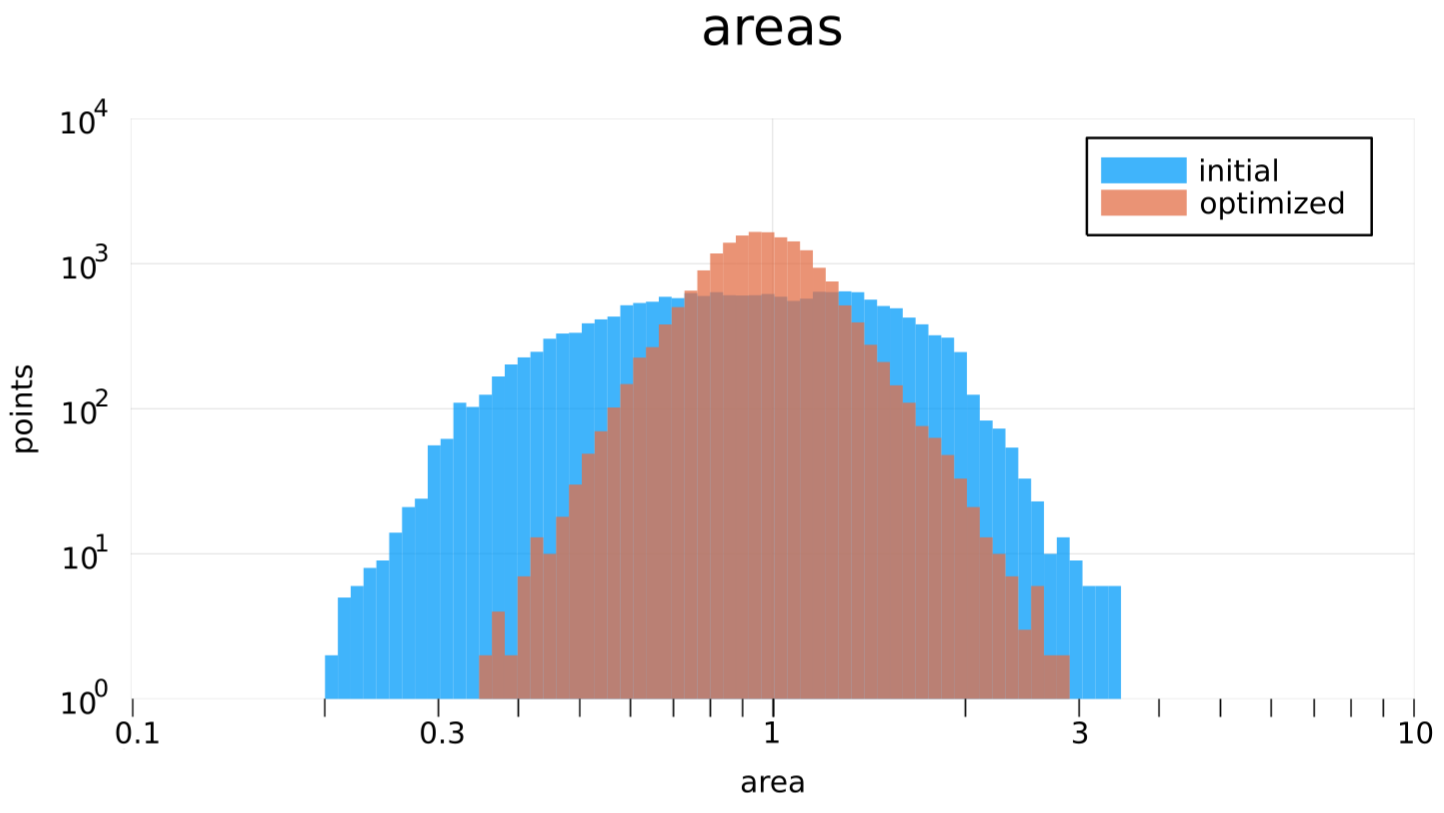}}
			\\
			%		\subfigure[]{\label{fig:errorcp}
				%			\includegraphics[width=0.75\textwidth]{./Poisson/numericalerror_modified.png}}
		\end{tabular}
		%	\caption{Logarithmic point-wise Riemannian length and area, and element-wise error histograms for (blue) initial and (orange) optimized quadratic triangular meshes.}
		\caption{Logarithmic Riemannian length and area density histograms for (blue) initial and (orange) optimized quadratic triangular meshes.}
		\label{fig:histpoisson}
	\end{figure}
	We show the physical meshes in Figure \ref{fig:poisson}, where they are colored according to the point-wise size-shape quality measure of Equation \eqref{eq:pointquality}.
	We illustrate the initial anisotropic mesh in Figure \ref{fig:poisson}\subref{fig:poisson0}. We observe that almost all elements are of medium quality. For this reason, in Figure \ref{fig:poisson}\subref{fig:epoisson0}, we observe several oscillations of the numerical error at the curved transition. The corresponding optimized mesh is shown in Figure \ref{fig:poisson}\subref{fig:poisson1}. We observe that, the elements are accumulated and match the metric stretching, alignment, and sizing at the sharp transition region. Accordingly, in Figure \ref{fig:poisson}\subref{fig:epoisson1}, we observe mitigated oscillations of the numerical error at the curved transition, when compared with the initial straight-sided mesh.
	Finally, we illustrate the metric matching improvement from the measure statistics in Table \ref{table:poisson} and from the density histograms of Figures \ref{fig:histpoisson}\subref{fig:lpcp} and \ref{fig:histpoisson}\subref{fig:apcp}, see the reasoning in Section \ref{sec:volumetric}.
	
	%In addition, we illustrate the numerical error improvement from the histogram of Figure \ref{fig:histpoisson}\subref{fig:errorcp}, see the reasoning of Section \ref{sec:error}.
	%In Table \ref{table:poisson}, we illustrate the size-shape quality, and Riemannian length and area measure statistics.
	%In addition, Figure \ref{fig:histpoisson} shows the point-wise Riemannian length and area histograms for the initial and optimized triangular quadratic meshes.
	%The statistics and histograms illustrate an improved metric matching of the optimized mesh versus the initial one, see Section \ref{sec:volumetric}.
	
	%\color{green}In Table \ref{table:poisson}, we show the statistics for elemental qualities (Equation \eqref{eq:qualityreg}) and Riemannian lengths and areas. The table allow us to compare between the initial and optimized meshes in terms of the target metric. We observe that the maximum, minimum, mean, and standard deviation become closer to unit values in almost all cases. That is, in general, all statistics are improved.
	
	%\color{orange}In Figure \ref{fig:histpoisson}, we show the point-wise Riemannian length and area of the initial and optimized triangular quadratic meshes. As in Section \ref{sec:volumetric}, we use a logarithmic scale to illustrate the different scales of the corresponding ratios. The logarithmic representation illustrates the behavior near the minimum, maximum, and geometric mean of the distribution.
	
	%mean initial: 1e-6, optimized: 3.81e-7
	
	\begin{table}[t!]
		\caption{Interpolation, approximation, and numerical $L^2$ errors of the initial adapted straight-edged and optimized quadratic meshes.}
		\label{table:interpolationpoisson}
		\centering
		\begin{tabular}{c c c c}
			\hline\noalign{\smallskip}
			Mesh&Interpolation error&Approximation error&Numerical error\\
			\noalign{\smallskip}\hline\noalign{\smallskip}
			Initial&0.0049&0.0035&0.0204\\
			Optimized&0.0023&0.0017&0.0126\\
			\noalign{\smallskip}\hline\noalign{\smallskip}
		\end{tabular}
	\end{table}
	\begin{figure}[t!]
		\centering
		\hspace{-0.5cm}
		\includegraphics[width=1.0\textwidth]{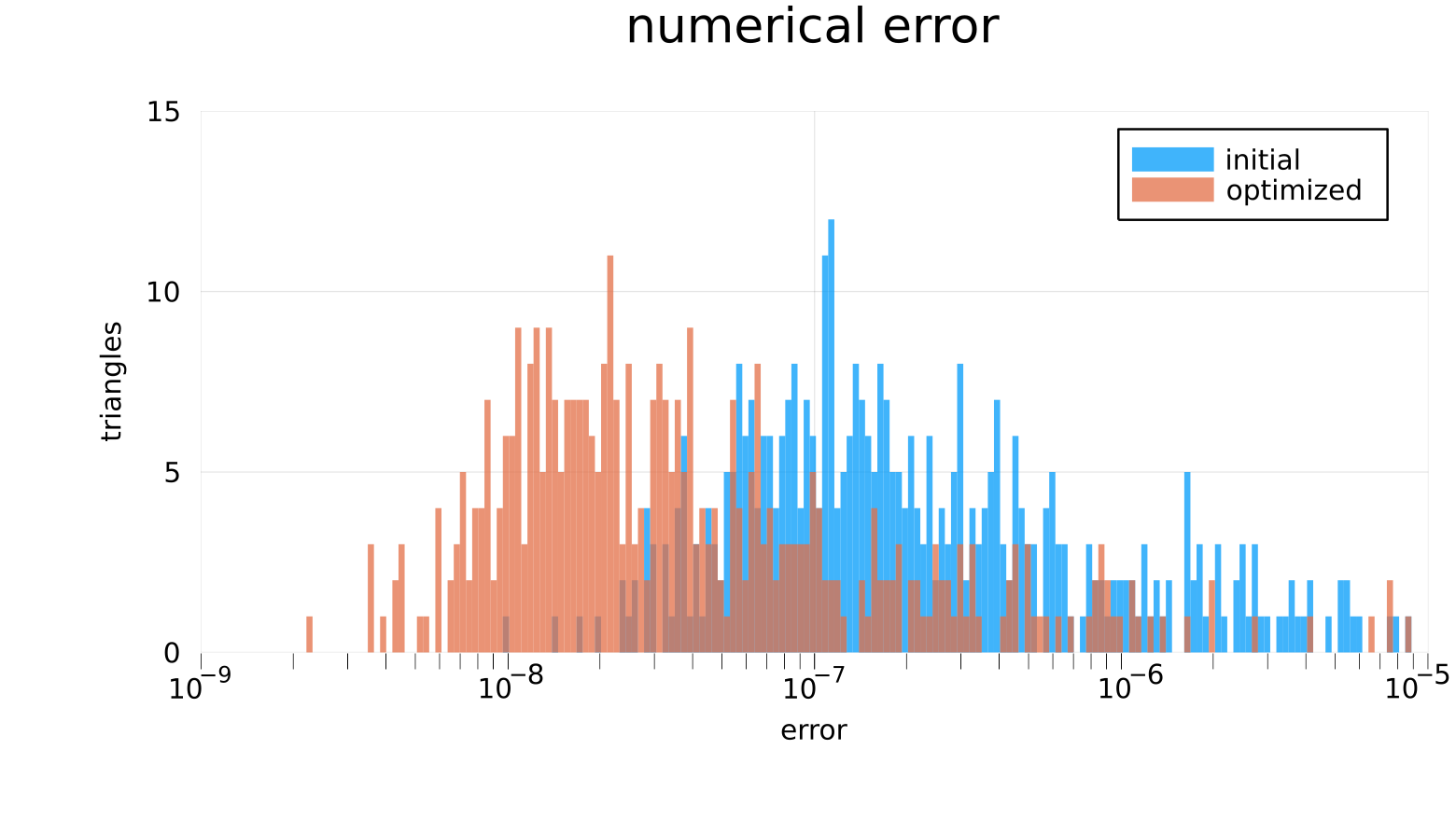}
		\caption{Logarithmic element-wise error histogram for (blue) initial and (orange) optimized quadratic triangular meshes.}
		\label{fig:histerrpoisson}
	\end{figure}
	The presented example shows how our method can be used to improve the error of a straight-edged mesh according to an input model. On the one hand, in Table \ref{table:interpolationpoisson}, we present the global interpolation, approximation, and numerical error of the initial and optimized meshes. We observe that the errors are improved almost two times for the optimized mesh. This is because the output mesh is optimized according to a metric accounting for an error estimator, reducing the numerical error of the solution.
	For the interpolation and approximation errors see the reasoning of Section \ref{sec:error}.
	On the other hand, to understand the distribution behavior, we illustrate the element-wise error in Figure \ref{fig:histerrpoisson}.
	We observe that the numerical error is shifted to the left, indicating a reduction of almost one order of magnitude for a considerable amount of elements. This shows the potential of curved $r$-adaptation to improve the accuracy of a numerical simulation.

	%We define the numerical $L^2$-error as $e_N := \|u - u_h\|_{L^2(\Omega)}$.
	%From Céa's lemma we have $e_N \leq \frac{C}{\alpha} e_A = \frac{C}{\lambda_1} e_A = \frac{C}{\pi^2} e_A$.
	
	%\color{teal}The presented example shows how our method can be used to improve the error of a straight-edged mesh according to an input model. In Table \ref{table:interpolationpoisson}, we present the global interpolation, approximation, and numerical error of the initial and optimized meshes. We observe that the errors are improved almost two times for the optimized mesh. This is because the output mesh is optimized according to a metric accounting for an error estimator, reducing the numerical error of the solution.
	
	%\color{blue}In addition, the presented example shows the capability of curved elements to capture sharp curved transition regions. We observe that, even if the straight-edged elements approximate the curved transition region, this is not sufficient. Only when we curve them, we reduce the interpolation, approximation, and numerical error.
	
	% see also Section \ref{sec:error}.
	%The global error illustrated in Table \ref{table:interpolationpoisson} is not sufficient.
	%Figure \ref{fig:histerrpoisson} illustrates the element-wise error distribution.
	%We observe that the numerical error is reduced by one order of magnitude for a considerable amount of elements.
	
	\section{Concluding remarks}\label{sec:conclusions}
	Next, we present the concluding remarks of this work. First, in Section \ref{sec:discussion}, we outline the main discussions. Second, in Section \ref{sec:futureWork}, we present an outline of the work that we have planned for the near future. Finally, in Section \ref{sec:subconc}, we present the main conclusions.
	
	\subsection{Discussion}\label{sec:discussion}
	Next, we include discussions corresponding to the comparison of our results with standard mesh curving, the challenges of the minimization, and the requisites to obtain satisfactory results.
	
	\paragraph{Adaptive versus standard mesh curving} Our adaptive results on solution accuracy are compatible with standard non-adaptive results. For meshes adapted to a target, we have demonstrated the potential of curved elements to improve solution accuracy. On the contrary, for meshes not adapted to a target, it is known that curved elements might be detrimental to solution accuracy \cite{engvall2020mesh}. Although these results on solution accuracy might seem contradictory, both results are compatible because they apply to different situations. In our approach, the mesh is adapted to a target, while on standard mesh curving, the mesh is not adapted to a target.
	
	\paragraph{Minimization challenges} Although the goal of this paper is to propose a new size-shape distortion measure, to illustrate its applicability we need to implement the distortion minimization procedure. This implementation presents several challenges we address in previous works. First, to reduce the computational cost, we use a specific-purpose nonlinear optimization solver \cite{aparicio2020severoochoa}. Second, to enhance convergence, we use a specific-purpose backtracking line-search procedure \cite{aparicio2020severoochoa}. Third, to deal with curved geometry and tangential node motion, we use an implicit geometry modeling approach \cite{aparicio2023combining}.
	
	\paragraph{Requisites for satisfactory results} To obtain satisfactory results minimizing the size-shape distortion, we need the following requisites. First, the initial mesh must feature enough resolution and anisotropy to capture the sharp features of the target metric. Furthermore, for metrics featuring sharp variations, we must use sufficient quadrature points. In both cases, it is important to use an optimization solver that robustly converges to a local minimum because higher orders and sharp metric transitions stiff the optimization problem.
	
	\subsection{Future work}\label{sec:futureWork}
	
%	The presented size-shape distortion minimization method has some advantages and limitations not explored in this work.
	
	In the near future, to improve the applicability of our method, we have planned to perform the following work regarding topology modifications, curved surfaces, worst distortion, and large-scale optimization.
	
	\paragraph{Topology modifications} To show improvements by optimizing our size-shape distortion, we have used a fixed mesh connectivity. Nevertheless, we think it would lead to further improvements if we would use topological mesh modifications. To this end, we have planned to incorporate our approach within a local cavity framework for mesh optimization.
	
	\paragraph{Curved surfaces} We have developed our size-shape distortion measure for meshes equipped with a metric and with a number of dimensions equal to the spatial dimensions. Nevertheless, our size-shape distortion measure might be extended to deal with curved surfaces embedded in a three-dimensional space. To this end, because for surfaces the Jacobian matrices of the element mappings are rectangular, we have planned to substitute the determinants of square matrices in Equation \eqref{eq:vol}, by the square root of the determinant of the transposed Jacobian times the Jacobian.
%	First, the method could be applied to mesh optimization on curved surfaces. This is because when the mapping between elements $E^\triangle$ and $E^P$ is an embedding, the distortion remains unaffected and the product involved in $\zSmetric$, presented in Equations \eqref{eq:frob&size} and \eqref{eq:vol}, remains valid and with the analogous geometric interpretation. However, since we cannot account for inverted embedded elements, the determinant of the distortion measure should be computed as in Equation \eqref{eq:frob&size} (in the pointwise sense) rather than as in Equation \eqref{eq:vol}.
	
	\paragraph{Worst distortion} To show improvements in the interpolation and approximation error, we have not needed to minimize the worst distortion. Nevertheless, in some cases, it might be interesting to optimize the worst distortion. Hence, we have planned to numerically approximate the supreme norm with an $L^p$-norm with a value of $p$ significantly greater than two.
	
	\paragraph{Large-scale optimization} In this work, we aim to propose a size-shape distortion measure and illustrate its applicability, but we do not seek to show large-scale applications. Nevertheless, for large-scale metric-aware applications, we have planned to modify our previously developed large-scale approaches for Euclidean curved mesh optimization \cite{ruiz2022automatic}.
	
	\subsection{Conclusions}\label{sec:subconc}
	
	The defined distortion measure is applied to curve straight-edged meshes to improve the node configuration according to the desired metric. To perform the distortion minimization we use the framework for high-order optimization presented in \cite{aparicio2019imr}. The numerical examples show optimized meshes with an improved stretching, alignment, and sizing according to the metric. This improvement leads in all cases to an increase of the minimum element mesh quality and a reduction of the standard deviation between the different element qualities.
	
	To independently measure whether the optimized mesh matches the input metric, we propose point-wise Riemannian densities and measures of the mesh entities equipped with the metric. These are the Riemannian edge length, surface area, and cell volume. The results show that the optimized meshes improve the length, area, and volume distributions in the metric sense. This illustrates that the distortion minimization enables meshes that effectively match the input metric.
	
	To illustrate the potential applications of the method, we also measure the numerical error for an input function. These are the interpolation and approximation errors of the function matched by the mesh. The results show that the optimized meshes reduce both the interpolation and approximation errors. Moreover, our particular example illustrates that the distortion minimization reduces the numerical errors by one order of magnitude for an initial adapted mesh and by two orders of magnitude for an initial isotropic mesh.
	In addition, we apply the distortion minimization for domains with curved boundaries.
	The results show that the mesh approximates the stretching, alignment, and sizing of the discrete metric while preserving the curved features of the boundary model.
	
	From the results, depending on the application, practitioners should choose between shape and size-shape distortions. In those applications where it is important to match not only the stretching and alignment but also the sizing determined by the metric, we should favor using a size-shape distortion. On the contrary, if only the stretching and alignment are relevant, we might prefer using a shape measure.
	
	For many results, we infer that for targets featuring curved features, curved meshes adapt to the target better than straight-edged meshes. In contrast, we should also infer that on those applications where the target metric presents ruled features, adapted straight-edged meshes might be more efficient than adapted curved meshes. Nevertheless, both for straight-edged and curved meshes, the optimization of the proposed size-shape distortion leads to meshes adapted better to the target metric.
	
	Our long term goal is to consider the distortion minimization for functions obtained from a numerical solution of a flow problem.
	Accordingly, we have presented preliminary results regarding the solution of a Poisson problem.
	In addition, to enable a fully adapted mesh, we would like to couple the distortion minimization procedure ($r$-adaptation) with topological mesh modification methods ($h$-adaptation).
	\section*{Acknowledgements}
	This project has received funding from the European Research Council (ERC) under the European Union's Horizon 2020 research and innovation programme under grant agreement No 715546. This work has also received funding from the Generalitat de Catalunya under grant number 2017 SGR 1731. The work of X. Roca has been partially supported by the Spanish Ministerio de Econom\'ia y Competitividad under the personal grant agreement RYC-2015-01633.
	\bibliography{references}
\end{document}

%% file: definitions.tex
\newcommand{\zR}{\mathds{R}}
\newcommand{\zdim}{d}
\newcommand{\zmetric}{\textbf{M}}
\newcommand{\zbmetric}{\hat{\textbf{M}}}

\newcommand{\zfield}{\textbf{F}}
\newcommand{\ztr}{\mathrm{tr}}
\newcommand{\zmaster}{E^M}

\newcommand{\zequilater}{E^{\triangle}}

\newcommand{\zideal}{E^I}

\newcommand{\zphysical}{E^P}

\newcommand{\zequilaterphysicalmap}{\zphi_E}
\newcommand{\zidealphysicalmap}{\zphi_E}
\newcommand{\zequilatermap}{\zphi_{\triangle}}
\newcommand{\zidealmap}{\zphi_I}
\newcommand{\zphysicalmap}{\zphi_P}
\newcommand{\zisotropicphysical}{E^{P_\triangle}}

\newcommand{\zJacobianisotropicphysical}{\textbf{D}\zphi_{P_{\triangle}}}
\newcommand{\zequilaterisotropicphysicalmap}{\zphi_{U}}

\newcommand{\zJacobianequilaterisotropicphysical}{\textbf{D}\zequilaterisotropicphysicalmap}

\newcommand{\zJacobianidealphysical}{\textbf{D}\zidealphysicalmap}
\newcommand{\zJacobianequilater}{\textbf{D}\zequilatermap}
\newcommand{\zJacobianideal}{\textbf{D}\zidealmap}
\newcommand{\zJacobianphysical}{\textbf{D}\zphysicalmap}

\newcommand{\zconjugate}{\left(\zJacobianphysical\ \zJacobianequilater^{-1}\right)^\mathrm{T}\ \zmetric\ \zJacobianphysical\ \zJacobianequilater^{-1}}
\newcommand{\zconjugatexi}{\textbf{A}(\zxi)^\mathrm{T}\ \zmetric(\zphysicalmap(\zxi))\ \textbf{A}(\zxi)}
\newcommand{\zrotation}{\textbf{R}(\theta)}

\newcommand{\zu}{\textbf{u}}

\newcommand{\zx}{\textbf{x}}
\newcommand{\zxi}{\boldsymbol{\xi}}
\newcommand{\zy}{\textbf{y}}

\newcommand{\zid}{\textbf{Id}}
\newcommand{\zphi}{\phi}

\newcommand{\zmesh}{\mathcal{M}}
\newcommand{\zbmesh}{\hat{\mathcal{M}}}

\newcommand{\zeuclideandistortionoperator}{\mathcal{N}\zequilaterisotropicphysicalmap}
\newcommand{\zeuclideandistortionoperatorreg}{\mathcal{N}_0\zequilaterisotropicphysicalmap}
\newcommand{\zeuclideanqualityoperatorreg}{\mathcal{Q}_0\zequilaterisotropicphysicalmap}
\newcommand{\zdistortionoperator}{\mathcal{N}\zequilaterphysicalmap}

\newcommand{\zshape}{\eta_{\mathrm{shape}}}
\newcommand{\zsize}{\eta_{\mathrm{size}}}

\newcommand{\zSmetric}{\textbf{S}_{\zmetric}}
\newcommand{\zsigmametric}{\sigma_{\zmetric}}